%% file: TPAMI-2025-03-0755.R2_Wang.tex
\documentclass[lettersize,journal]{IEEEtran}

\usepackage{amsthm}
\newtheorem{defn}{\textbf{Definition}}
\newtheorem*{prob_state}{\textbf{Problem Statement}}

\newtheorem{claim}{\textbf{Claim}}

\usepackage{color}
\usepackage{graphicx}
\usepackage{subfig}
\usepackage{amsmath}
\usepackage{amssymb}
\usepackage{mathrsfs}
\usepackage[ruled, vlined, linesnumbered]{algorithm2e}
\usepackage[hidelinks]{hyperref}
\usepackage{cite}
\usepackage{makecell}
\usepackage[numbers]{natbib}
\usepackage{array}
\usepackage{tikz}
\usetikzlibrary{trees}
\usepackage{tabularx,colortbl}
\usepackage{threeparttable}
\usepackage{booktabs}
\usepackage{multirow}
\usepackage{longtable}
\usepackage{rotating}
\usepackage[capitalize,noabbrev]{cleveref}
\crefname{equation}{Eq.}{Eqs.}
\usepackage{ragged2e}
\usepackage{braket}

\usepackage{adjustbox}
\usepackage{enumitem}
\usepackage{tcolorbox}
\usepackage{mdframed}
\usepackage{fancybox}

\usepackage{floatrow}
\usepackage{diagbox}

\usepackage{soul}

\newcommand{\meanstd}[2]{#1{\tiny$\pm$#2}}

\definecolor{DeepPink}{HTML}{FF1493}
\definecolor{Orchid}{HTML}{DA70D6}
\definecolor{Magenta}{HTML}{FF00FF}
\definecolor{Fuchsia}{HTML}{FF00FF}
\definecolor{LavenderPink}{HTML}{FFB6C1}
\definecolor{verylightgray}{rgb}{0.9, 0.9, 0.9}
\definecolor{lightred}{rgb}{1,0.8,0.8}

\begin{document}


\title{BlindU: Blind Machine Unlearning without Revealing Erasing Data}

\author{Weiqi~Wang,~\IEEEmembership{Member,~IEEE},
	Zhiyi~Tian,~\IEEEmembership{Member,~IEEE},
	Chenhan~Zhang,~\IEEEmembership{Member,~IEEE},\\
	and Shui~Yu,~\IEEEmembership{Fellow,~IEEE}

	\thanks{This paper is partially supported by Australia ARC LP220100453 and ARC DP240100955. The authors would like to thank the anonymous reviewers for their valuable comments and review. \emph{(Corresponding author: Zhiyi Tian.)} }
	\IEEEcompsocitemizethanks{\IEEEcompsocthanksitem W. Wang, C. Zhang, and S. Yu are with the School of Computer Science, University of Technology Sydney, Australia.
		\IEEEcompsocthanksitem Z. Tian is with the School of Cyber Science and Engineering, Southeast University, China.\protect\\
		E-mail: {\{weiqi.wang, shui.yu\}@uts.edu.au}, \{zhiyi.tian, chzhang\}@ieee.org
	}
}

\markboth{This paper has been accepted for publication in IEEE Transactions on Pattern Analysis and Machine Intelligence}%
{Shell \MakeLowercase{\textit{et al.}}: A Sample Article Using IEEEtran.cls for IEEE Journals}


\maketitle

\input{Contents/0_abstract}

\input{Contents/1_intro}

\input{Contents/2_background}

\input{Contents/4_approach}

\input{Contents/5_experiments}

\input{Contents/6_summary}

\footnotesize
\bibliographystyle{IEEEtranN}
\bibliography{MCFU.bib}

\begin{IEEEbiography}[{\includegraphics[width=1in,height=1.25in,clip,keepaspectratio]{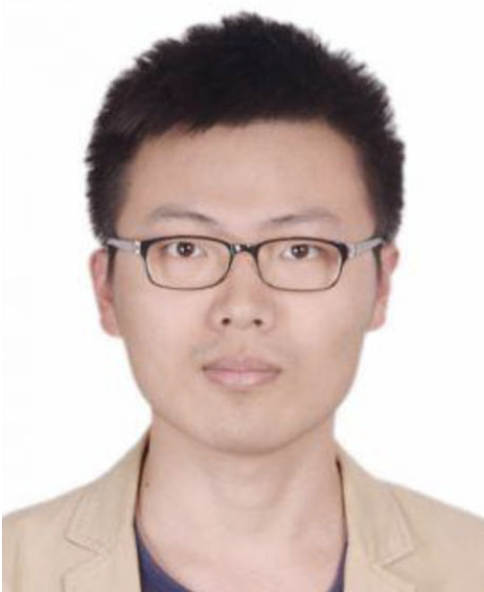}}]{Weiqi Wang}(IEEE M'24)
	received his Ph.D. degree from University of Technology Sydney, Australia, in 2024. He currently is a postdoctoral research associate of University of Technology Sydney, Australia, advised by Prof. Shui Yu. He previously worked as a senior algorithm engineer at the Department of AI-Strategy, Local consumer services segment, Alibaba Group. He has been actively involved in the research community by serving as a reviewer for prestige journals such as ACM Computing Surveys, IEEE Communications Surveys and Tutorials, IEEE TIFS, IEEE TDSC, IEEE TIP, IEEE TMC, IEEE Transactions on SMC, and IEEE IOTJ, and international conferences such as CVPR, WWW, ICLR, IEEE ICC, and IEEE GLOBECOM. His research interests are the security and privacy in machine learning.
\end{IEEEbiography}


\begin{IEEEbiography}[{\includegraphics[width=1in,height=1.25in,clip,keepaspectratio]{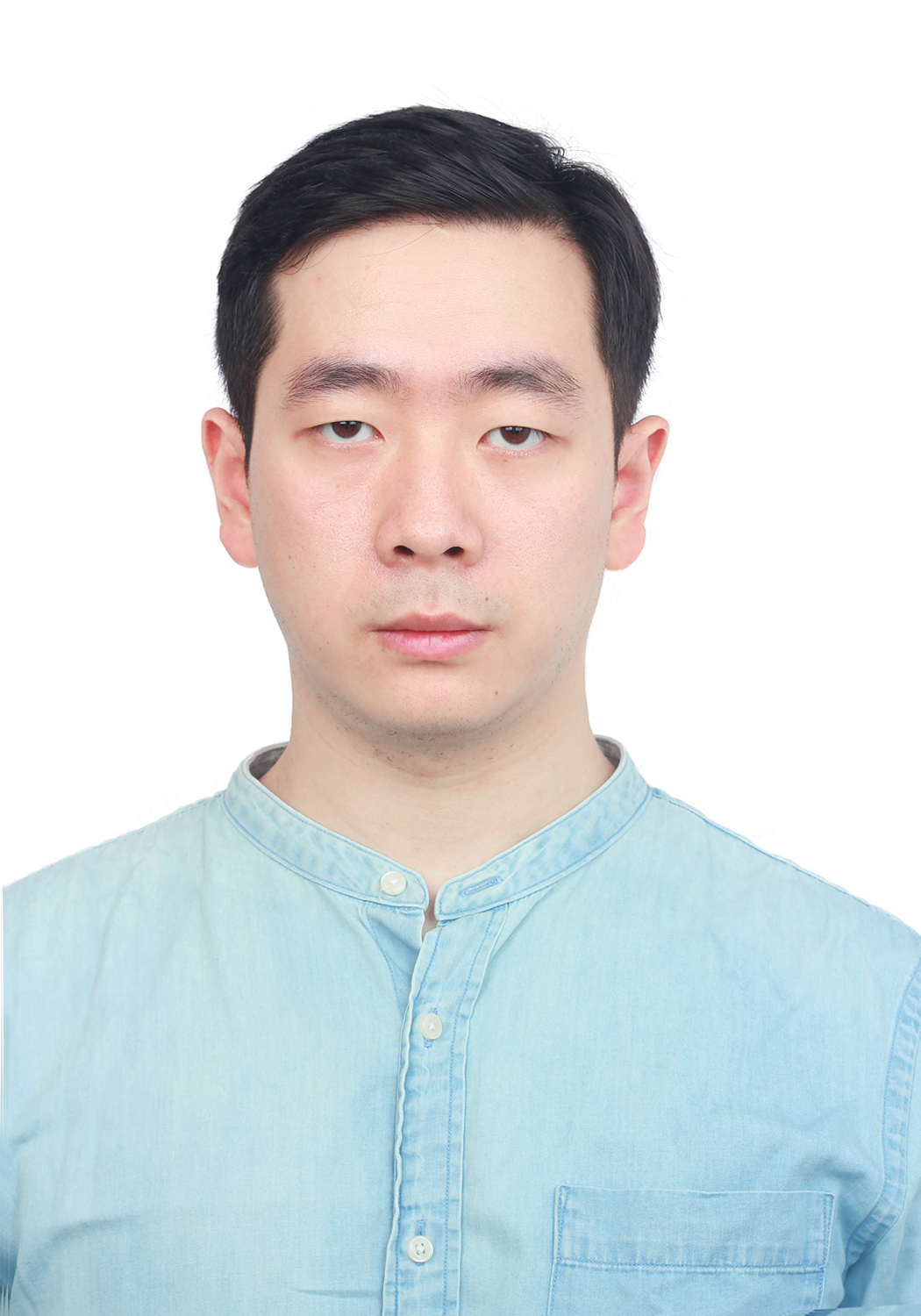}}]{Zhiyi Tian} (IEEE M'24) received the B.S. degree and the M.S. degree from Sichuan University, China, in 2017 and 2020, respectively. He obtained his Ph.D. degree in 2024 from University of Technology Sydney, Australia, where he was advised by Prof. Shui Yu. He currently is a research associate of Southeast University. His research interests include security and privacy in deep learning, semantic communications. He has been actively involved in the research community by serving as a reviewer for prestige journals, such as ACM Computing Surveys, IEEE Communications Surveys and Tutorials, IEEE TIFS, IEEE TDSC; and international conferences, such as IEEE ICC and IEEE GLOBECOM.
\end{IEEEbiography}

\begin{IEEEbiography}[{\includegraphics[width=1in,height=1.25in,clip,keepaspectratio]{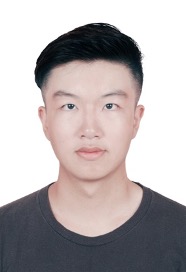}}]{Chenhan Zhang} (IEEE S'19 M'24) obtained his Ph.D. from University of Technology Sydney, Australia, in 2024, where he was advised by Prof. Shui Yu. Before that, he obtained his B.Eng. (Honours) from University of Wollongong, Australia, and  M.S. from City University of Hong Kong, Hong Kong, in 2017 and 2019, respectively. He is currently a postdoctoral research fellow at University of Technology Sydney. His research interests include security and privacy in graph neural networks and trustworthy spatiotemporal cyber physical systems. He has been actively involved in the research community by serving as a reviewer for prestige venues such as ICLR, IJCAI, INFOCOM, IEEE TDSC, IEEE IoTJ, ACM Computing Survey, and IEEE Communications Surveys and Tutorials.
\end{IEEEbiography}

\begin{IEEEbiography}[{\includegraphics[width=1in,height=1.25in,clip,keepaspectratio]{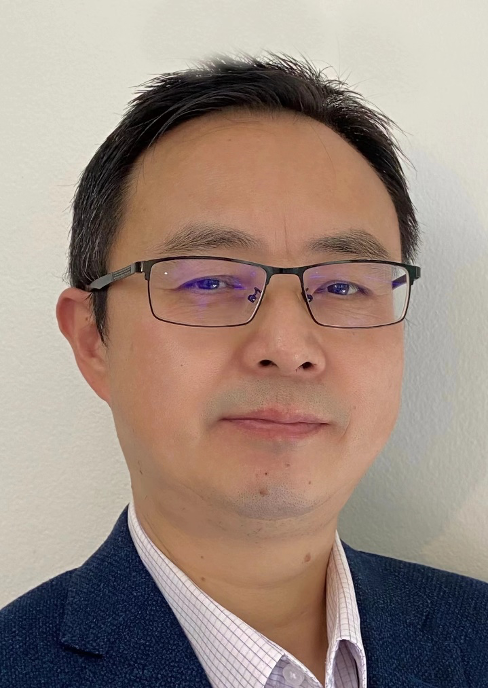}}]{Shui Yu} (IEEE F'23) obtained his PhD from Deakin University, Australia, in 2004. He is a Professor of School of Computer Science, Deputy Chair of University Research Committee, University of Technology Sydney, Australia. His research interest includes Mathematical AI, Cybersecurity, Network Science, and Big Data. He has published seven monographs and edited two books, more than 650 technical papers at different venues. His current h-index is 86. Professor Yu promoted the research field of networking for big data since 2013, and his research outputs have been widely adopted by industrial systems, such as Amazon cloud security. He is currently serving the editorial boards of IEEE Communications Surveys and Tutorials (Area Editor), IEEE Transactions on Cognitive Communications and Networking, and IEEE Transactions on Dependable and Secure Computing. He is a Distinguished Visitor of IEEE Computer Society, and an elected member of the Board of Governors of IEEE Communications Society. He is a member of ACM and AAAS, and a Fellow of IEEE.
\end{IEEEbiography}

\end{document}

%% file: Contents/0_abstract.tex
\begin{abstract}
	\justifying
Machine unlearning enables data holders to remove the contribution of their specified samples from trained models to protect their privacy. However, it is paradoxical that most unlearning methods require the unlearning requesters to firstly upload their data to the server as a prerequisite for unlearning. These methods are infeasible in many privacy-preserving scenarios where servers are prohibited from accessing users' data, such as federated learning (FL). In this paper, we explore how to implement unlearning under the condition of not uncovering the erasing data to the server. We propose \textbf{Blind Unlearning (BlindU)}, which carries out unlearning using compressed representations instead of original inputs. BlindU only involves the server and the unlearning user: the user locally generates privacy-preserving representations, and the server performs unlearning solely on these representations and their labels. For the FL model training, we employ the information bottleneck (IB) mechanism. The encoder of the IB-based FL model learns representations that distort maximum task-irrelevant information from inputs, allowing FL users to generate compressed representations locally. For effective unlearning using compressed representation, BlindU integrates two dedicated unlearning modules tailored explicitly for IB-based models and uses a multiple gradient descent algorithm to balance forgetting and utility retaining. While IB compression already provides protection for task-irrelevant information of inputs, to further enhance the privacy protection, we introduce a noise-free differential privacy (DP) masking method to deal with the raw erasing data before compressing. Theoretical analysis and extensive experimental results illustrate the superiority of BlindU in privacy protection and unlearning effectiveness compared with the best existing privacy-preserving unlearning benchmarks. 
\end{abstract}

%% file: Contents/1_intro.tex
\section{Introduction}
\IEEEPARstart{M}{achine} unlearning, as a solution to the ``right to be forgotten'', is proposed to protect users' privacy by allowing users to erase the contribution of specified data from trained machine learning (ML) models. When a user exercises this right~\cite{mantelero2013eu}, such as requesting the removal of sensitive data from an ML application, the server must ``unlearn'' his/her specified data from trained models~\cite{cao2015towards}. The user is referred to as the ``unlearning user'' or ``unlearning requester'' in the remainder of the paper. Paradoxically, while machine unlearning aims to protect user privacy, most unlearning methods require unlearning requesters to upload the erasing data to the ML server as a prerequisite for implementing unlearning~\cite{bourtoule2021machine,sekhari2021remember}. This operation limits its practical application because many users may be unwilling to disclose data due to privacy concerns. For example, in privacy-preserving ML scenarios, the servers collaborate with users to train models by using federated learning (FL)~\cite{sun2020lazily,sun2022decentralized} or secure multi-party computation (MPC)~\cite{mohassel2017secureml} mechanisms. These servers are strictly limited to accessing individual data. Exposing sensitive data to the server for unlearning can jeopardize the privacy of users.



There are a few studies exploring privacy-preserving machine unlearning approaches to protect erasing data privacy, primarily in the context of FL. These FL unlearning methods still utilize FL frameworks and protect the privacy of erasing data by uploading gradients instead of the original data~\cite{liu2021federaser,liu2022right,wang2023bfu,lin2024incentive,su2023asynchronous}. {However, gradients based on erasing data for unlearning present a greater risk of privacy breaches compared to those derived from the entire local dataset for FL \cite{salem2020updates,melis2019exploiting,boenisch2023curious}. This is because machine unlearning requests usually only contain a single sample or several samples, making the corresponding gradients more vulnerable to reconstruction attacks~\cite{salem2020updates,boenisch2023curious}.}

 
Considering this research gap, we propose the research question: \textit{``In privacy-sensitive scenarios such as FL, how can we achieve privacy-preserving machine unlearning with strong unlearning effectiveness under the strict constraint that erased data must not be disclosed to the server?''} The critical challenge lies in desensitizing the erased data for the server while maintaining information precision for unlearning.


Exploring how to utilize traditional encryption or differential privacy (DP) techniques as potential solutions for safeguarding erased data in unlearning is challenging. Encryption-based ML techniques are considered time-consuming and often impractical for most deep-learning scenarios~\cite{gilad2016cryptonets,xu2019cryptonn,kung2017collaborative}. Adding DP noise~\cite{arachchige2019local} will disturb the original data, making it difficult to determine whether the model unlearns the erased data or the noise.

\begin{figure}[t]
	\centering
	\includegraphics[width=0.95\linewidth]{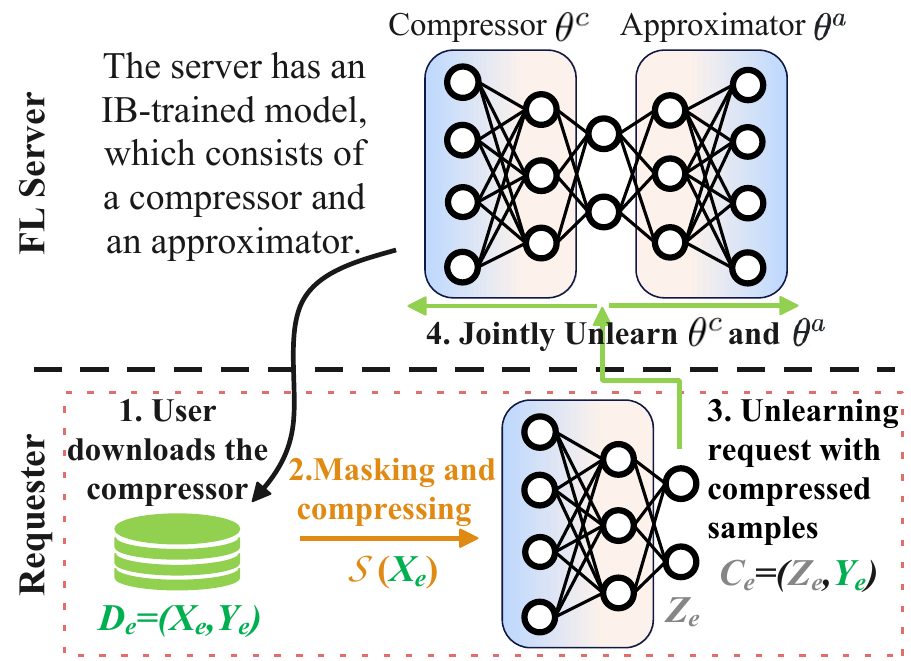}
	\vspace{-2mm}
	\caption{BlindU only involves the FL server and the unlearning user. The unlearning user first downloads the compressor from the server to deal with his/her erasing data $D_e$. Before compressing the erasing data, a DP masking method can be implemented through a sampling strategy $\mathcal{S}$. Then, the user uploads the masked and compressed data to the server. BlindU incorporates two dedicated unlearning modules that jointly unlearn the IB-based FL model using this compressed data.
	\vspace{-2mm}	
}
	\label{fig:funlandmcfu2}
\end{figure}

By contrast, we try to investigate privacy-preserving unlearning for models trained using an information bottleneck (IB) mechanism~\cite{tishby2000information,achille2018information}. The structure of an IB model is illustrated as the upper half in \Cref{fig:funlandmcfu2}, including a compressor and an approximator. The IB method is easy to implement in ML models as it only adds one loss function for the representation layer and is commonly used in personalized FL \cite{liu2023bayesian,zhang2022personalized,makhija2025bayesian,issa2024rve}. If a FL model is trained using IB, it will learn representations that squeeze out task-irrelevant information by the compressor (encoder) while preserving task-relevant information by the approximator. The compressing property of IB models enables the protection of the input, and the compressed representations would be utilized for unlearning.


In this paper, to answer the research question, we begin with formalizing a privacy-preserving machine unlearning problem, i.e., preventing the exposure of erasing data to servers meanwhile ensuring effective unlearning. We propose a Blind Unlearning (BlindU) approach to solve this issue, as illustrated in \Cref{fig:funlandmcfu2}. In BlindU, the erased raw data is first masked and compressed to new representations on the user side, and then two unlearning modules are implemented to jointly unlearn the compressor and approximator of the IB model using the compressed representations on the server side.


The BlindU approach ensures dual privacy protection through IB compression and a differential privacy (DP) masking, both of which are easily executed on the user side. Since the FL model is trained using IB methods, users who request unlearning can directly use the compressor of the FL model to locally generate the compressed data. The IB compressor ensures the representation only retains task-relevant information, effectively discarding and protecting task-irrelevant information of the unlearning data. To further protect the privacy of task-relevant information remaining in the representation, we design a noise-free DP masking strategy before compressing. We implement two sampling strategies to sample data features and mask the remaining unsampled features, providing the feature-level DP protection without adding noise. Then, the user uploads the masked and compressed data to the server for unlearning.


BlindU performs unlearning on the server side using only the compressed representations uploaded by the unlearning user. The procedure is decoupled from the original training and does not rely on the FL framework for retraining. BlindU introduces two customized unlearning modules that jointly unlearn the compressor and the approximator of the trained IB model. For the approximator, the basic idea is to minimize the mutual information between the compressed representations and their corresponding labels. For the compressor, the server has no access to the raw inputs corresponding to the compressed data, so we instead minimize the mutual information between the erased representations and auxiliary representations produced by the compressor on auxiliary (or randomly generated) inputs. We cast the compressed representation-based unlearning as a constrained optimization problem and employ a Multiple Gradient Descent Algorithm (MGDA) to adaptively balance forgetting and utility retaining.


We evaluate BlindU from both privacy protection and unlearning effectiveness perspectives. For evaluating privacy protection, we conduct two kinds of privacy leakage attacks: the privacy reconstruction attacks~\cite{salem2020updates,melis2019exploiting,Hu2024sp,zhang2023conditional} and the membership inference attacks~\cite{chen2021machine,kurmanji2024towards}, on various unlearning methods to compare defense performance. Unlearning effectiveness is assessed based on data removal impact and model utility preservation. The results show that BlindU significantly enhances privacy protection, increasing the privacy reconstruction mean square error (MSE) from $57.76$ to $421.9$ on CIFAR10 compared to existing methods. With the achievement in privacy protection, our methods have the best removal effect and keep a similar or slight superiority in model utility preservation and efficiency compared with the state-of-the-art unlearning methods.


In summary, our contributions are fourfold.

\begin{itemize}[itemsep=0pt, parsep=0pt, leftmargin=*]
	\item We formalize the problem of erasing data privacy-preserving machine unlearning based on the IB method framework and propose BlindU to solve it. Dual privacy protection through compressing and DP masking is adopted in BlindU to generate new representations, protecting both task-irrelevant and relevant information. 
	\item We customize two unlearning modules to jointly unlearn the compressor and approximator of an IB-trained model, using only the compressed representation. An MGDA algorithm is employed to balance forgetting and utility retaining in BlindU optimization.
	\item We conduct a formal privacy protection analysis and extensive experiments to validate our method. Different privacy leakage attacks are conducted to evaluate the effect of privacy protection. The results demonstrate the superiority of BlindU in erasing data privacy protection and unlearning effectiveness compared to existing solutions. 
	\item The source code and the artifact of BlindU are released at \url{https://github.com/wwq5-code/BlindU}, which creates a new tool for protecting the privacy of erasing data in machine unlearning and sheds light on the design of future privacy-preserving unlearning methods.
\end{itemize}

The rest of the paper is structured as follows. We review related work in Section \ref{rw}. We provide the preliminary knowledge and formalize an erasing data privacy-preserving machine unlearning problem in \Cref{pr,problem_s}. We introduce the solution, BlindU, in detail in \Cref{sec_MCFU}. A formal privacy protection analysis of the proposed method is conducted in Section~\ref{theoretical}. \Cref{ex_setting,ex} presents our experimental settings, results, and comparison with related work. Finally, we summarize the paper in Section \ref{summary}.



%% file: Contents/2_background.tex
\section{Related Work}
\label{rw}


Existing unlearning methods can be roughly categorized to \textit{fast retraining} and \textit{approximate unlearning} methods. The fast retraining methods are extended from naive retraining and aim to reduce the computational cost of retraining new models by redesigning the learning algorithms and storing the intermediate parameters in the learning process \cite{bourtoule2021machine,hu2024eraser,yanarcane2022unlearning}. The approximate unlearning aims to implement unlearning only by using the trained model and erasing samples to approach the model retrained on the remaining dataset~\cite{wang2023machine,sekhari2021remember,guo2019certified,nguyen2020variational}. 

All these studies are trying to tackle the basic problem of unlearning, i.e., how to erase the contribution of specified samples from a trained model. They assume the ML server knows the erasing samples from the user's unlearning request \cite{wang2023machine,nguyen2020variational}. Some of them moreover require the ML server to access the whole training dataset \cite{guo2019certified,bourtoule2021machine}. However, uploading the erasing data raises potential privacy threats in privacy-preserving learning scenarios.

\noindent
\textbf{Privacy-Preserving Studies about Unlearning.} 
A few studies pay attention to the problem of unlearning data's exposure to servers and explore privacy-preserving unlearning with a limitation of servers' access to users' data~\cite{liu2022right,wang2023bfu,liu2021federaser,su2023asynchronous}. These studies are commonly in FL and called federated unlearning.
Some federated unlearning methods \cite{liu2021federaser} tried to unlearn a user's whole influence from the trained FL model. They stored all users' uploaded parameters on the server side and estimated the unlearning user's contribution to the model based on these stored parameters. This way, the FL server can execute unlearning without requiring user interaction. However, it will significantly harm the original model's utility and is infeasible for a user who wants only to unlearn a small part of his/her local dataset. 

Unlike unlearning a user's total contribution, the authors of \cite{liu2022right,wang2023bfu} investigated how to unlearn the user-specified samples in FL. 
They proposed fast retraining methods based on the Hessian matrix and Bayesian inference, but the disadvantage is that the FL server should reactivate all users to participate in retraining, which is impractical and inefficient in the real world. To solve this issue, \citeauthor{lin2024incentive}~\cite{lin2024incentive,su2023asynchronous} dynamically selected a cluster of users for unlearning, significantly improving the efficiency in asynchronous scenarios. All these methods achieve privacy protection by uploading gradients instead of the raw data. However, gradients computed solely using the erasing data for unlearning pose a higher risk of privacy leakage compared to the standard FL training gradients on all local data~\cite{salem2020updates,boenisch2023curious,carletti2025sok,dimitrov2024spear}. A general and robust erasing data privacy-preserving unlearning method is still necessary and urgent.


\section{Preliminary}
\label{pr}

\subsection{Basic Notations}
We summarize the primary notations used in the paper, as presented in \Cref{notation}.
In typical FL scenarios, the server has access to the entire training dataset $D=\{D_1,D_2,...,D_K\}=(X,Y)$, which may consist of $K$ local dataset $D_K=(X_K,Y_K)$ from $K$ users. $X$ and $Y$ are the input and label of the training data. Suppose a user $u$ wishes to unlearn his/her erasing data $D_e=(X_e,Y_e) \subseteq D_{K_u}  \subseteq D$, we call the user an ``unlearning user'' or ``unlearning requester''. Theoretically, there is a remaining dataset $D_r=(X_r, Y_r)$, making $D=(X, Y) = (X_r, Y_r)  \bigcup (X_e,Y_e)$ and $(X_r, Y_r) \bigcap (X_e,Y_e) = \emptyset$. Besides input $X$ and label $Y$, $Z$ denotes the representation learned by IB models. For an IB model $\mathcal{M}(\theta^c, \theta^a)$, it includes a compressor $\theta^c$ and an approximator $\theta^a$. A Lagrange multiplier $\beta$ is used to control the compression during the IB model training. In our paper, we use $\mathcal{U}$ to denote the unlearning algorithms, $\mathcal{C}$ to denote the privacy-preserving mechanism, and $\mathcal{S}$ to denote the sampling strategy for noise-free DP masking.

\begin{table}[t]
	\scriptsize
	\centering  
	\caption{ { Basic notations}  \vspace{-2mm}}
	\label{notation}
	\resizebox{\linewidth}{!}{ 
		\begin{tabular}{c || l }  
			\toprule[0.12em]
			{ Notations }  & { \makecell[l]{Descriptions} } \\ 
			\midrule  
			{ $D$ }  & { \makecell[l]{the entire training dataset includes inputs and labels ($X,Y$);}  }  \\    
			{ $D_e, D_r$} &  { the erased dataset and the remaining dataset of $D$ } \\
			{ $X, Y$ } &  { \makecell[l]{the inputs $X$ and labels $Y$ of trianing data; } }    \\     
			{ $Z$ } &  { the learned representation of IB models } \\
			{ $\theta^c$}  & { the parameters of compressor of an IB model }  \\
			{ $\theta^a$} & { the parameters of approximator of an IB model }  \\
			{ $\mathcal{M} (\theta^c, \theta^a)$} & { the IB model with compressor and approximator}  \\
			{ $\beta$} & { the Lagrange multiplier (compression ratio) of IB}  \\
			{ $\mathcal{U}$ } &  { \makecell[l]{the unlearning algorithm} } \\     
			{ $\mathcal{C}$} & { \makecell[l]{the privacy-preserving mechanism} } \\  
			{ $\mathcal{S}$ } & {  \makecell[l]{the sampling strategy for DP masking}  }\\    
			\bottomrule[0.12em]
	\end{tabular}}
\end{table}

\subsection{Information Bottleneck} \label{IB_intro}



IB objective can be described as follows,
\begin{equation} \label{IB_eq} 
	\vspace{-2mm}
 \mathcal{L}_{\mathrm{IB}} = \underbrace{\beta\, I(Z;X)}_{\mathcal{L}_{\mathrm{com}}}
	+ \underbrace{\big(- I(Z;Y)\big)}_{\mathcal{L}_{\mathrm{app}}}, 
\end{equation}
where $I(Z; X)$ is the mutual information between the compressed representation $Z$ and inputs $X$. $I(Z; Y)$ denotes the mutual information between $Z$ and $Y$. $\beta$ is the Lagrange multiplier that controls the compression ratio of $X$.

An example of IB model structure is illustrated in the upper half of \Cref{fig:funlandmcfu2}. An IB-trained model splits the standard training process into two components. Firstly, it has a compressor $\theta^c$ to compress the information from inputs $X$ into a compact representation $Z$. Secondly, it employs an approximator $\theta^a$ to predict the target approximation based on the representation $Z$. The learned compressed representation $Z$ should satisfies two conditions. First, $Z$ is sufficient for targets $Y$, which the approximator learns. Second, $Z$ discards information not relevant to the targets in $X$. The compressed representation loss function is $\mathcal{L}_{\mathrm{com}} = I(Z;X)$ and the corresponding approximation loss function is $\mathcal{L}_{\mathrm{app}} = - I(Z;Y)$.

Directly optimizing the traditional IB objective is infeasible in deep neural networks as it is hard to calculate the exact mutual information \cite{alemi2016deep}. Hence, existing works \cite{alemi2016deep,achille2018information,bang2021explaining} expanded the two mutual information items and proposed two variational distributions, $q(Z)$ and $q(Y|Z)$, to obtain the loss function upper bound. They optimize the learning process by minimizing the upper bound as variation inference \cite{fox2012tutorial}. 
Assume a distribution $q(Z)$ where the components of the space $Z$ are mutually independent, which is $q(Z) = \prod_j q_j(z_j)$. Finally, an IB loss objective Eq.~\eqref{IB_eq} can be expanded in a per-sample way as
\begin{equation} \label{eq:total_loss_q}
	\begin{small}
		\begin{aligned}
			\mathcal{L} = \mathcal{L}_{\mathrm{com}} + \mathcal{L}_{\mathrm{app}}  = \frac{1}{N}\sum_{i=1}^{N} \Big[ 			\underbrace{ \beta \text{KL}[p_{\theta^c}(Z|x_i)||  \prod_{i}^{|Z|} q_i(z_i)]}_{\mathcal{L}_{\mathrm{com}}} +\\
			+  \underbrace{\mathbb{E}_{z \sim p_{\theta^c}(Z|x_i)}[-\log \ p_{\theta^a}(y_i|z)]}_{\mathcal{L}_{\mathrm{app}}}
			\Big]. 
		\end{aligned}
	\end{small}
\end{equation} 
This is a trainable surrogate for \Cref{IB_eq}. The KL term discourages unnecessary detail of $X$, and the negative log-likelihood is the usual supervised loss that keeps information in $Z$ that helps predict $Y$. \citeauthor{alemi2016deep} \cite{alemi2016deep} has proved that \Cref{eq:total_loss_q} can be derived with the same mathematical form as VAEs \cite{kingma2014auto} that implemented using Bayesian inference. 
The FL frameworks using Bayesian inference (the same mathematical form as IB derived) are commonly studied, which are typically suitable in personalized federated learning and heterogeneous settings~\cite{liu2023bayesian,zhang2022personalized,makhija2025bayesian,issa2024rve}.



\section{Problem Statement} \label{problem_s}


Since our privacy-preserving unlearning solution is proposed based on the IB framework, we directly assume that the FL service model has been trained using an IB method as \Cref{eq:total_loss_q} based on the full dataset $D=(X,Y)$, and the trained model is denoted as $\mathcal{M}(\theta^c,\theta^a)$. Normally, after receiving the erasure request, i.e., unlearning $D_e=(X_e, Y_e)$, the server will execute an unlearning algorithm $\mathcal{U}$ to remove the trace of $D_e$ from the trained model $\mathcal{M}(\theta^c,\theta^a)$, resulting the unlearned model $\mathcal{M}_u(\theta^c, \theta^a)$. The unlearned model is expected to find the new representation $Z_u$ approximated to the representation of a retraining model based only on the remaining dataset, i.e., discarding the knowledge of erased data $D_e$ while keeping the knowledge of the remaining data $D_r$. The standard retrained representation can be obtained by retraining the IB model based on the remaining dataset as 
\begin{equation} \label{retraining_eq}
\begin{aligned}
\min \  & I(X_r;Z) \\
\text{s.t. } \ & I(X_r;Y_r) - I(Z;Y_r) = I(X_r;Y_r|Z)=0.
\end{aligned}
\end{equation}
The retraining representation $Z$ in \Cref{retraining_eq} is the gold standard that the unlearned representation $Z_u$ of $\mathcal{M}_u(\theta^c, \theta^a)$ aims to achieve. Existing work \citep{wang2023machine} has shown that further minimizing $I(X_e;Z)$ and $I(Y_e;Z)$ can remove the contribution of $D_e$. Their method samples mini-batches from both $D_e$ and $D_r$ and uses a multi-objective optimization to balance the forgetting and utility preservation gradients so that the update direction is aligned with the IB retraining objective \Cref{retraining_eq}. However, this method needs the server to have access to erasing samples $D_e$, which directly exposes the privacy to the server.





In existing federated unlearning methods~\cite{liu2022right,liu2021federaser,wang2023bfu}, the uploaded gradients of the unlearning users can be seen as a privacy-preserving mechanism $\mathcal{C}$, transforming the erasing data into a privacy-preserving form and preventing malicious entities from inferring private information from gradients.
However, it is important to note that gradients contain data information, which can lead to privacy leakage \cite{jin2021cafe,boenisch2023curious}. This risk is exacerbated in unlearning scenarios, as unlearning requests typically target a small subset of samples, even a single sample, rather than the entire local dataset. Unlearning a smaller number of samples reduces the gradients' protective effect \citep{zhu2019deep,geiping2020inverting,yin2021see,dimitrov2024spear,carletti2025sok}. Considering the requirements for effective unlearning and a privacy-preserving mechanism $\mathcal{C}$, we can formally define the problem as follows.
\begin{prob_state}\label{eq_eu_with_C}
(Privacy-preserving machine unlearning to prevent exposure of erasing data)
Suppose the FL server now has a trained IB model $\mathcal{M} (\theta^c, \theta^a)$, and the IB model learned the representation $Z$ on $D$. Let $D_e=(X_e, Y_e)$ be the erasing dataset of an unlearning user and $D_e \subseteq D$. Theoretically, a remaining dataset $D_r$ exists, but the FL server cannot access any users' private data.
The goal of the privacy preserving machine unlearning problem is to design an unlearning method $\mathcal{U}$ for the FL server and a data privacy-preserving mechanism $\mathcal{C}$ for the unlearning user, such that: 
\begin{enumerate}[itemsep=0pt, parsep=0pt, leftmargin=*]
\item $\mathcal{C}$ can transform $D_e$ into a privacy-preserving form, $\mathcal{C}(D_e)$, making the FL server cannot infer sensitive information about $D_e$ from $\mathcal{C}(D_e)$.
\item {The unlearning algorithm $\mathcal{U}$ can unlearn based on the transformed data $\mathcal{C}(D_e)$ and make the unlearned IB representation $Z_u$ equal or approximate to the retrained representation, i.e., similar property as $Z$ in \Cref{retraining_eq}. }
\end{enumerate}
\end{prob_state}



The key challenge lies in that this mechanism should make $\mathcal{C}(D_e)$ insensitive to the server while still retaining sufficient information for effective unlearning. In this work, we choose to use compressive property of IB as $\mathcal{C}$ and design corresponding unlearning algorithm $\mathcal{U}$ based on the compressed data, which aligns with the IB framework.




%% file: Contents/4_approach.tex
\section{Blind Machine Unlearning}
\label{sec_MCFU}


\subsection{Overview of BlindU}


We propose the Blind Unlearning (BlindU) method to solve the above privacy-preserving unlearning problem that ensures the erased data is unseen from the server. BlindU only involves the FL server and unlearning user as shown in \Cref{fig:funlandmcfu2}, where the privacy-preserving mechanism is first executed on the user side to prepare the compressed representation, and the unlearning is then implemented on the server side. In this section, we will first introduce the BlindU unlearning framework based on compression, and then introduce the noise-free DP masking for privacy enhancement.




\subsection{Blind Unlearning based on Compression}

\noindent
\textbf{Compressed Erasing Data Preparation.}
Before implementing unlearning for ML service models, the first step is to prepare the erasing data by the unlearning user. Given that the model is well trained using the IB method, the compressor is now capable of squeezing out as much as possible task-irrelevant information from the input data while maintaining as much as possible task-relevant information. The unlearning user only needs to download the compressor (first half of the model) from the server and uses this compressor to distort his/her sensitive erasing data locally before unlearning. 

The only limitation is that the compression ratio is predetermined when the original model is trained. As a result, the user cannot select the compression rate at this stage. Fortunately, users have the option to select different masking strategies to adjust the level of differential privacy protection, which we will discuss in more detail later. We denote the compressed data of the erasing data as $Z_e = \mathcal{C}(X_e)$. After compressing, the user uploads the distorted representation and corresponding labels $C_e = (Z_e, Y_e)$ to the server for unlearning.

\noindent
\textbf{Unlearning Compressor.} 
Existing methods unlearn an IB model by further minimizing the mutual information between the learned presentation $Z$ and the unlearning data $D_e$, which requires accessing the unlearning data \cite{wang2023machine}. However, in our setting for privacy requirements, the FL server only receives the compressed data $Z_e$ and the corresponding label $Y_e$. Unlearning the compressor $\theta^c$ with only the output $Z_e$ but no information about the corresponding $X_e$ is a challenging task. 





We cannot simplify this problem to merely unlearning a single class from the model. Our objective is to unlearn the information specifically related to $Z_e$ and $X_e$, rather than unlearning the entire class associated with $Z_e$. Fortunately, the original purpose of the compressor is to discard the task-irrelevant information about inputs from the representation. To further eliminate the information about $Z_e$ and $X_e$ from the trained compressor $\theta^c$, we can minimize the mutual information between $Z_e$ and other representations generated by other inputs during model training.

We first construct a small auxiliary dataset, which generates a few samples for different labels and can be denoted as $(X_a, Y_a)$. We can even generate a random $X_a$ to implement unlearning on compressor $\theta^c$, but for better accuracy preservation on $\theta^a$, we construct several general samples from each class as an auxiliary dataset. We do not need to construct too many auxiliary samples. For ease of implementation, we generate the same size as the erasing dataset in our experiment. 

We implement unlearning on the compressor by minimizing the mutual information between the compressed erasing representation $Z_e$ and the model representation $Z_{\theta^c(X_a)}$, which can be denoted as $I(Z_e; Z_{\theta^c(X_a)})$, where $Z_{\theta^c(X_a)}$ is outputted by $\theta^c$ with the input $X_a$. 
To simplify notation, we also denote $Z_{\theta^c(X_a)}$ as $Z_a$.
By minimizing this mutual information, we can remove the information about $Z_e$ from the compressor. We describe and expand the mutual information loss function of unlearning the compressor, which can be transformed into the Kullback-Leibler (KL) divergence form as below,
\begin{equation} \label{rep_unl} \small
	\begin{aligned}
		\mathcal{L}_{\text{com} }^u =  I(Z_e;Z_{a}) &=  \int d z_e \ dz_a \ p(Z_e, Z_{a})  \log \frac{p(Z_e, Z_{a}) }{p(Z_e) p(Z_{a}) }  \\
		&=\text{KL} [p(Z_e,Z_{a})||p(Z_e) p(Z_{a})].
	\end{aligned}
\end{equation} 
The mutual information is equivalent to the KL-divergence between the joint, $p(Z_e,Z_a)$, and the product of the marginals $p(Z_e) p(Z_{a})$. However, it is hard to compute the exact mutual information of the above equation, and variational methods are not suitable in this scenario. Inspired by MINE \cite{belghazi2018mutual}, we find a solution to calculate the approximate mutual information.
Similar to MINE, since the KL-divergence admits the Donsker-Varadhan representation \cite{donsker1983asymptotic} and the $f$-~divergence representation \cite{nguyen2010estimating,nowozin2016f}, we can expand the Eq.~\eqref{rep_unl} in a tight estimation. 
Here, we use the Donsker-Varadhan representation \cite{donsker1983asymptotic} of the KL-divergence as an example to expand the Eq.~\eqref{rep_unl}. 
Let $\mathcal{F}$ to be the family of functions $T_{\theta}: \mathcal{Z}_e \times \mathcal{Z}_a \to \mathbb{R}$ parametrized by a deep neural network with parameters $\theta \in \Theta$. Therefore, we can achieve the bound for Eq. \eqref{rep_unl} as 
\begin{equation} \label{mi_estimator}  \small
	I(Z_e;Z_a) \geq I_{\Theta}(Z_e;Z_a) = \sup_{\theta \in \Theta} \mathbb{E}_{p(Z_e,Z_a)}[T_{\theta}] - \log(\mathbb{E}_{p(Z_e)p(Z_a)} [e^{T_{\theta}}]),
\end{equation}
where $I_{\Theta}(Z_e;Z_a)$ is the neural information measure. 
We can calculate the estimation of the mutual information in Eq. \eqref{mi_estimator} using gradient ascent based on the empirical samples from $p(Z_e,Z_a)$ and $p(Z_e)p(Z_a)$. 

\begin{figure}[t]
	\centering
	\includegraphics[width=0.94\linewidth]{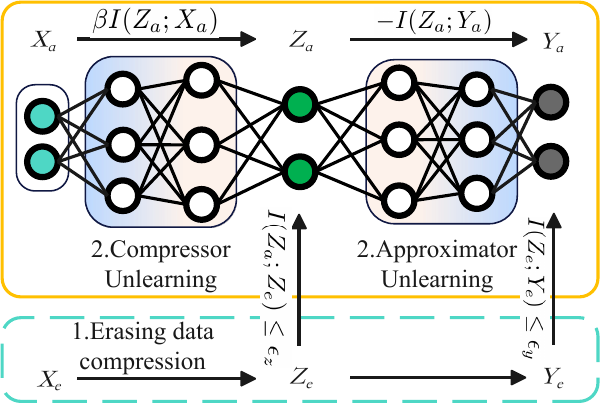}
	\vspace{-2mm} 
	\caption{
The unlearning user first downloads the IB compressor of the service model and compresses the erasing data, as shown in the green dashed box. Then, the compressed representations and labels are uploaded to the server for unlearning with \Cref{unlearning_with_constraint}, as shown in the yellow solid box.
	}
	\label{fig:mcfudetail}
\end{figure}


\noindent
\textbf{Unlearning Approximator.}
Compared to compressor unlearning with only the compressed representation $Z_e$ (the output of the compressor), unlearning the approximator will be easier as we have both the representation and the labels (input and output of approximator). An easy way is to apply existing variational Bayesian unlearning methods \cite{nguyen2020variational}, which can be conducted based only on erasing data. Alternatively, we have identified a straightforward approach that aligns with the IB theory, which involves directly minimizing the mutual information between $Z_e$ and $Y_e$ to unlearn the erasing data from the approximator. 
\begin{equation} \label{approximator_unl}
		\begin{aligned}
		\mathcal{L}_{\text{app}}^u &= I(Y_e;Z_e) 
	\end{aligned}
\end{equation}

Although the method is easy to implement, simply directly minimizing $I(Y_e; Z_e)$ may easily cause model utility degradation. Moreover, directly minimizing \Cref{rep_unl} may also decrease the compressor utility. To mitigate the utility degradation of the IB model and constrain the unlearned representation, we formalize the compressive unlearning as constrained IB retraining on the auxiliary data $(X_a, Y_a)$ to preserve the retained information.

\noindent
\textbf{Solving Blind Unlearning as Constrained IB Retraining.} 
With the above unlearning mechanisms for both compressor and approximator of the trained IB model, we can formalize the blind unlearning as constrained IB retraining on the auxiliary data $D_a = (X_a, Y_a)$ to preserve the retained information. We do not require $D_a$ to match the retained distribution exactly; $D_a$ only needs to approximate it at the class and feature level so that optimizing on $D_a$ preserves the predictive structure on the retained data. In practice, such an auxiliary dataset can be constructed by model inversion \citep{yin2021see,carletti2025sok} on the FL server side. We minimize the IB risk on $D_a$ subject to (1) representation-level decoupling from the erased representations and (2) label-information erasure on the erased representations. Formally,
\begin{equation}  \label{unlearning_with_constraint}
	\min_{(\theta^c, \theta^a)} \mathcal{L}_{\text{IB}} \big( (\theta^c, \theta^a); D_a \big) \text{ s.t. }  I(Z_a; Z_e) \leq \epsilon_z, I(Z_e;Y_e) \leq \epsilon_y.
\end{equation}
However, determining the optimal static values for $\epsilon_z$ and $\epsilon_y$ is non-trivial; a value too large fails to enforce unlearning, while a value too small destroys model utility. To address this, we treat the problem as a weighted-sum objective for both retraining $\mathcal{L}_{\text{IB}} \big( (\theta^c, \theta^a); D_a \big)$ and forgetting $\mathcal{L}_{\text{forget}} = I(Z_a; Z_e) + I(Z_e;Y_e)$, which can be described as 
\begin{equation}  \label{pareto_unl}
	\min_{\alpha} || \alpha \nabla \mathcal{L}_{\text{IB}} + (1-\alpha) \nabla \mathcal{L}_{\text{forget}}||^2_2,
\end{equation}
where $\alpha  \in [0,1]$. We employ Multiple Gradient Descent Algorithm (MGDA) \citep{sener2018multi} to find a Pareto stationarity of retraining and forgetting, and $\alpha$ can be calculated as adaptive values before each round update via an anlytical solution, $\alpha = \max(\min(\dfrac{ ( \nabla\mathcal{L}_{\text{IB}} - \nabla \mathcal{L}_{\text{forget}} )^{\mathsf{T}} \nabla\mathcal{L}_{\text{IB}} }{ || \nabla \mathcal{L}_{\text{forget}}  - \nabla \mathcal{L}_{\text{IB}} ||_2^2},1),0)$. In the ideal case where there exists a model that both (i) minimizes IB risk on the retained distribution and (ii) achieves perfect unlearning, the Pareto-weighted solution of \Cref{pareto_unl} does not perform worse (in IB risk on retained distribution) than retraining on $D_r$. In the general case where the constraints are incompatible with perfect IB optimality, the weight-sum objective traces a Pareto frontier that explicitly quantifies the unlearning and retaining trade-off.

\subsection{The Noise-Free Differential Privacy Masking}
	
	

	

	

The previously introduced unlearning method operates on compressed representations that discard task-irrelevant information, thereby providing privacy protection. Our experiments show that this compression alone can already effectively defend against privacy inference and reconstruction attacks \citep{salem2020updates,Hu2024sp,chen2021machine}. However, the user cannot select the compression rate during unlearning (as it was set in the original IB training process), and the representation still attempts to retain maximum task-relevant information from the inputs. We propose noise-free DP masking strategies as a plug-in module applied before compressing to further protect input information, enabling dual privacy protection.



From \cite{he2022masked}, we know that a good masking strategy can promote model performance. Moreover, from \cite{sun2020federated}, we know different random sampling strategies will guarantee different DP protection for data. Inspired by these two, we introduce two sampling schemes, \textbf{randomly sampling with replacement} and \textbf{without replacement}, which provide different DP protection without noise injection. We choose sampling strategies rather than direct DP noise injection because noise injection changes the information of the erasing data. Unlearning samples with noise injection would make it difficult to determine if the model unlearns the true information of the erasing data or just the noise.

\begin{claim} [\textbf{($\epsilon, \delta$)-differential privacy masking of random sampling with replacement}] \label{dp_theorem_with_replacement}
	Given a data point with $n$ features, randomly sampling with replacement achieves ($k \cdot \ln\frac{n+1}{n}$, $1-(\frac{n-1}{n})^{k}$)-differential privacy, where $k$ is the size sampled feature subset. 
\end{claim}

\begin{claim} [\textbf{($\epsilon, \delta$)-differential privacy masking of random sampling without replacement}] \label{dp_theorem_without_replacement}
	Given a data point with $n$ features, randomly sampling without replacement achieves ($\ln\frac{n+1}{n+1-k}$, $\frac{k}{n}$)-differential privacy, where $k$ is the size sampled feature subset.
\end{claim}



\begin{figure}[t]
	\centering
	\includegraphics[width=0.97\linewidth]{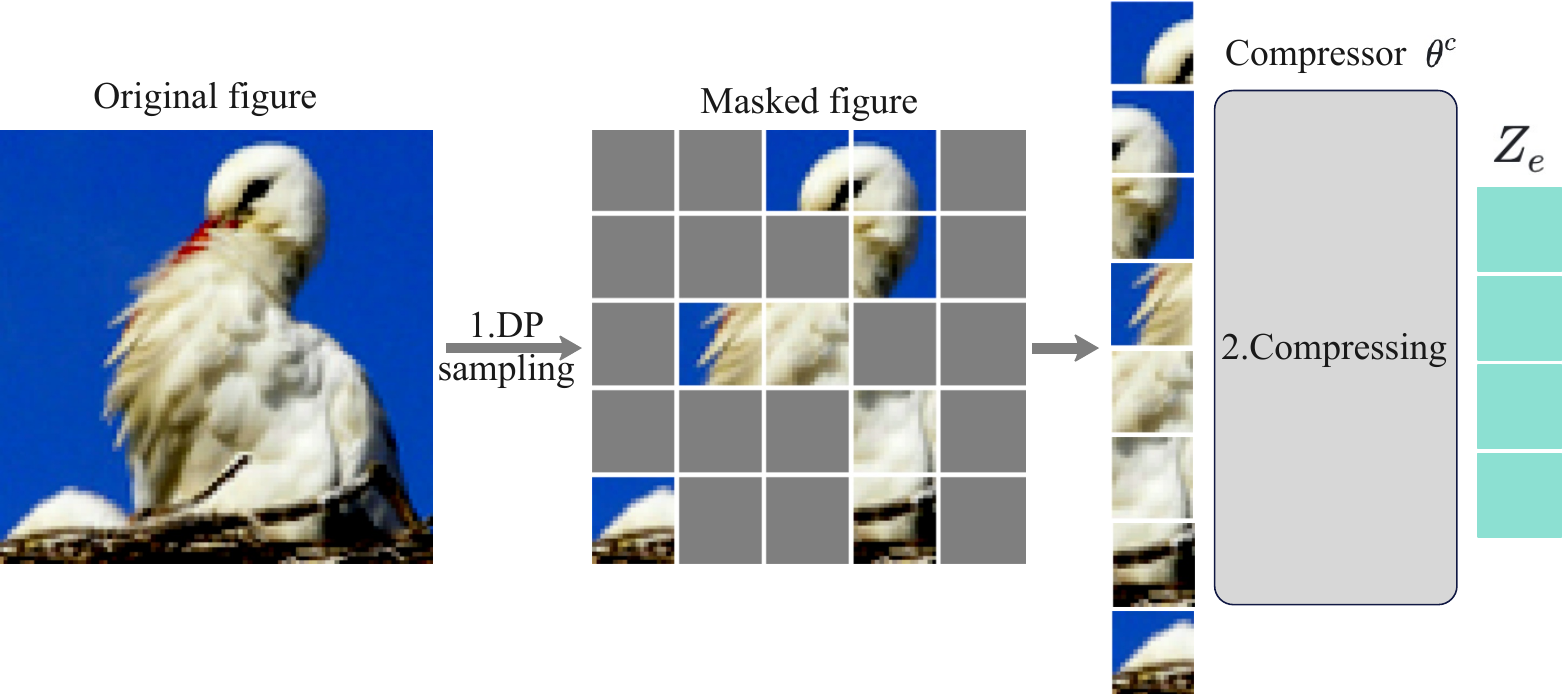}
	\vspace{-2mm}
	\caption{An example of noise-free DP masking on image data. 
	}
	\label{fig:dpsamplingmasking}
\end{figure}

We proved that Claim \ref{dp_theorem_with_replacement} and Claim \ref{dp_theorem_without_replacement} meet differential privacy \citep{DworkR14} in \Cref{privacy_protection_dp}. After executing the sampling strategy (Claim \ref{dp_theorem_with_replacement} or Claim \ref{dp_theorem_without_replacement}), we obtain a subset of several original data features and mask other unselected features. For image data, we can set the unselected features black or grey as shown in \Cref{fig:dpsamplingmasking}. For other data, like tabular data, we can set the unselected features to none. After sampling, the user feeds the masked samples to the compressor $\theta^c$ to distort the target-irrelevant information. Finally, the masked and compressed representations are uploaded as the unlearning request to the server, enabling dual privacy protection.



\section{Privacy Protection Analysis}
\label{theoretical}




In this section, we will analyze the privacy protection from compressing and DP masking, respectively, and the privacy-utility tradeoff when combining these two techniques.



\subsection{Privacy Protection by Compressing}

\begin{claim} [\textbf{Task-irrelevant information protection}] \label{task_ir_pos}
	Suppose we have the inputs $X$, the task target $Y$, and the representation $Z$ that is compressed for $Y$ based on $X$ using IB methods. The information of $X$ can be divided into two independent parts, the task-relevant information $X_{\mathrm{re}}$ and the task-irrelevant information $X_{\mathrm{ir}}$, denoting as $X=\{X_{\mathrm{re}}, X_{\mathrm{ir}}\}$,  the mutual information $I(X_{\mathrm{ir}}; Z)$ is upper bounded by $I(Z;X) - I(Z;Y)$:
	\begin{equation}\label{privay_upper}
		I(X_{\mathrm{ir}}; Z) \leq I(Z;X) - I(Z;Y).
	\end{equation}
\end{claim} 
We can prove Claim \ref{task_ir_pos} in a manner similar to the proof of Proposition 1 in \cite{yu2021recognizing} by using the Markov chain. Hence, we omit the detailed proof here. 

Usually, the IB objective is trained as $\mathcal{L}_{\mathrm{IB}} = \beta I(Z;X) - I(Z;Y)$ with a Lagrange multiplier $\beta$ that controls the balance between compression and utility. In the classical information bottleneck theory \citep{tishby2000information}, different value of $\beta$ parameterize the Pareto frontier between $I(Z;X)$ and $I(Z;Y)$: increasing $\beta$ moves the optimal compressor towards representations with smaller $I(Z;X)$ (stronger compression) and, beyound the flat region of curve, inevitably smaller $I(Z;Y)$ (reduced predictive utility). In our setting, we inherit this interpretation: larger $\beta$ encourages compressors with smaller $I(Z;X)$ and therefore a tighter upper bound on $I(X_{\mathrm{ir}}; Z)$, but overly large $\beta$ eventually harms $I(Z;Y)$ and accuracy.



\subsection{Privacy Protection by DP Masking} \label{privacy_protection_dp}

We proposed to mask the input data using two sampling strategies, Claim \ref{dp_theorem_with_replacement} and Claim \ref{dp_theorem_without_replacement}, which can guarantee feature-level DP protection without any noise injection. A classic DP definition can be found in \citep{DworkR14}. Our two sampling strategies, sampling with and without replacement, meet $(\epsilon, \delta)$-differential privacy guarantee as the DP definition in \citep{DworkR14}. We provide an example proof of Claim \ref{dp_theorem_with_replacement} as follows.


\begin{proof}[Proof of Claim \ref{dp_theorem_with_replacement}]
	We assume executing sampling strategy $\mathcal{S}$ on two neighboring feature sets $X$ and $X'$ in the feature space $\mathcal{X}$. Each feature set contains $n$ dimension. Assume $s$ is the sampled subset with $k$ replacement sampling from $X$. The sampling subsets join domains is $\mathbb{S}$ where $s \in \mathbb{S}$. Then, for a random subset $s$, we have, 
	\begin{equation}
		\Pr( \mathcal{S}(X)=s) =   \begin{cases}
			\frac{1}{|X|^k}, &\text{if } s\in \Gamma(X) ,\\
			0, &\text{otherwise. }
		\end{cases}
	\end{equation}
	\begin{equation}
		\Pr( \mathcal{S} (X')=s) =   \begin{cases}
			\frac{1}{|X'|^k}, &\text{if } s\in \Gamma(X') ,\\
			0, &\text{otherwise. }
		\end{cases}
	\end{equation}
Since $X$ and $X'$ are neighbouring sets, $X=X' \cup \{u\}$ and $X'=X \cup \{u\}$ are the only two cases, where $u$ is the different feature between the two feature sets.

\noindent \textbf{Case 1 of $(X'=X \cup \{u\})$:}  Due to $X \subseteq X'$, we have, 
\begin{equation}
	\Pr(\mathcal{S}(X) \in \Gamma (X)) =  1,
\end{equation}
\begin{equation}
	\Pr(\mathcal{S}(X') \in \Gamma (X)) =  \frac{ |X|^k}{|X'|^k}.
\end{equation} 
Let $P$ is a subset of $\Gamma (X') \setminus \Gamma(X)$, since the only difference $u$ is not included in $X$, therefore,
\begin{equation}
	\Pr(\mathcal{S}(X) \in P) =  \Pr(\mathcal{S}(X) \in \Gamma (X') \setminus \Gamma(X)) = 0.
\end{equation} 
Assume $s$ is composed by two disjoint subsets, i.e., $s = s_{X} \cup s_{ X'\setminus X}$, where $s_{X} \subseteq  \Gamma (X)$  and $s_{ X'\setminus X} \in  \Gamma (X') \setminus  \Gamma (X)$, then we have
\begin{equation}
	\begin{aligned}
		   \Pr( \mathcal{S}(X )\in s) \leq \Pr(\mathcal{S}(X') \in s) \cdot  (\frac{n+1}{n})^k \\
	\end{aligned} 
\end{equation}

\noindent \textbf{Case 2 of $(X=X' \cup \{u\})$:} Due to $X' \subseteq X$, then we have, 
\begin{equation}
	\Pr(\mathcal{S}(X) \in \Gamma (X’)) =   \frac{ |X’|^k}{|X|^k},
\end{equation}
\begin{equation}
	\Pr(\mathcal{S}(X') \in \Gamma (X')) = 1.
\end{equation} 
Let $P$ is a subset of $\Gamma (X') \setminus \Gamma(X)$, then we have 
\begin{equation}
	\begin{aligned}
		\Pr(\mathcal{S}(X) \in P) & \leq \Pr(\mathcal{S}(X) \in \Gamma (X') \setminus \Gamma(X))  \\
		&\leq 1 - (\frac{n-1}{n})^k.
	\end{aligned}
\end{equation}
Assume $s$ is composed by two disjoint subsets, i.e., $s = s_{X'} \cup s_{ X\setminus X'}$, where $s_{X'} \subseteq  \Gamma (X')$  and $s_{ X\setminus X'} \in  \Gamma (X) \setminus  \Gamma (X')$, then we have 
\begin{equation}
	\begin{aligned}
	\Pr(\mathcal{S}(X) \in s) \leq \Pr(\mathcal{S}(X') \in s) \cdot ( \frac{n-1}{n})^k +  (1- (\frac{n-1}{n})^k)   \\
	\end{aligned} 
\end{equation}
Merge the \textbf{Case 1} and \textbf{Case 2} together, we have $e^{\epsilon} = \max((\frac{n+1}{n})^k, (\frac{n-1}{n})^k) = (\frac{n+1}{n})^k $ and $\delta = \max(0, 1- (\frac{n-1}{n})^k) = 1- (\frac{n-1}{n})^k$. Therefore, $\mathcal{S}$ with replacement satisfies $(k \ln \frac{n+1}{n}, 1- (\frac{n-1}{n})^k)$-differential privacy.
And \textbf{Claim} \ref{dp_theorem_without_replacement} can be proved in a similar way.
\end{proof}
The designed feature-level DP sampling mechanisms achieve DP by random sampling rather than noise injection. These two Claims guarantee that even if attackers gain access to the compressed representation, they cannot confidently determine whether a specific feature was part of the input data, meeting the DP requirement as \citep{DworkR14}.


\subsection{Dual Privacy Protection Analysis} \label{dual_privacy_protection}

We now analyse how the IB compression parameter $\beta$ and the masking sampling rate (SR) jointly influence privacy and utility in an idealized setting, where SR$=\frac{k}{n}$.

\noindent
\textbf{Dual Privacy Protection.} 
Let $X$ be the input, $Y$ the task label, and $X = (X_{\mathrm{re}}, X_{\mathrm{ir}})$ the decomposition into task-relevant and task-irrelevant parts. The DP masking module with sampling rate (SR) produces a masked input $\tilde X = \mathcal{S}_{\text{SR}}(X)$, and the IB compressor with weight $\beta$ produces a representation $Z = \theta^c_{\beta}(\tilde X)$. Our pipeline is presented as follows:
\begin{equation} \label{markov_pipeiling}
	X \xrightarrow{\text{DP masking (SR)}} \tilde X \xrightarrow{\text{IB compressor ($\beta$)}} Z \xrightarrow{\text{approximator}} \hat Y. 
\end{equation}
Based on Claim \ref{task_ir_pos} and the training procedure, we can define the IB leakage bound 
\begin{equation}
	\Phi(\beta, \text{SR}) := I(Z_{\beta,\text{SR}};X) - I(Z_{\beta, \text{SR}};Y),
\end{equation}
which upper-bounds $I(X_{\mathrm{ir}}; Z_{\beta, \text{SR}})$.

On the DP side, Claim \ref{dp_theorem_with_replacement} and Claim \ref{dp_theorem_without_replacement} show that, for a data point with $n$ features, the masking module with sampling rate SR$=\frac{k}{n}$ achieves $(\varepsilon(\text{SR}),\delta(\text{SR}))$-DP at the feature level. By the post-processing property of differential privacy \citep{DworkR14}, the entire pipeline of \Cref{markov_pipeiling} inherits the same $(\varepsilon(\text{SR}),\delta(\text{SR}))$-DP guarantee. 
Hence for every pair $(\beta, SR)$ we obtain a dual privacy descriptor 
\begin{equation}
\text{Dual}(\beta, \text{SR}) = \big( \Phi(\beta, \text{SR}),\ \varepsilon(\text{SR}),\ \delta(\text{SR}) \big),
\end{equation}
where smaller values mean stronger privacy (smaller MI bound and tighter DP parameters).

\noindent
\textbf{Utility Side.} 
Utility is measured by how well $Z$ predicts Y. We take $\text{Uti}(\beta,\text{SR}) = I(Z_{\beta,\text{SR}};Y)$. Assuming a uniform prior over the labels $\mathcal{Y}$, Fano's inequality implies that for any approximator $\theta^a$ based on $Z$, 
\begin{equation}
	\text{err}(\beta, \text{SR}) \;\ge\; 1 - \frac{I(Z_{\beta,\text{SR}};Y) + \log 2}{\log |\mathcal{Y}|},
\end{equation} 
so larger utility $\text{Uti}(\beta,\text{SR})$ (higher $I(Z;Y)$) is necessary for low classification error.

Putting these pieces together, the pair $(\beta, \text{SR})$ provides dual privacy protection: moving towards larger $\beta$ and smaller SR monotonically improves the dual privacy descriptor $\text{Dual}(\beta, \text{SR})$ (smaller MI bound and smaller $\varepsilon,\delta$), but it also degrades the maximal achievable utility to some extent. In our experiments (Section VIII.D), we complement this partial theoretical characterization by sweeping $\beta$ and SR and plotting both privacy indicators and utility indicators.

%% file: Contents/5_experiments.tex
\section{Experimental Setting} \label{ex_setting}

\noindent
\textbf{Datasets.}
We conducted extensive experiments on four representative image datasets: MNIST, CIFAR10, CIFAR100, and TinyImageNet, which cover a wide range of object categories with different learning complexities. Moreover, considering the main purpose of unlearning is to protect users' ``right to be forgotten,'' we conduct experiments on a practical tabular dataset, Adult, which is consisted of users' private features such as age, education, and race. Unlearning the contribution of these specified private samples from trained models will be a common scenario to exercise users' right to be forgotten. 


\noindent
\textbf{Models.}
We select three distinct architectures to construct the IB model: a 5-layer multi-layer perceptron (MLP) with ReLU activations, a ResNet-18, and a Vision Transformer (ViT). For MNIST, we train an IB model with two 5-layer MLPs, with a learning rate $\eta = 0.001$ and batch size 20. For CIFAR10, CIFAR100, and TinyImageNet, we employ two ResNet-18 networks with $\eta=0.0005$ and batch sizes of 100 (CIFAR) and 200 (TinyImageNet). Additionally, we evaluate a hybrid IB model comprising a ResNet-18 compressor and a ViT approximator on MNIST and CIFAR10. All models are implemented using Pytorch, and experiments are done on a cluster with four NVIDIA 1080ti GPUs.

\noindent
\textbf{Compared Benchmarks.}
We compare our methods with state-of-the-art privacy-preserving unlearning methods, including the representative Hessian-matrix-based federated unlearning in FL (HBFU)~\cite{liu2022right} and Bayesian federated unlearning (BFU)~\cite{wang2023bfu}. We choose HBFU and BFU because they achieve the best model utility during unlearning, with the assistance of all users to participate in retraining. In contrast, \cite{lin2024incentive,su2023asynchronous} focuses on finding an appropriate user cluster method to improve efficiency, which evidently degrades the model utility. 

Moreover, we include a typical machine unlearning method, variational Bayesian unlearning (VBU) \cite{fu2022knowledge, nguyen2020variational}. The VBU method implements unlearning based only on erasing data without relying on the remaining dataset. As a result, it can be seen as a baseline unlearning approach with no privacy protection for the unlearning data. We have not chosen the exact unlearning methods~\cite{cao2015towards,bourtoule2021machine,yanarcane2022unlearning} to compare because these methods rely on accessing the whole training dataset, which is infeasible in the privacy-preserving setting.


BlindU and VBU are not significantly influenced by the FL environment because unlearning is implemented on the server side, and no users participate in the unlearning training process. However, the HBFU and BFU methods are based on the FL framework and needs all users to assist in implementing unlearning collaboratively. We set 10 users with independent and identically distributed~(IID) local datasets to help train the original FL model and set 3 unlearning users for HBFU and BFU with the same total size of erasing samples as BlindU and VBU. This setting is the ideal FL scenario, and HBFU achieves the best unlearning performance \cite{liu2022right}. 

%



\noindent
\textbf{Metrics for Privacy Protection Evaluation.}
We evaluate the privacy protection of the proposed methods in two privacy leakage attacks: the privacy reconstruction attacks \cite{salem2020updates, melis2019exploiting} and the membership inference attacks (MIA) \cite{chen2021machine,salem2020updates}. Specifically, for privacy reconstruction attacks, we conduct the experiments of reconstruction attacks \cite{salem2020updates} to recover the unlearned samples on MNIST and CIFAR10 to evaluate whether BlindU can improve the prevention against these attacks. 
We evaluate the privacy protection effectiveness by the reconstruction's Mean Square Error (\textbf{MSE}). For membership inference attacks, we rely on the traditional \textbf{AUC} metric to measure the absolute performance of the membership inference according to \cite{chen2021machine}. Additionally, we use the mutual information (\textbf{MI}) of $I(Z_e;X_e)$ to evaluate how much information of $Z_e$ related to $X_e$ is exposed to the server.




\noindent
\textbf{Metrics for Unlearning Effectiveness Evaluation.}
To effectively evaluate the unlearning effect, we refer to a common method \cite{hu2022membership}, adding the backdoor triggers to the erasing samples for the original model training. Then, we execute the unlearning methods to erase the backdoor. After unlearning, we verify whether the backdoor can attack the unlearned model to evaluate the unlearning effect. Specifically, we use models' \textbf{accuracy} on the test dataset and \textbf{backdoor accuracy} \cite{hu2022membership} on the erasing dataset to evaluate the unlearning effectiveness. We also conduct unlearning experiments for multiple class normal unlearning requirements (randomly selected benign samples), and use accuracy on the test dataset and unlearning dataset to evaluate the unlearning effect. Moreover, we evaluate the efficiency of unlearning methods by their \textbf{running time}.

\section{Performance Evaluation}
\label{ex}

In this section, we conduct experiments to answer the following research questions (RQ) to evaluate BlindU:
\begin{itemize}[itemsep=0pt, parsep=0pt, leftmargin=*]
	\item \textbf{RQ1}: How does the proposed BlindU perform on unlearning efficacy and efficiency compared with the state-of-the-art machine unlearning methods? (See \Cref{unlearning_evaluation,normal_unl_eval})
	\item \textbf{RQ2}: How does the proposed BlindU perform on the privacy protection for erased samples, as compared with the state-of-the-art privacy-preserving unlearning methods? (See \Cref{privacy_evaluation})
	\item \textbf{RQ3}: How do different hyperparameters, specifically the compressive ratio $\beta$ and masking sampling ratio \text{SR}, influence the unlearning effecitiveness and privacy protection performance of BlindU? (See \Cref{privacy_evaluation,ablation_study_eval})
\end{itemize}

\begin{figure*}[t]
	\centering
	\subfloat{ 	\label{fig:mnistbackaccercurve}
		\includegraphics[scale=0.25]{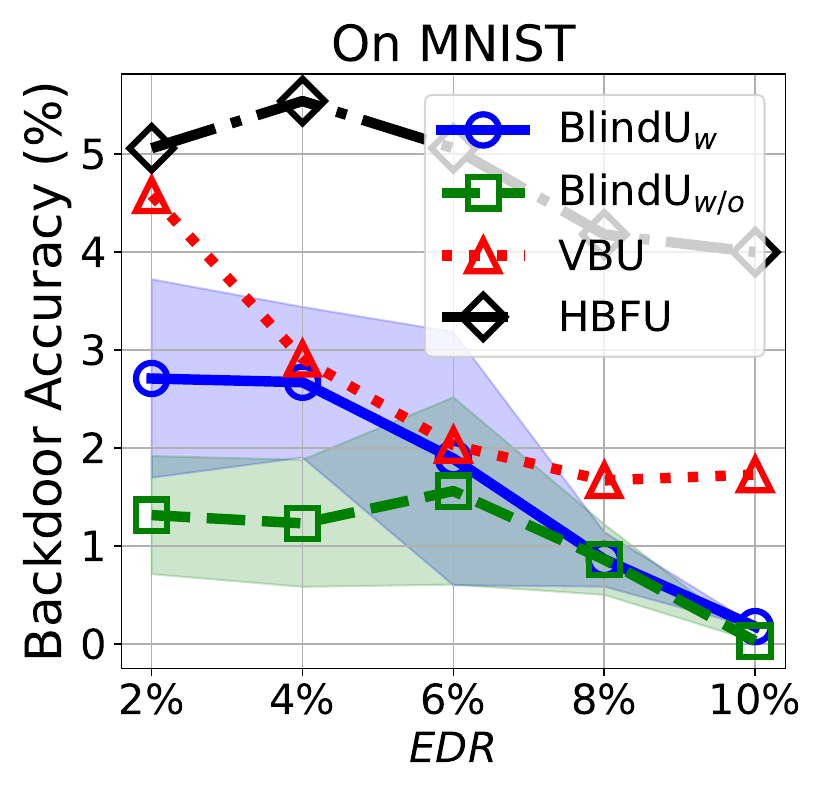}
	}
	\hspace{1mm}
\subfloat{ 	\label{fig:cifarbackaccercurve}
	\includegraphics[scale=0.25]{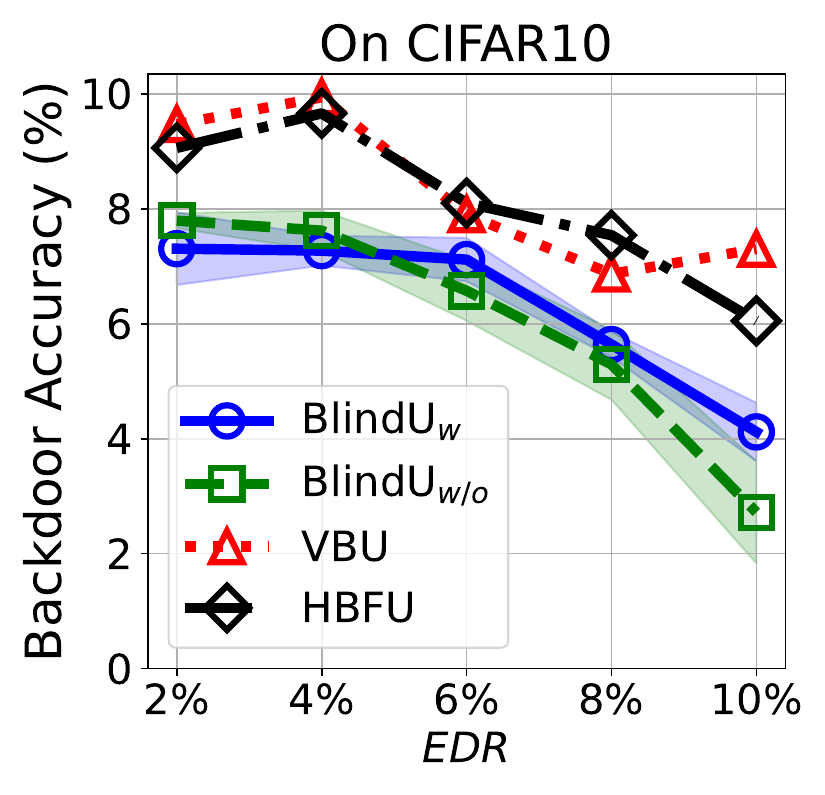}
}	\hspace{1mm}
\subfloat{ 	\label{fig:cifar100backaccercurve}
\includegraphics[scale=0.25]{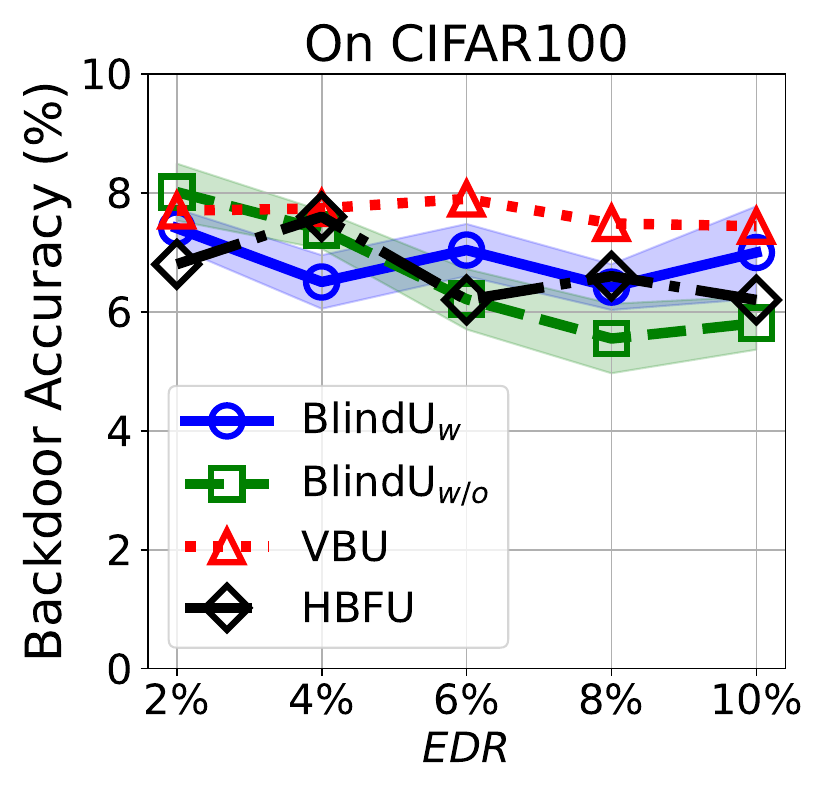}
}	\hspace{1mm}
\subfloat{  	\label{fig:tinyibackaccercurve}
\includegraphics[scale=0.25]{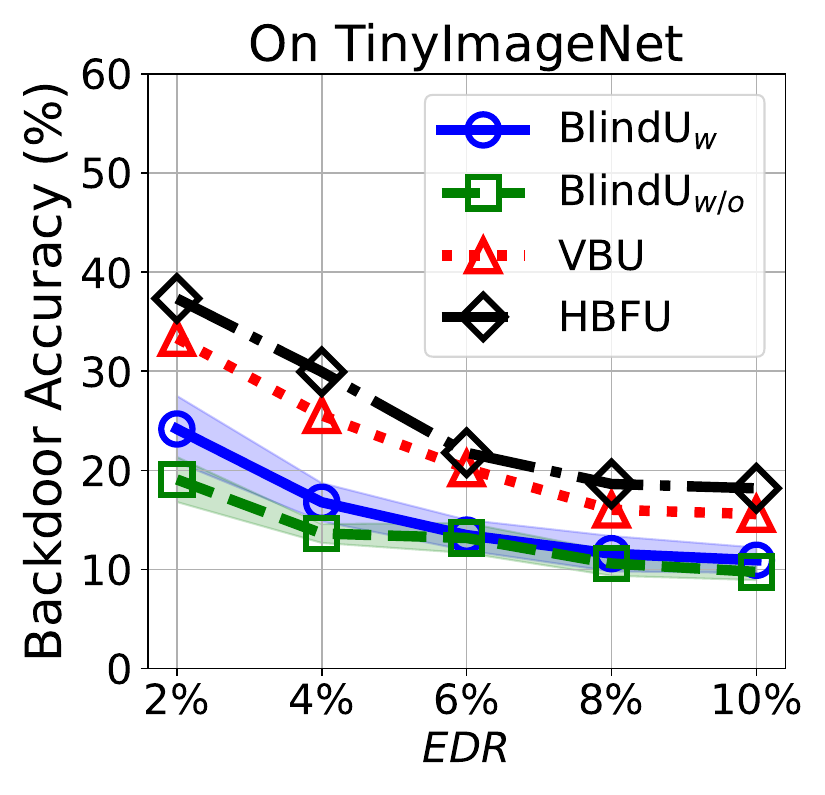}
}
\\	\vspace{-3mm}
 \subfloat{ \label{fig:mnistaccercurve}
 	\includegraphics[scale=0.25]{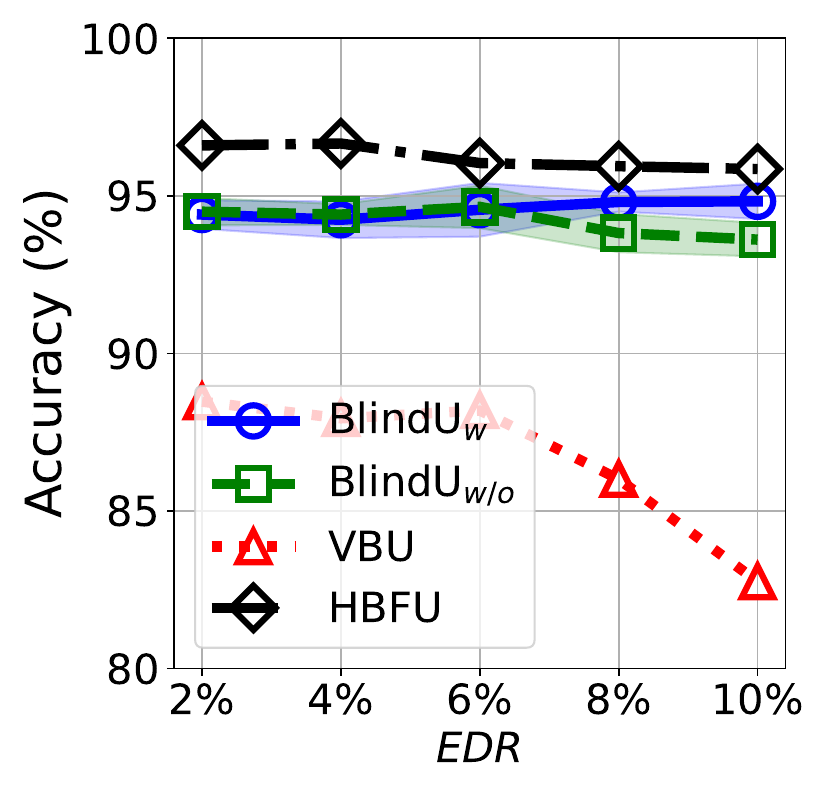}
 }
	\hspace{1mm}
	\subfloat{ \label{fig:cifaraccercurve}
		\includegraphics[scale=0.25]{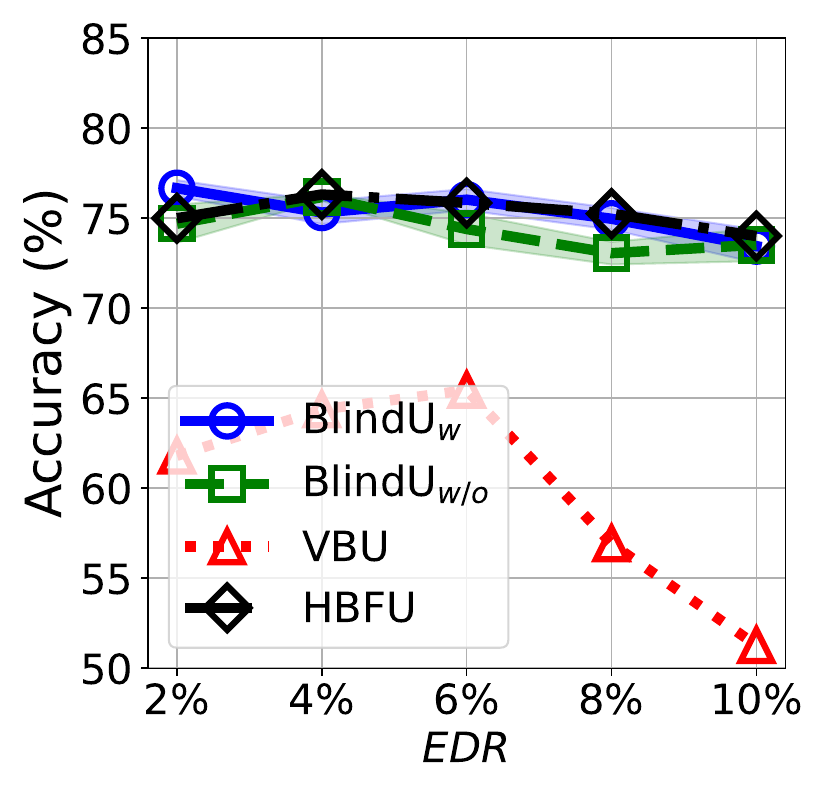}
	}
	\hspace{1mm}
	\subfloat{ 	\label{fig:cifar100accercurve}
		\includegraphics[scale=0.25]{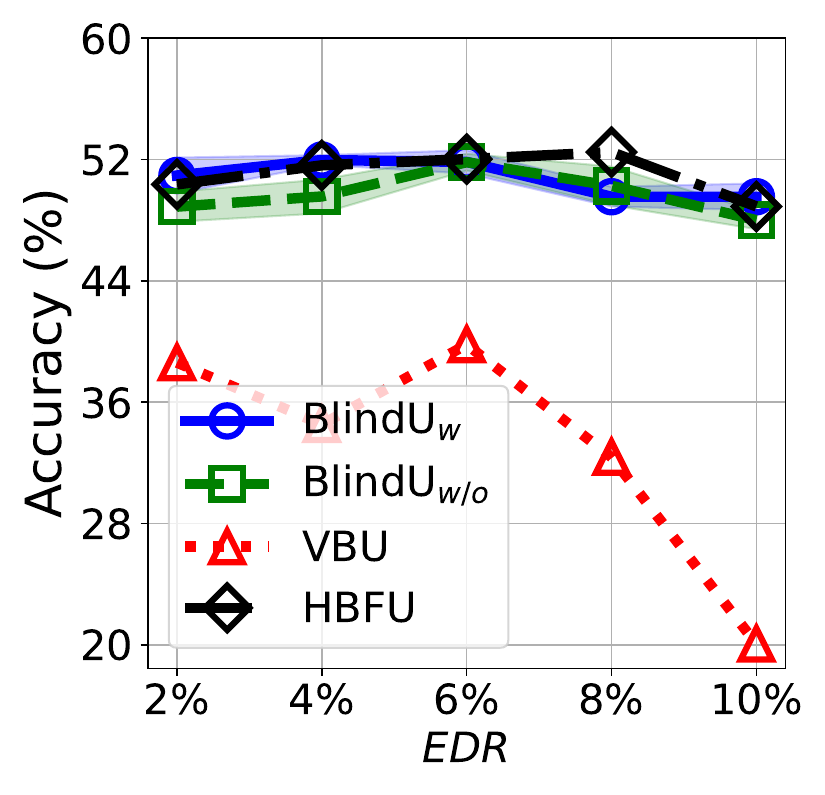}
	} 
	\hspace{1mm}
	\subfloat{ 	\label{fig:tinyiaccercurve}
		\includegraphics[scale=0.25]{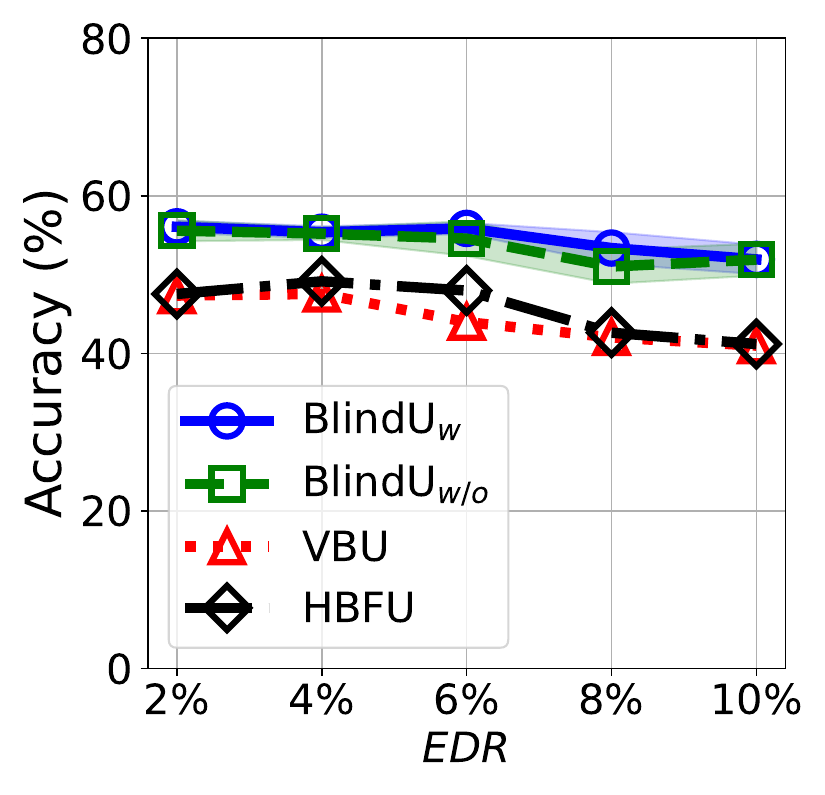}
	}
	\vspace{-3mm}
	\caption{Evaluations of effect about different \textit{EDR}.	\vspace{-2mm}
	} 
	\label{evaluation_of_edr} 
\end{figure*}

\subsection{Evaluation of Unlearning Effectiveness} \label{unlearning_evaluation}

We first evaluate the unlearning effectiveness by unlearning backdoored samples with different erasing data ratios (\textit{EDR}), which is commonly set from $2\%$ to $10\%$. We implement BlindU using two sampling strategies: random sampling with replacement BlindU$_{\text{w}}$ and random sampling without replacement BlindU$_{\text{w/o}}$. The compared methods are HBFU \citep{liu2022right} and VBU \citep{nguyen2020variational}.


\noindent
\textbf{Setup.} 
We evaluate different methods on MNIST, CIFAR10, CIFAR100, and TinyImageNet, shown in \Cref{evaluation_of_edr}. During the training process, we fixed the distorting ratio $\beta=0.001$ on MNIST, $\beta=0.01$ on CIFAR10 and CIFAR100, and $\beta=0.0001$ on TinyImageNet. Moreover, we set the DP sampling ratio $\textit{SR}=60\%$, which means we sample $60\%$ of the input features and mask the remaining $40\%$ features before compressing. We conducted experiments 5 times with the random seeds in [0, 1, 2, 3, 4] and reported the mean and standard deviation. We only report the mean value of each metric of HBFU and VBU for a clear illustration in figures.


\noindent
\textbf{Overall Unlearning Effectiveness.} 
In the first row of \Cref{evaluation_of_edr}, the results demonstrate that the effectiveness of backdoor erasure, in which the backdoor accuracy is mostly below $10\%$, while the backdoor accuracy is more than $90\%$ on the trained model. The backdoor accuracy exhibits a consistent downward trend across all datasets as the \textit{EDR} increases. This reduction confirms that the unlearning methods effectively scale to remove larger portions of malicious data, as the backdoor accuracy remains significantly low—mostly below $10\%$ on MNIST, CIFAR10, and CIFAR100. Notably, BlindU$_{\text{w/o}}$ achieves the best data erasure effect in almost all datasets and \textit{EDR} settings; this superiority is particularly evident on the complex TinyImageNet dataset, where BlindU${_\text{w/o}}$ reduces backdoor accuracy to approximately $10\%$, whereas the baseline methods HBFU and VBU struggle to drop below $20\%$ even at higher \textit{EDR} levels. 
	
The second row of \Cref{evaluation_of_edr} demonstrates model utility after unlearning. Our BlindU achieves comparable accuracy across the test datasets. While HBFU yields the highest accuracy on the simple MNIST dataset, its performance degrades on more complex datasets such as TinyImageNet, where it falls below BlindU. It is worth noting that HBFU’s accuracy preservation stems from its reliance on all FL users to assist in the unlearning process; normal users continue to execute original FL training on their local datasets. By aggregating these gradients, the global unlearning model is adjusted to maintain utility. However, this approach comes at a cost: reducing backdoor accuracy becomes more challenging because the gradients from normal users dilute the unlearning update.

 \begin{figure}[t]
	\centering
	\vspace{-2mm}
	\subfloat{  \label{fig:mnistclientmutualinfodetail}
		\includegraphics[scale=0.25]{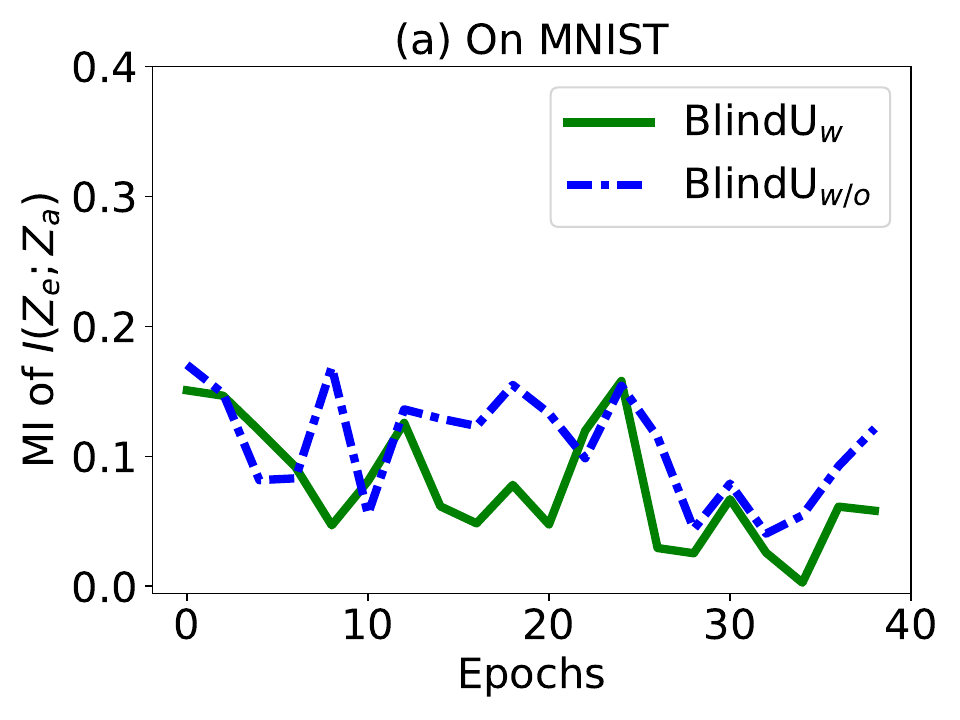}
	}
	\subfloat{ 	\label{fig:cifarmutualinfodetailacc}
		\includegraphics[scale=0.25]{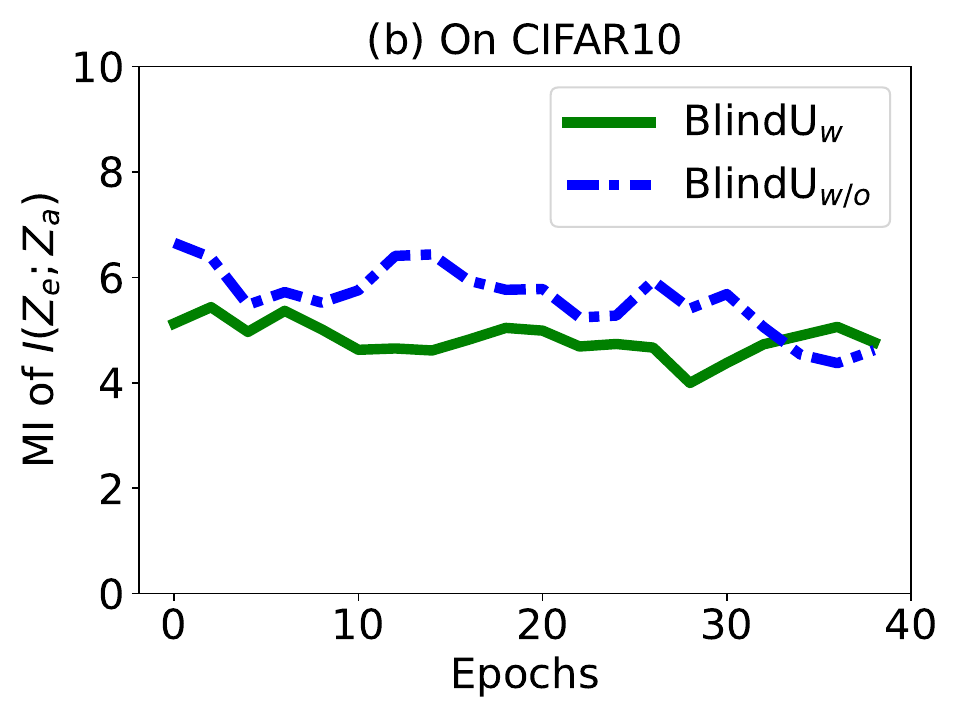}
	}
	\vspace{-2mm}
	\caption{Loss changes of $I(Z_e;Z_a)$ during Unlearning. 	\vspace{-2mm}}
	\label{MI_changes}
\end{figure}

\noindent
\textbf{Mutual Information Loss Behaviour in BlindU.} 
Since we minimize $I(Z_e;Z_a)$ to eliminate the information about the erased sample from the unlearned model representation, we record the changes in the mutual information loss during the unlearning training process, as presented in \Cref{MI_changes}. The \textit{EDR} is set as $6\%$ here, and other parameters are normal setting as introduced at the beginning. The degradation of mutual information on MNIST and CIFAR10 during the unlearning process demonstrates that our unlearning methods truly remove the key information in $Z_e$ of the erased samples from the unlearned representation.

\begin{figure}[t]
	\centering
	\subfloat{ 	\label{fig:mnistepochaccbaccotherform}
		\includegraphics[scale=0.23]{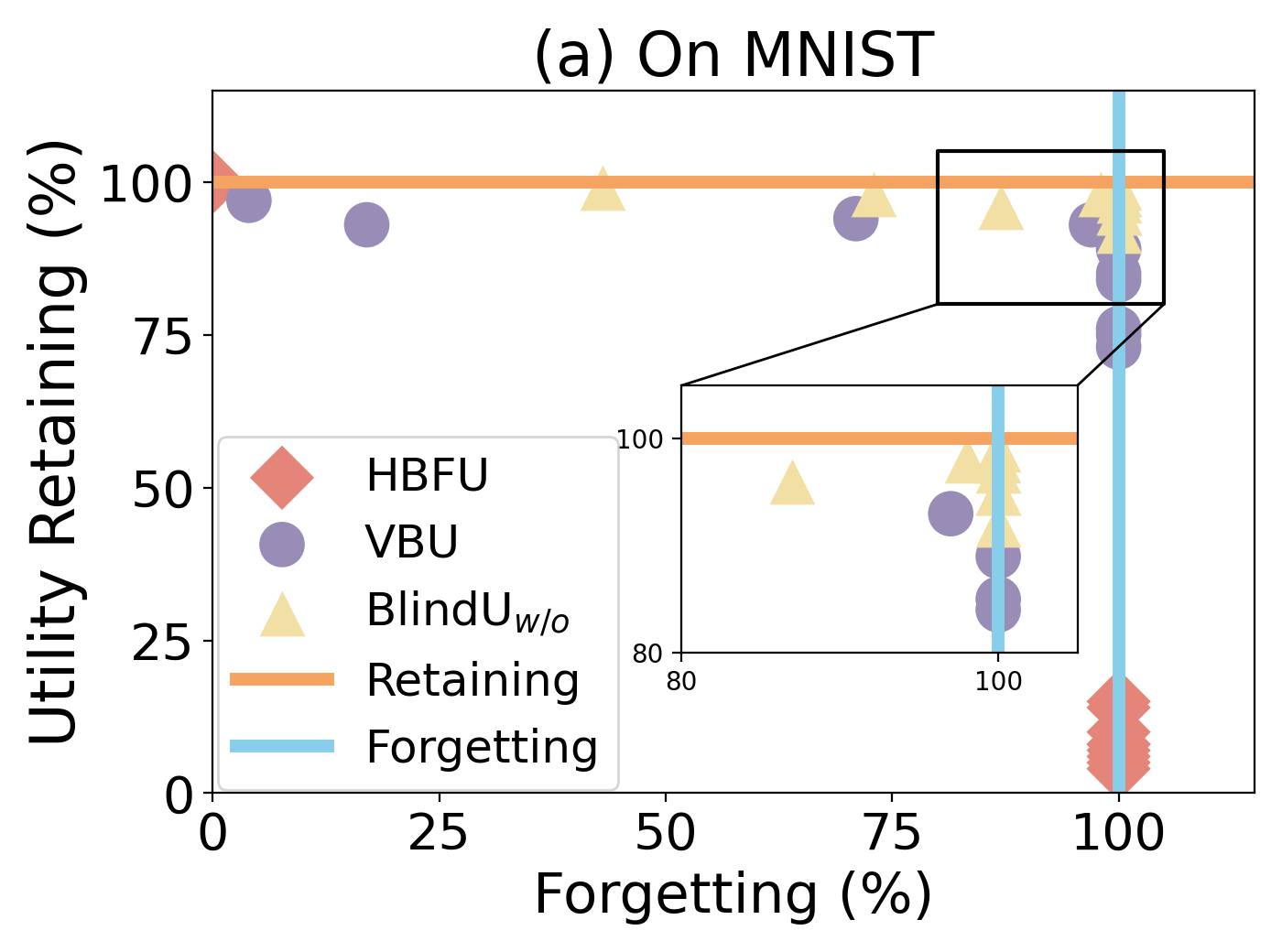}
	}
	\hspace{-2mm}
	\subfloat{ 	\label{fig:cifarepochaccbaccotherform}
		\includegraphics[scale=0.23]{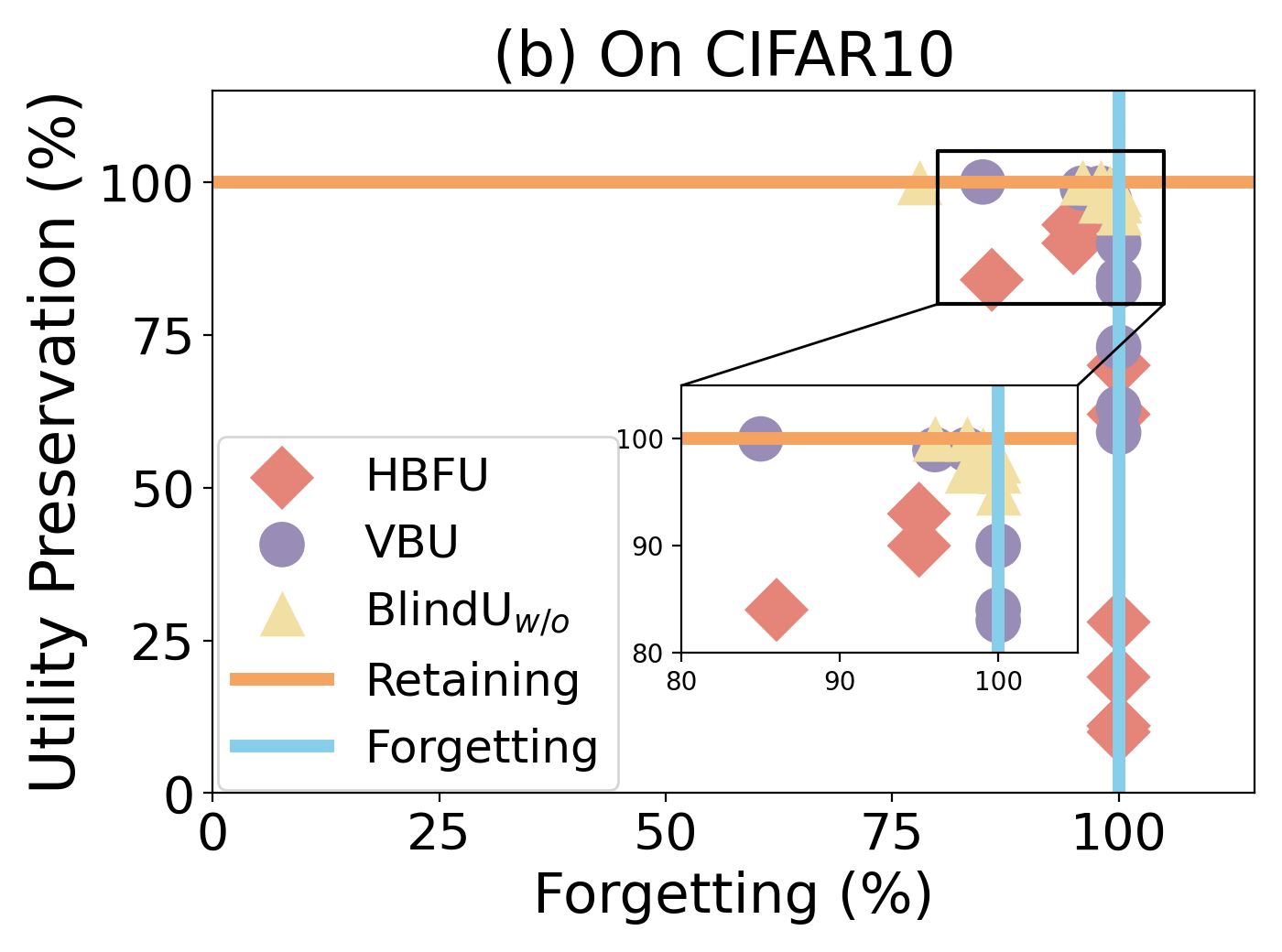}
	}
	\vspace{-2mm}
	\caption{Optimization results on during unlearning. The utility retaining is the model accuracy on the remaining dataset and $\texttt{Forgetting} =  1 - \texttt{backdoor accuracy}$.
	}
	\label{Pareto_change}
\end{figure}

\noindent
\textbf{Optimization during Unlearning.}
As we solving the final BlindU problem by MGDA for two objectives, retaining and forgetting. We record the optimization results on remaining and forgetting datasets during unlearning. The results of HBFU, VBU, and BlindU${_\text{w/o}}$ when $\textit{EDR}=6\%$ are reported in \Cref{Pareto_change}. The horizontal axis reports $\texttt{Forgetting} =  1 - \texttt{backdoor accuracy}$, while the vertical axis reports Utility Retaining, i.e., the model accuracy on the remaining data. Each marker traces intermediate checkpoints during unlearning for HBFU, VBU, and BlindU. The zoomed insets highlight the high-forgetting and high-utility region near the upper-right corner, where BlindU achieves the optimal balance of backdoor removal and retaining accuracy preservation, much better than HBFU and VBU. 


\begin{table}[t]
	\scriptsize
	\caption{ Effectiveness of Different Model Structures. \vspace{-2mm}
	}
	\label{test_of_different_model_structure}
	\resizebox{\linewidth}{!}{
		\setlength\tabcolsep{3pt}
		\begin{tabular}{c|c|cc|cc}
			\toprule
			& \multirow{2}{*} {Metrics} & \multicolumn{2}{c}{ResNet-18}& \multicolumn{2}{c} {ViT} \\
			\cmidrule(r){3-4}   \cmidrule(r){5-6}
			& & BlindU$_{\text{w}}$  & BlindU$_{\text{w/o}}$    & BlindU$_{\text{w}}$  & BlindU$_{\text{w/o}}$   \\
			\midrule 
			\multirow{2}{*} {   \rotatebox{90}{MNIST} } 
			& Bac. Acc. (\%) (before unlearning) & \meanstd{99.98}{0.01}     & \meanstd{99.98}{0.01}      & \meanstd{99.98}{0.01}    & \meanstd{99.98}{0.01}      \\   
			& Bac. Acc. (\%) (after unlearning)    & \meanstd{1.89}{1.29}  & \meanstd{1.56}{0.95}    & \meanstd{4.40}{1.12}   &   \meanstd{1.72}{0.83}  \\
			& Accuracy (\%) (after unlearning)   & \meanstd{94.55}{0.85}    &  \meanstd{94.65}{0.67}   &  \meanstd{96.38}{0.34}    & \meanstd{95.05}{0.46}   \\ 
			\midrule 
			\multirow{2}{*} {  \rotatebox{90}{CIFAR10} } 
			& Bac. Acc. (\%) (before unlearning) & \meanstd{99.93}{0.05}  & \meanstd{99.93}{0.05}   & \meanstd{99.94}{0.05}       & \meanstd{99.94}{0.05}   \\   
			&Bac. Acc. (\%) (after unlearning)    &  \meanstd{7.11}{0.37}  	 &  \meanstd{6.57}{0.51}   &  \meanstd{ 2.49}{0.81}      &  \meanstd{ 4.57}{0.98}   \\
			&Accuracy (\%) (after unlearning)  &  \meanstd{75.83}{0.94}   &  \meanstd{74.39}{0.77}  &   \meanstd{ 76.14}{0.32}        &  \meanstd{ 74.95}{0.39}   \\  
			\bottomrule[1pt]
	\end{tabular}}
\end{table}


\noindent
\textbf{Evaluation of Different Model Structures.} 
To assess the versatility of our framework, we evaluated the unlearning effectiveness across different network architectures. Specifically, we tested the performance using ResNet-18 and Vision Transformers (ViT) as the underlying backbones of IB approximator. The results on MNIST and CIFAR10, presented in \Cref{test_of_different_model_structure}, demonstrate that our method effectively reduces backdoor accuracy while preserving main task utility across both architectures. This indicates that the BlindU framework is scalable and robust to different model structures.

 \begin{figure}[t]
	\centering
	\subfloat{	\label{fig:mnistrterbar} 
		\includegraphics[scale=0.25]{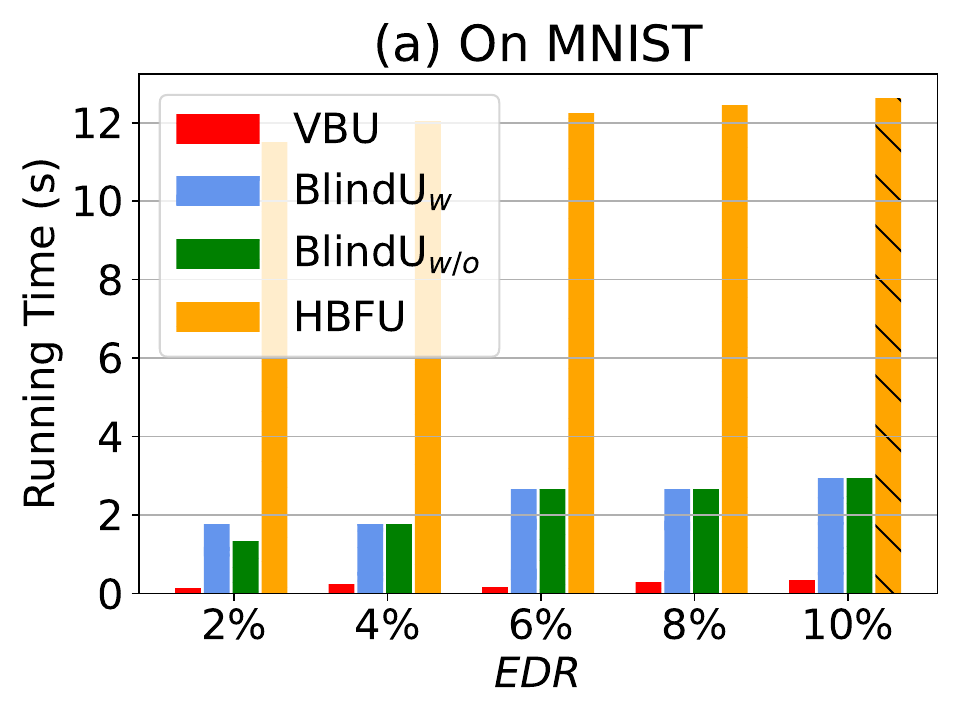}
	}
	\subfloat{ \label{fig:cifarrterbar} 
		\includegraphics[scale=0.25]{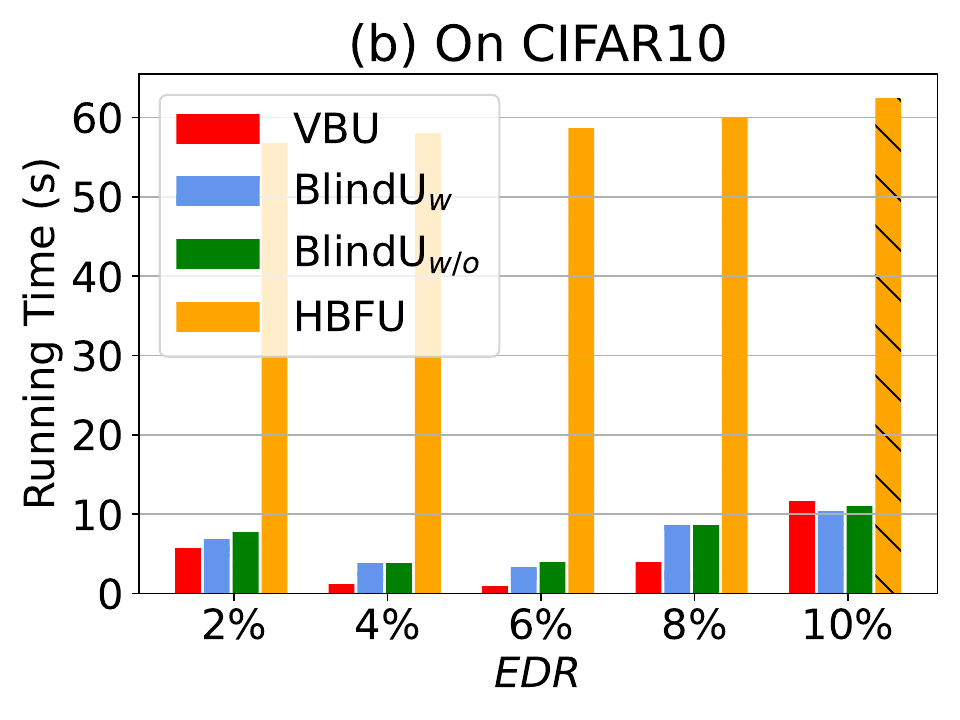}
	}
	\vspace{-2mm}
	\caption{Evaluations of efficiency about different \textit{EDR}. 
	}
	\label{fig_runningtime_er}
\end{figure}

\noindent
\textbf{Unlearning Efficiency.}
\Cref{fig_runningtime_er} shows the running time of different unlearning methods. As \textit{EDR} increases, all unlearning methods do take more time to erase the specified samples. VBU, BlindU$_{\text{w}}$ and BlindU$_{\text{w/o}}$ achieve a speedup of over $5\times$ compared to HBFU on all datasets. HBFU performs worse than VBU and our proposed BlindU because it needs to calculate the hessian matrix and relies on all users to help with retraining. VBU performs best efficiently because it calculates only based on the erasing dataset.







\subsection{Comparison between Unlearning Backdoored and  Multiple-Class Normal IID Data} \label{normal_unl_eval}


Previous experiments have demonstrated the effectiveness of unlearning backdoored samples, offering a direct measure of unlearning performance. However, in real-world settings, most unlearning requests likely involve normal independent and identically distributed (IID) data. In this section, we randomly select $\textit{EDR}=6\%$ normal IID samples as the erasing data, and compare the differences between unlearning backdoored samples and normal IID samples in unlearning.

\begin{figure}[t]
	\vspace{-2mm}
	\centering
	\subfloat[\footnotesize Unlearning backdoored samples]{ \label{fig:mnistepochbackacctemp}
		\hspace{-4mm}
		\includegraphics[scale=0.255]{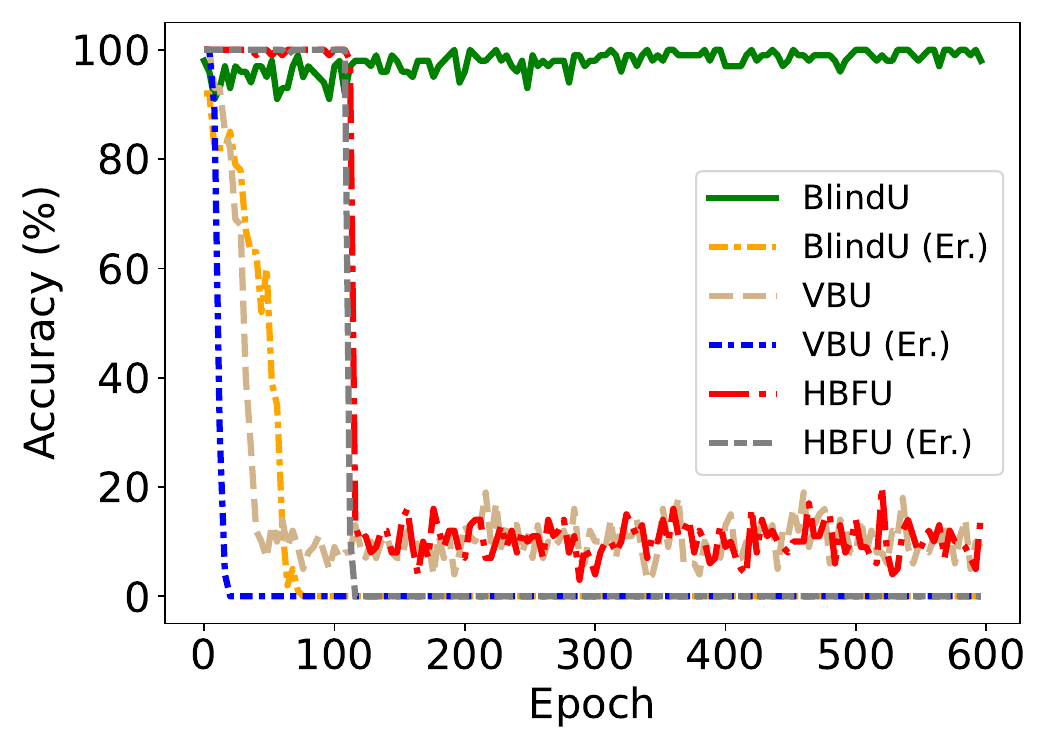}
	}
	\subfloat[\footnotesize Unlearning normal IID samples]{ 	\label{fig:mnistepochnobackacctemp}
		\hspace{-4mm}
		\includegraphics[scale=0.255]{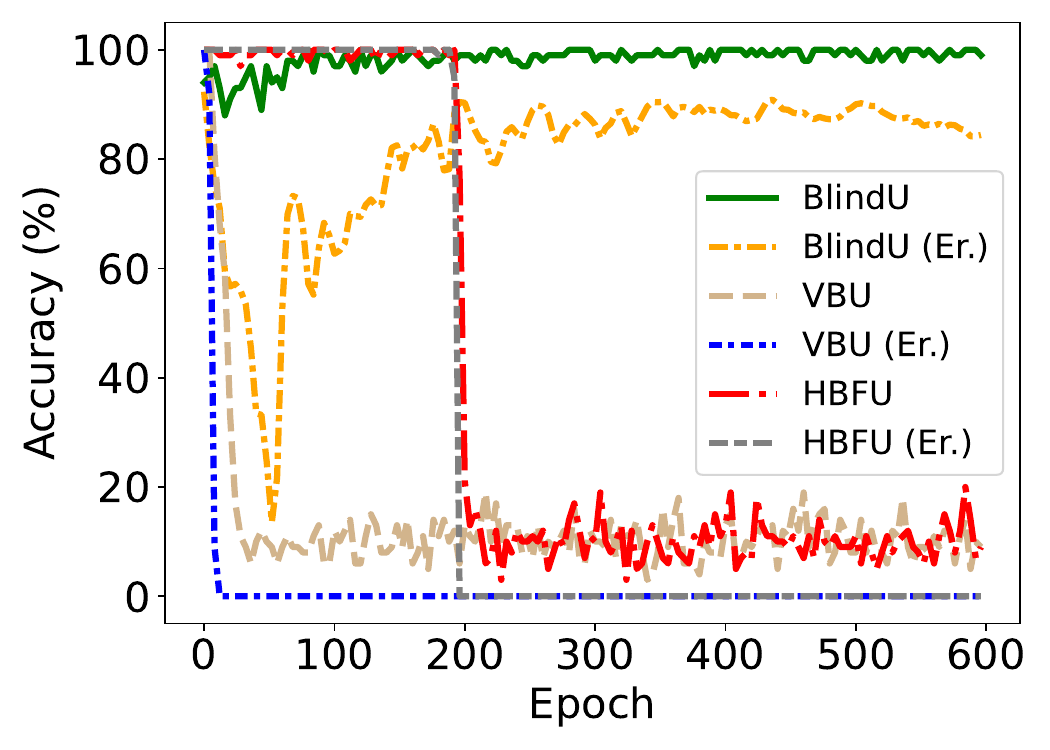}
	}
	\vspace{-2mm}
	\caption{Detailed changes of accuracy on the remaining and erasing (abbreviated Er.) datasets during unlearning training on MNIST.
	} 
	\label{changes_compare} 
\end{figure}

\noindent
\textbf{Difference in Training Process.}
\Cref{changes_compare} compares the changes in model accuracy on the remaining and erasing datasets when unlearning backdoored and normal IID samples on MNIST, where the detailed changes of HBFU are conducted on the unlearning user side. As shown in \Cref{fig:mnistepochbackacctemp}, all unlearning methods effectively remove the backdoor, but model accuracy declines significantly during unlearning. Only BlindU maintains satisfactory accuracy throughout the training process because we solve the unlearning problem as constrained IB retraining. In contrast, both VBU and HBFU show minimal differences between the remaining and erasing datasets, making it difficult to determine an optimal stopping point. This indicates that VBU and HBFU struggle to achieve unlearning without compromising model utility. BlindU, however, consistently maintains accuracy while effectively removing the backdoor at any stage of training.

When unlearning the normal IID data in \Cref{fig:mnistepochnobackacctemp}, HBFU and VBU perform similarly to unlearning backdoored samples. The model accuracy on the remaining dataset declines rapidly as the decrease of model accuracy on the erasing dataset. In contrast, BlindU decreases model accuracy on the erasing data at the beginning, accompanied by a slight model accuracy reduction on the remaining dataset. As the training processes, the main retraining terms in BlindU recovers the accuracy, and finally, the model converges at around $99\%$ accuracy on the remaining dataset and around $80\%$ on the erasing dataset. The gap is around $20\%$, much higher than the changes' gaps of HBFU and VBU.



\noindent
\textbf{Final Results Comparison.}
We compare the performance of unlearning backdoored versus normal IID samples on MNIST, CIFAR10, and Adult with $\textit{EDR}=6\%$, as presented in \Cref{table_comparison_of_bac_and_normal_IID}. In the backdoor scenario, BlindU consistently achieves effective erasure with minimal utility loss, whereas HBFU fails to unlearn the backdoor on the Adult dataset ($99.99\%$ accuracy on $D_e$). When transitioning to unlearning normal IID samples, the challenge to model utility becomes more pronounced. For instance, on MNIST, we stop the training of HBFU and VBU when the accuracy on the erasing dataset reaches approximately $17\%$; even at this threshold, both methods suffer significant degradation in test accuracy compared to BlindU. These results indicate that unlearning normal samples poses a greater risk to the original model's utility, a challenge that BlindU mitigates more effectively than the baselines.

\begin{table}[t]
	\scriptsize
	\caption{ Unlearning backdoored and normal IID samples. \vspace{-2mm}}
	\label{table_comparison_of_bac_and_normal_IID}
	\resizebox{\linewidth}{!}{
		\setlength\tabcolsep{3pt}
		\begin{tabular}{c|c|cc|cc}
			\toprule
			& \multirow{2}{*} {Methods} & \multicolumn{2}{c} {Unlearning backdoored samples}& \multicolumn{2}{c} {Unlearning normal samples} \\
			\cmidrule(r){3-4}   \cmidrule(r){5-6}
			& & Acc. (\%) on testset  & Acc. (\%) on $D_e$   &Acc. (\%) on testset		  		 & Acc. (\%) on $D_e$    \\
			\midrule 
			\multirow{4}{*} { \rotatebox{90}{MNIST} } &HBFU     & \textbf{ \meanstd{95.86}{0.47} }   &  \meanstd{5.05}{0.92}     &\meanstd{86.64}{0.52}         & \meanstd{17.35}{0.81}  \\
			& VBU    &\meanstd{88.32}{0.73}     &  \meanstd{2.02}{0.41}   & \meanstd{89.19}{0.77}       & \meanstd{17.7}{0.57}    \\
			&BlindU$_{\text{w}}$   & \meanstd{94.55}{0.85}    & \meanstd{1.89}{1.29}  & \textbf{ \meanstd{93.01}{0.65} }        & \meanstd{17.19}{0.85}    \\  
			&BlindU$_{\text{w/o}}$       & \meanstd{94.65}{0.67}      	 & \textbf{ \meanstd{1.56}{0.95} }	   &\meanstd{91.38}{0.72}    & \textbf{ \meanstd{14.00}{1.08}  }  \\
			\midrule 
			\multirow{4}{*} { \rotatebox{90}{CIFAR10} } &HBFU     & \textbf{ \meanstd{75.83}{0.52} }    &  \meanstd{8.11}{0.45}     & \meanstd{71.32}{0.64}         &  \meanstd{ 68.28}{0.83}   \\
			& VBU    &\meanstd{65.45}{1.64}   &  \meanstd{7.92}{0.55}     & \meanstd{ 69.16}{0.83}      & \textbf{  \meanstd{  63.93}{1.32} } \\
			&BlindU$_{\text{w}}$   & \meanstd{75.83}{0.94}    & \meanstd{7.11}{0.37}  & \textbf{ \meanstd{ 75.21}{0.67}   }        & \meanstd{ 68.02}{0.79}   \\  
			&BlindU$_{\text{w/o}}$       & \meanstd{74.39}{0.77}    	 & \textbf{ \meanstd{6.57}{0.51} }  &  \meanstd{ 73.89}{0.73}     &  \meanstd{ 67.66}{0.86}      \\
			\midrule 
			\multirow{4}{*} { \rotatebox{90}{Adult} } &HBFU     & \textbf{ \meanstd{85.45}{0.43}  }    & \meanstd{99.99}{0.0}   & \meanstd{83.03}{0.58}        & \meanstd{80.89}{0.73}   \\
			& VBU    & \meanstd{64.64}{0.34}      &  \meanstd{8.45}{0.35}      &\meanstd{51.32}{1.03}        & \textbf{ \meanstd{7.82}{2.33}  } \\
			&BlindU$_{\text{w}}$   & \meanstd{84.34}{0.53}      &  \meanstd{9.44}{0.41}    & \textbf{ \meanstd{ 84.08}{0.36}   }        & \meanstd{ 44.33}{7.37}    \\  
			&BlindU$_{\text{w/o}}$       & \meanstd{84.05}{0.46}      	 & \textbf{ \meanstd{7.25}{0.43}  }	   &\meanstd{ 84.05}{0.58}     & \meanstd{ 34.02}{2.87}   \\
			\bottomrule[1pt]
	\end{tabular}}
\end{table}

\subsection{Evaluation of Privacy Protection} \label{privacy_evaluation}

BlindU offers dual privacy protection through compression and DP sampling, as theoretically analyzed in \Cref{theoretical}. In this section, we first validate the specific differential privacy guarantees achieved by our sampling-based masking. Subsequently, we demonstrate the robustness of BlindU against state-of-the-art privacy reconstruction and membership inference attacks \cite{salem2020updates,Hu2024sp,chen2021machine} in the context of machine unlearning. 

\begin{figure}[t]
	\centering
	\vspace{-2mm}
	\hspace{-4mm}
	\subfloat{ \label{mnist_acc_er_curve} 
		\includegraphics[scale=0.278]{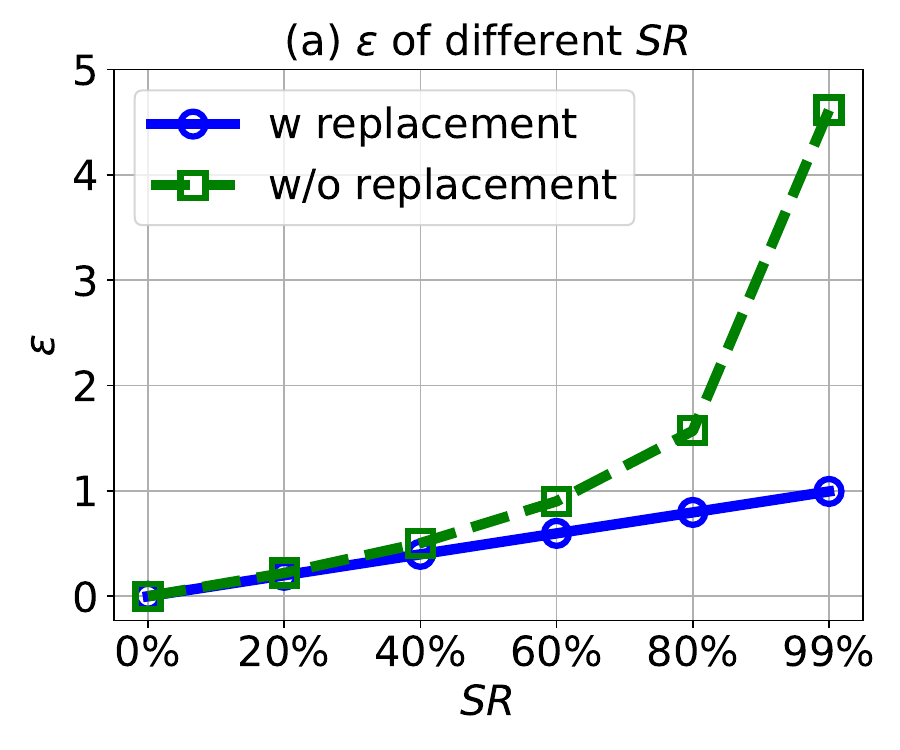}  
	}
	\hspace{-2mm}
	\subfloat{ \label{cifar_acc_er_curve} 
		\includegraphics[scale=0.278]{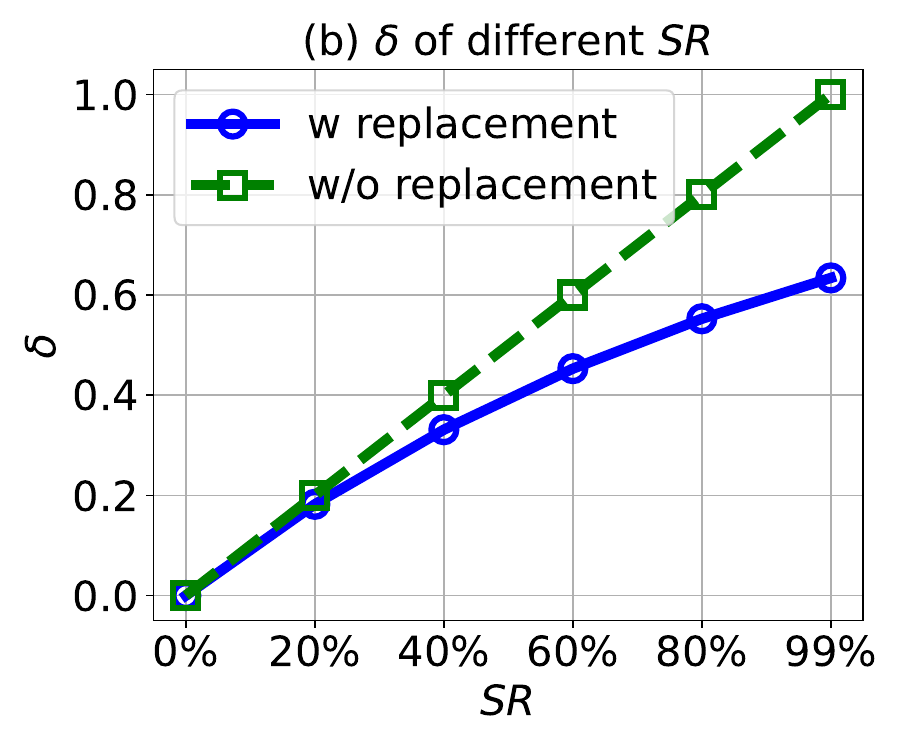}  
	}
	\vspace{-2mm}
	\caption{$\epsilon$ and $\delta$ comparison between without (abbreviated as w/o) replacement and with (abbreviated as w) replacement. \vspace{-2mm}} 
	\label{epsilon_delta} 
\end{figure}

\begin{table}[t]
	\caption{ Sampling with (w) and without (w/o) replacement on MNIST. \vspace{-2mm} }
	\label{sampling_results_on_mnist}
	\centering
	\resizebox{\linewidth}{!}{
		\setlength\tabcolsep{3pt}
	\begin{tabular}{c|c|c|c|c|c}
		\toprule
		\diagbox[width=7em]	 	 &${\textit{SR}}$ &$\epsilon$&$\delta$&Bac. Accuracy (\%)& Model Accuracy (\%)\\
		\midrule[0.8pt] 
		HBFU&$100\%$&$+\infty$&0&\meanstd{5.08}{1.13} & \meanstd{96.04}{0.32}  \\
		{DP-HBFU }& {$100\%$} & {2 }&{0}& \meanstd{23.66}{3.39} &  \meanstd{75.94}{1.64}\\
		\midrule 
		{BFU} & {$100\%$ }& {$+\infty$}& {0} &\meanstd{8.52}{0.96} &  \meanstd{96.38}{0.31} \\
		{DP-BFU} & {$100\%$} & {2} & {0 } & \meanstd{27.31}{3.53}   & \meanstd{81.73}{2.17}  \\
		\midrule 
		VBU & $100\%$ &$+\infty$ & 0 &   \meanstd{2.03}{0.25} &  \meanstd{88.19}{0.76}  \\
		{DP-VBU} & {$100\%$} & {2} & {0 } & \meanstd{2.33}{0.36}   & \meanstd{48.02}{4.37}   \\
		\midrule 
		BlindU$_{\text{w}}$ & $20\%$ & 0.199 & 0.182 & \meanstd{3.31}{0.51}  &  \meanstd{95.86}{0.30} \\
		BlindU$_{\text{w}}$  & $40\%$ & 0.398 & 0.331 & \meanstd{2.87}{0.67}   &  \meanstd{95.30}{0.24}  \\
		BlindU$_{\text{w}}$  & $60\%$ & 0.597 & 0.453 & \meanstd{1.89}{1.20} &  \meanstd{94.56}{0.85} \\
		\midrule 
		BlindU$_{\text{w/o}}$ & $20\%$ & 0.221 & 0.2 & \meanstd{2.11}{0.49} &  \meanstd{94.67}{0.34} \\
		BlindU$_{\text{w/o}}$  & $40\%$ & 0.504 & 0.5 & \meanstd{2.14}{0.50}&  \meanstd{93.93}{0.43} \\
		BlindU$_{\text{w/o}}$  & $60\%$ & 0.902 & 0.8 & \meanstd{1.56}{0.95} &  \meanstd{94.65}{0.68} \\
		\bottomrule[1pt]
	\end{tabular}}
\end{table}

\subsubsection{DP Achievement by Masking with Sampling}


Claims \ref{dp_theorem_with_replacement} and \ref{dp_theorem_without_replacement} indicate that different sampling strategies will result in different DP protection levels, i.e., different $\epsilon$ and $\delta$. Moreover, the sampled feature sizes will also determine different $\epsilon$ and $\delta$. Assuming the original features dimension of a single data point is $n$, and the sampled features size is $k$ where $k \leq n$, we define the sampling rate as $\textit{SR}=\frac{k}{n}$. Figure \ref{epsilon_delta} reflects the relationship of \textit{SR}, and $\epsilon$ and $\delta$, and it shows that sampling with replacement always has better DP protection than sampling without replacement.



\Cref{sampling_results_on_mnist} shows the unlearning effectiveness about backdoor accuracy (abbreviated as bac. acc.) and model accuracy (abbreviated as model acc.) after unlearning of different \textit{SR} on MNIST. An MNIST fixed-size image contains 28 × 28 pixels, i.e., 784 features. We sample $\textit{SR} = \frac{k}{n}$ of features to show the original value and mask the remaining unsampled features. In \Cref{sampling_results_on_mnist}, random sampling with replacement, BlindU$_{\text{w}}$, achieves better privacy protection at any sampling rate \textit{SR}, i.e., BlindU$_{\text{w}}$ achieves a lower $(\epsilon,\delta)$. 
{At the same time, BlindU$_{\text{w}}$ achieves a worse backdoor elimination effect than BlindU$_{\text{w/o}}$, as the compressed data of BlindU$_{\text{w/o}}$ contains more information. Moreover, we supplement a new federated unlearning comparison, BFU \cite{wang2023bfu}, and we add baselines of directly injecting DP-noise into the gradient (DP-HBFU and DP-BFU) and erased data (DP-VBU). Here, we set $\epsilon =2$ for DP-HBFU, DP-BFU, and DP-VBU, much larger than the $\epsilon$ value achieved by DP masking. However, the DP noise heavily hinders the unlearning effect and model utility. The highest accuracy degradation is around $40\%$.}

\subsubsection{Defense against Reconstruction Attacks}




To evaluate the privacy protection of the compressed representation (BlindU) and FL gradients (HBFU), we conduct the privacy reconstruction attack referring \cite{salem2020updates,Hu2024sp}.
Specifically, in \cite{salem2020updates,Hu2024sp}, the authors defined the difference between ``two'' versions of models as ``posterior difference'', $\delta$, and implemented inference and reconstruction attacks based on $\delta$. We treat the FL gradients of HBFU and compressed data of BlindU as the ``difference'' $\delta$, which is the input of the attacking model. Except for the input difference, the structure of the attacking model remains the same. When we evaluate the impact of $\beta$, we fix the \textit{SR} value, and vice versa.



\begin{figure}[t]
	\centering
	\hspace{-5mm}
	\subfloat[\footnotesize Reconstruction of different $\beta$]{ \label{fig_mnistbetarecontruction}  \rotatebox{90}{ \hspace{10mm}	\scriptsize{On MNIST} }
		\includegraphics[scale=0.25]{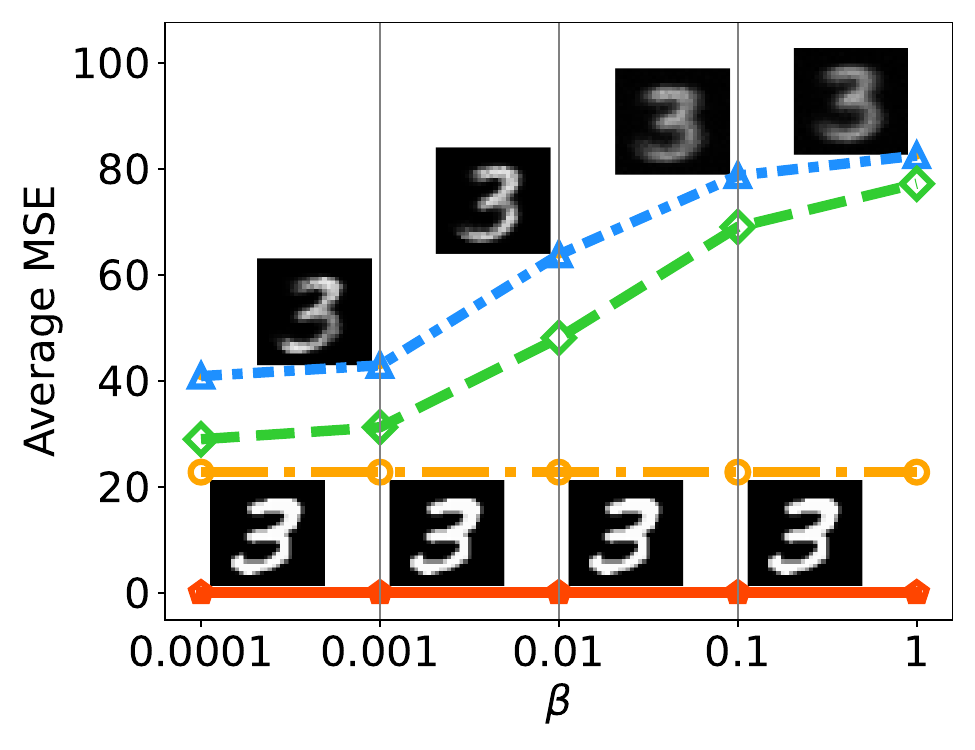} }
	\hspace{-1.5mm}
	\subfloat[\footnotesize Reconstruction of different ${\it SR}$]{ \label{fig_mnistsrrecontruction}
		\includegraphics[scale=0.25]{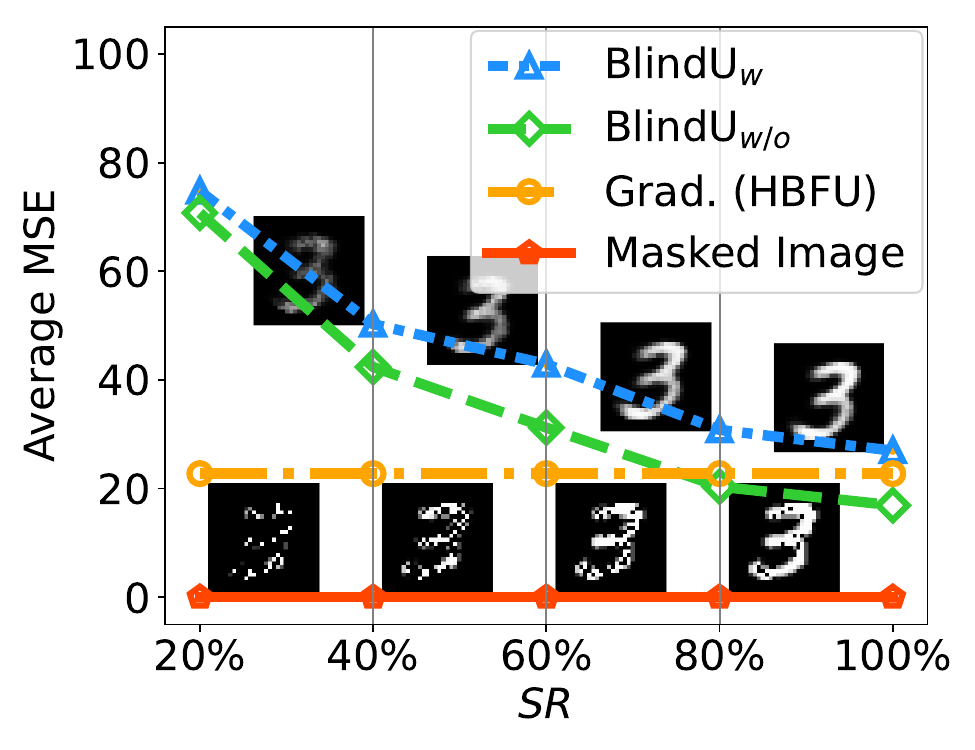} }
	\\
	\hspace{-5mm}
	\subfloat[\footnotesize Reconstruction of different $\beta$]{ \label{fig:cifarbetarecontruction}  \rotatebox{90}{ \hspace{10mm}	\scriptsize{On CIFAR10} }
		\includegraphics[scale=0.25]{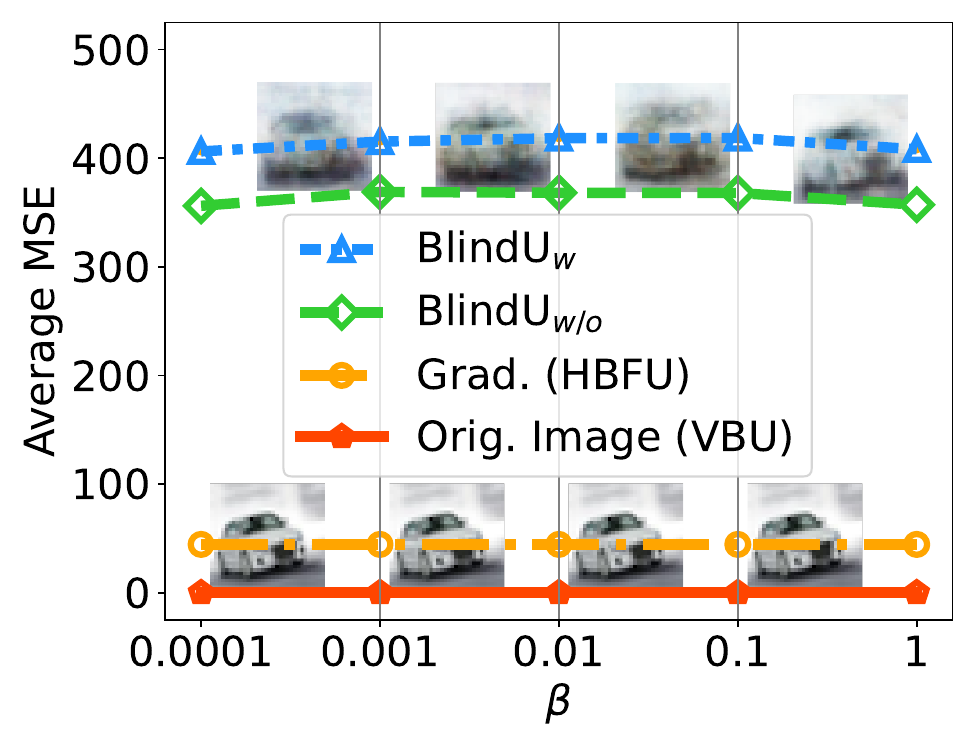} }
	\hspace{-1.5mm}
	\subfloat[\footnotesize Reconstruction of different ${\it SR}$]{ \label{fig:cifarsrrecontruction}
		\includegraphics[scale=0.25]{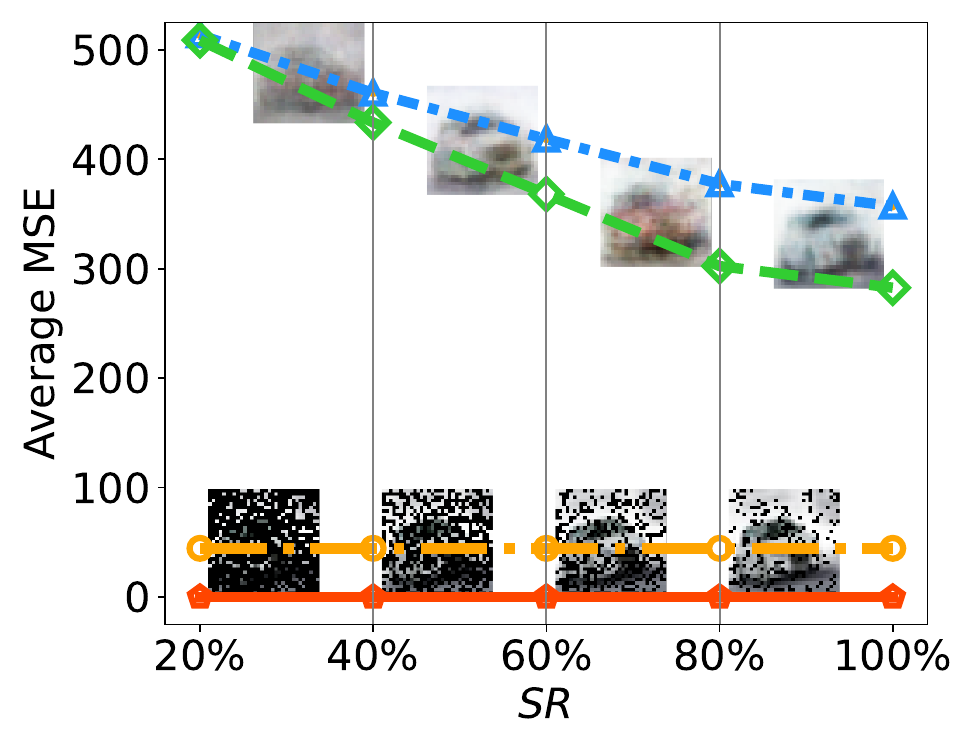} }
	\caption{Reconstruction MSE on MNIST and CIFAR10 based on the masked and compressed data, the training gradients (abbreviated as Grad.) and original (abbreviated as Orig.) data.  \vspace{-4mm}
	} 
	\label{reconstrut_on_mnist} 
\end{figure}

The reconstructing results on MNIST and CIFAR10 are shown in \Cref{reconstrut_on_mnist}. We show the average reconstruction MSE of recovering the original images based on the compressed data with two sampling strategies and based on the gradients of HBFU. In \Cref{fig_mnistbetarecontruction,fig:cifarbetarecontruction}, the difficulty of reconstruction based on compressed data increases when $\beta$ increases, corroborating our previous assertion. 
BlindU$_{\text{w}}$ and BlindU$_{\text{w/o}}$ always achieve better privacy protection than HBFU. 
Moreover, the reconstruction attacking results of different sampling rates \textit{SR} are shown in \Cref{fig_mnistsrrecontruction,fig:cifarsrrecontruction}. When \textit{SR} is small, fewer features are sampled, and more are masked. The compressed representation contains less information for reconstruction, leading to a higher reconstruction MSE. And the sampling with replacement, BlindU$_{\text{w}}$ has a better protection than BlindU$_{\text{w}/\text{o}}$. These results are consistent with the previous analysis.

\subsubsection{Defense against Membership Inference Attacks}

\begin{figure} [t]
	\centering
	\subfloat{ 	\label{fig:inferaccbetaoncifar10} \rotatebox{90}{ \hspace{9mm}	\scriptsize{On MNIST} }
		\includegraphics[scale=0.26]{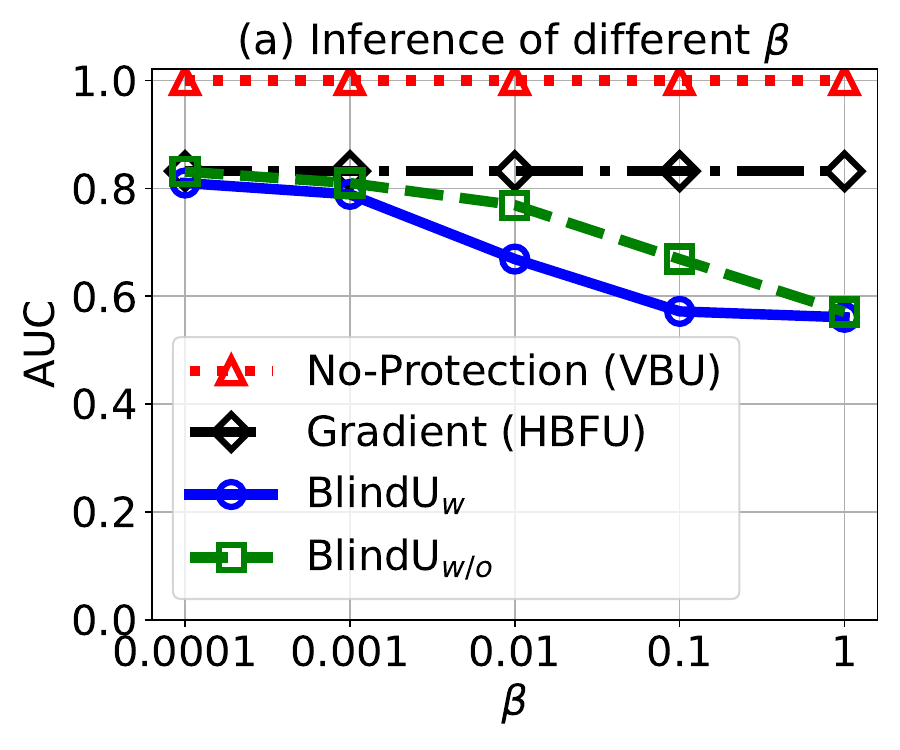} 
	}
	\subfloat{ 	\label{fig:inferaccsroncifar10}
		\includegraphics[scale=0.26]{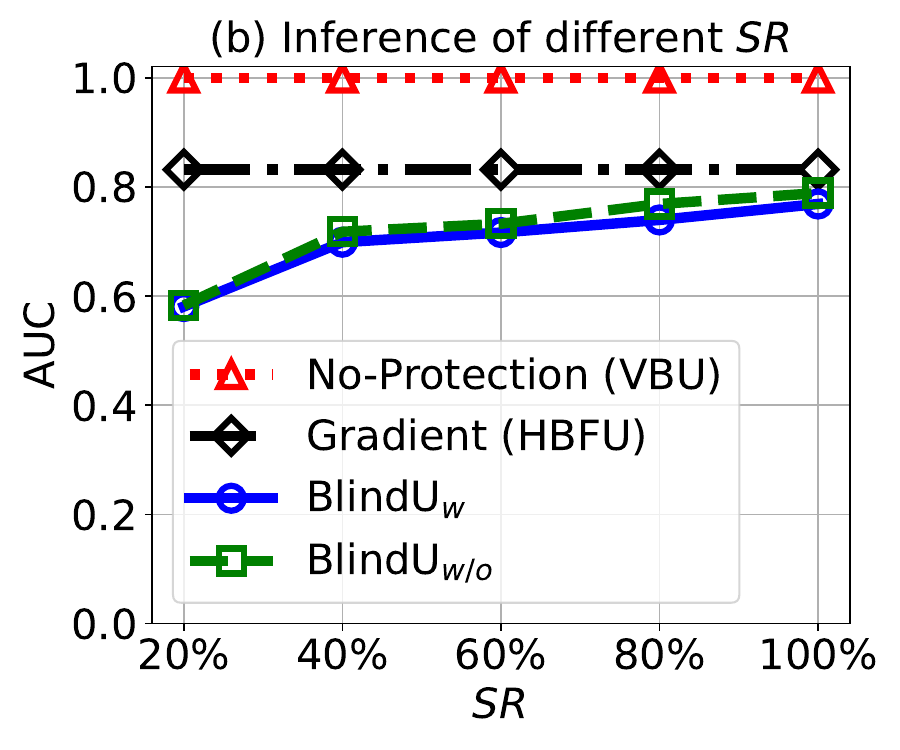} 
	} \\
	\vspace{-3mm}
	\subfloat{ 	\label{fig:inferaccbetaonmnist}  \rotatebox{90}{ \hspace{9mm}	\scriptsize{On CIFAR10} }
		\includegraphics[scale=0.26]{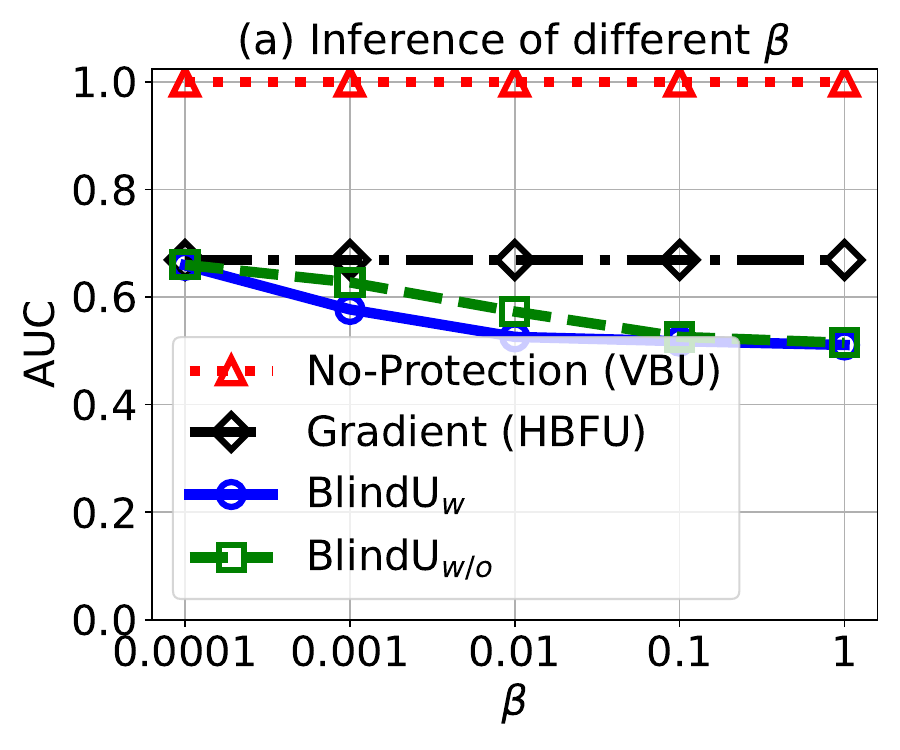}
	}
	\subfloat{ 	\label{fig:inferaccsronmnist}
		\includegraphics[scale=0.26]{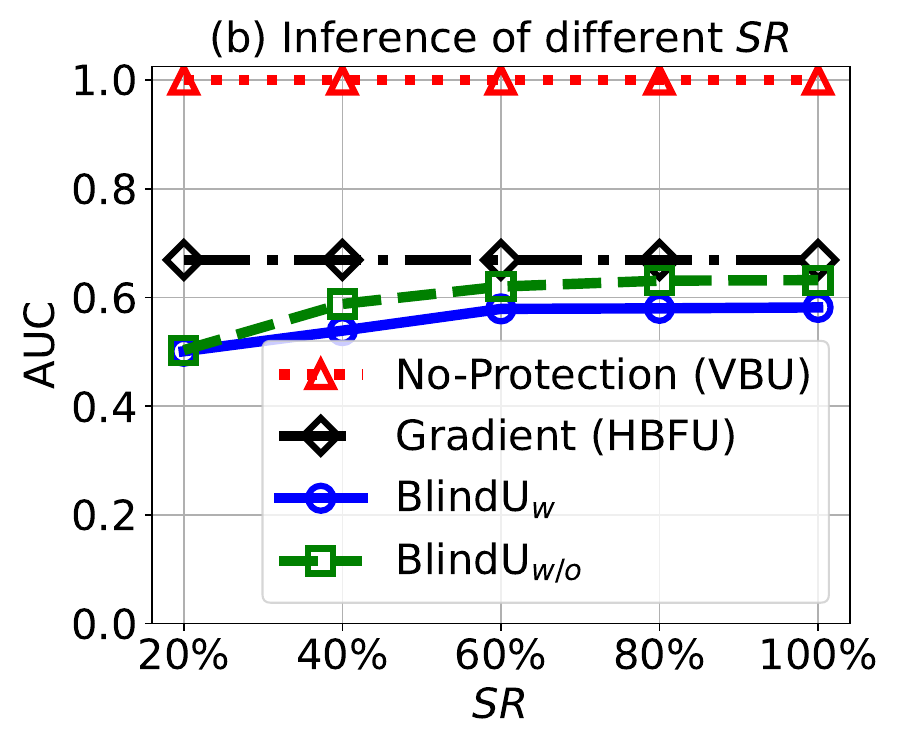}
	}
	\vspace{-2mm}
	\caption{{Membership inference AUC on MNIST and CIFAR10 from the compressed data, the training gradients (HBFU). \vspace{-2mm}
		}
	} 
	\label{MI_mnist_cifar} 
\end{figure}

We also conduct the membership inference attacks according to \cite{chen2021machine,salem2020updates} to infer the private membership information. Specifically, following \cite{chen2021machine}, we infer whether a target sample is part of the training set of the original model but out of the training set of the unlearned model. The inference attacks is conducted based on the model gradients (HBFU) and the masked-compressed representation (BlindU). The results are demonstrated in \Cref{MI_mnist_cifar}. Similar to the analysis of reconstruction attacks, a larger compressive ratio $\beta$ guarantees better privacy protection, decreasing the attacking AUC as $\beta$ increases. A smaller \textit{SR} guarantees better DP protection as fewer features are sampled, ensuring a low AUC when \textit{SR} is small.



\subsubsection{Ablation Study of DP Masking Module} 
\begin{figure} [t]
	\centering
	\subfloat[ \footnotesize On MNIST]{  	\label{fig:ablationmsebetaonmnist}
		\includegraphics[scale=0.26]{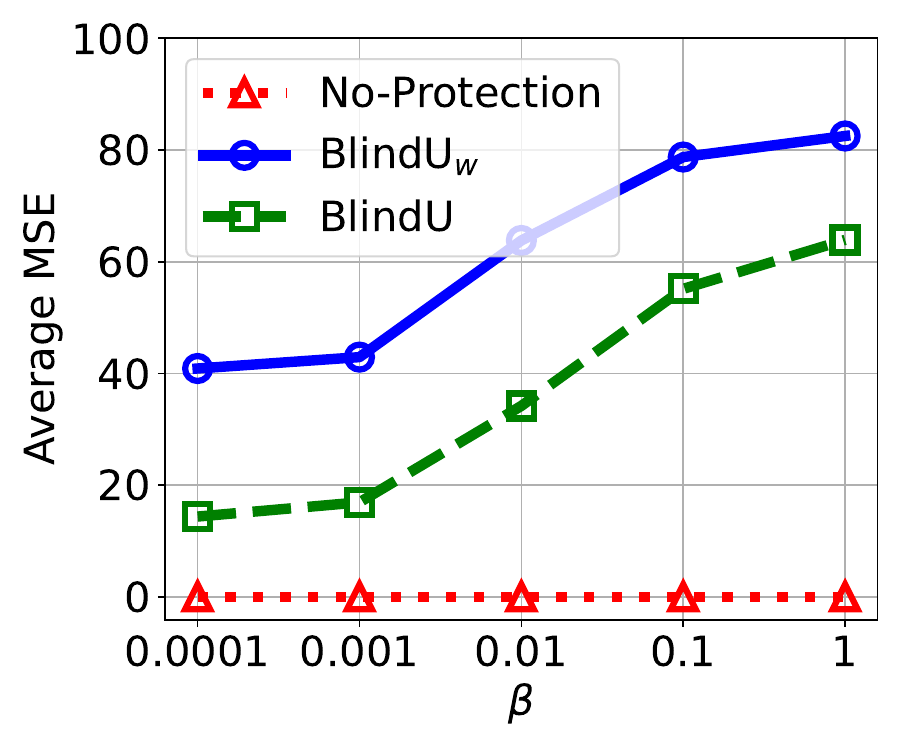}
	}
	\subfloat[ \footnotesize On CIFAR10 ]{  	\label{fig:ablationmsebetaoncifar10}
		\includegraphics[scale=0.26]{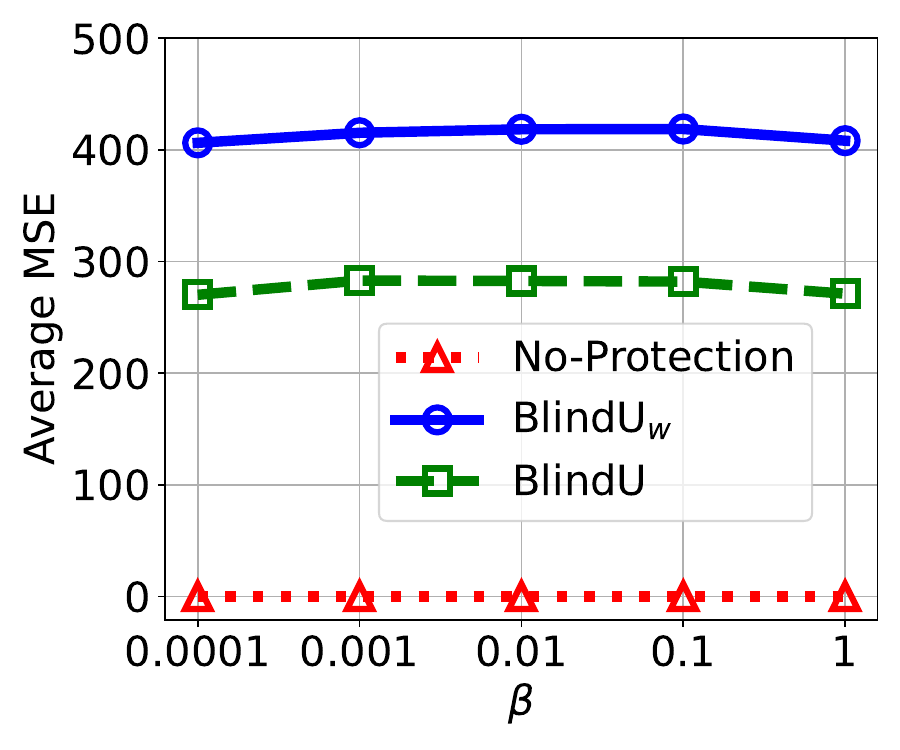}
	}
	\vspace{-2mm}
	\caption{Ablation study of defense against reconstruction attacks about the pure BlindU that only contains the compressive protection module and the BlindU$_{\text{w}}$ that includes both compressive protection module and DP masking module.
	} 
	\label{ablation_reconstruction_MSE} 
\end{figure}

We also conduct ablation studies to analyze the impact of the two components of BlindU, specifically the DP masking module and the compression module, on privacy protection. Since the core of BlindU is the compression-based unlearning mechanism, evaluating the DP masking module in isolation is not very informative, as masking alone cannot implement unlearning. Therefore, we test the pure BlindU that only contains the compression module and BlindU$_{\text{w}}$ that includes both compression and DP masking modules. We set $\textit{SR}=60\%$ for the DP masking, which ensures $(0.597,0.452)$-DP protection. The results of defending against reconstruction attacks are presented in \Cref{ablation_reconstruction_MSE}.
	

In \Cref{ablation_reconstruction_MSE}, both BlindU and BlindU$_{\text{w}}$ demonstrate significant improvements in defending against reconstruction attacks. On the MNIST dataset, even at small values of $\beta$, BlindU provides a comparable defense to that of the DP masking module, with its effectiveness increasing substantially as $\beta$ increases. Similarly, on CIFAR10, the pure BlindU shows a marked improvement in defense as $\beta$ progresses from 0.0001 to 1. Overall, both modules contribute to privacy protection, with the compressive protection module generally offering superior defense.

\subsection{Ablation Study for Unlearning Effectiveness} \label{ablation_study_eval}


Previous experiments have analyzed the privacy protection of the compressive rate $\beta$ and the masking sampling rate $\textit{SR}$. In this section, we analyze the influence of unlearning of these variables, and we will fix the value of other variables when we conduct experiments to evaluate one variable. 





\begin{figure}[t]
	\centering
	\hspace{-5mm}
	\subfloat{ \label{fig:mnistmibetacurve} \rotatebox{90}{\hspace{7mm}	\scriptsize{ On MNIST} }
		\includegraphics[scale=0.205]{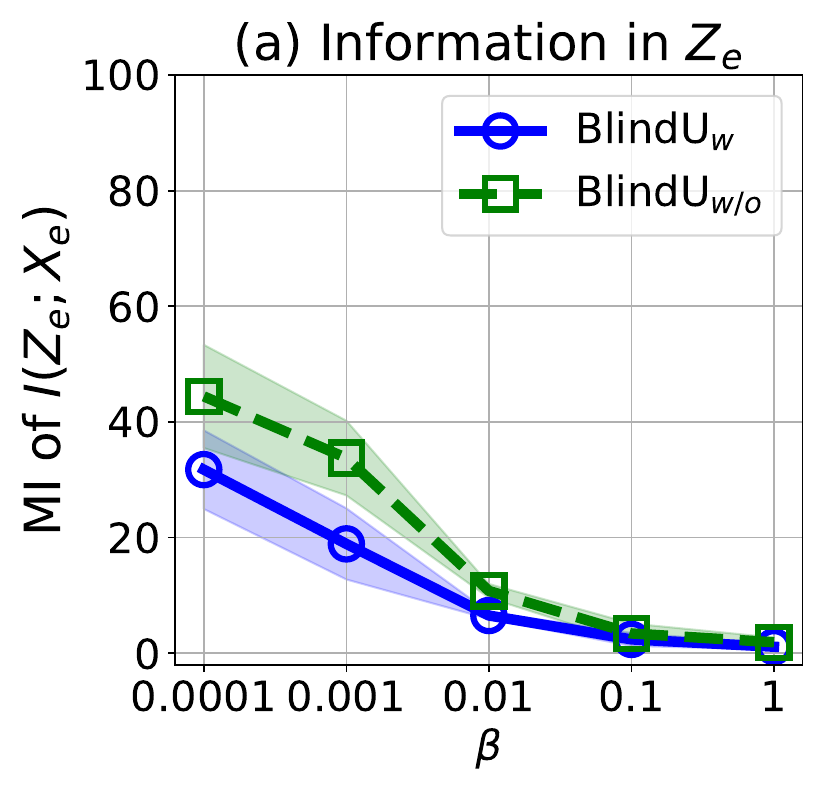}
	}			
	\hspace{-5mm}
	\subfloat{ 	\label{fig:mnistbackaccbetacurve}
		\includegraphics[scale=0.205]{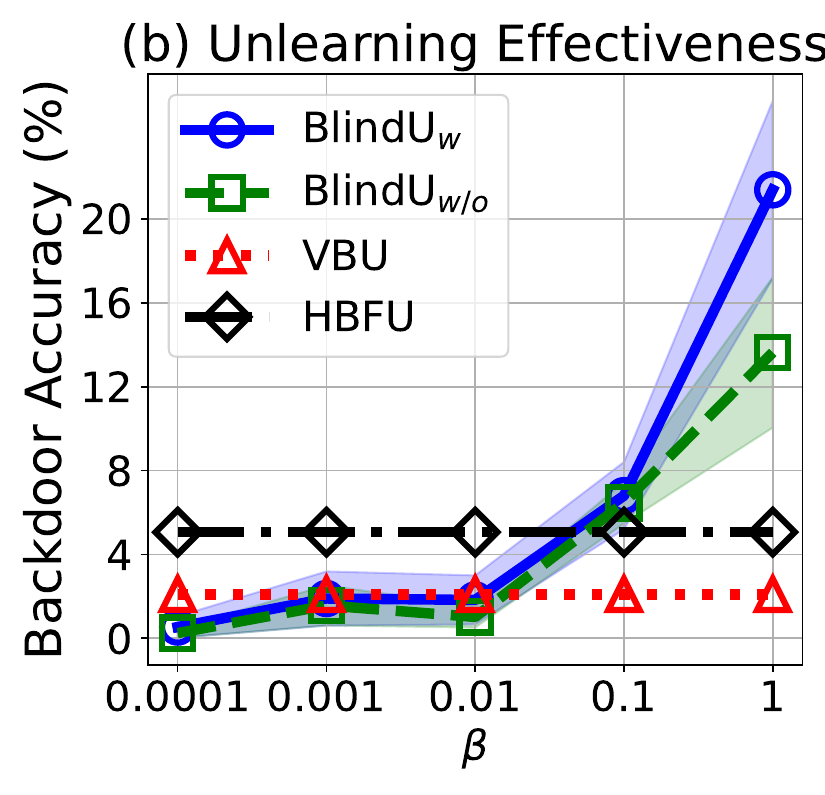}
	}
	\hspace{-5mm}
	\subfloat{ \label{fig:mnistaccbetacurve}
		\includegraphics[scale=0.205]{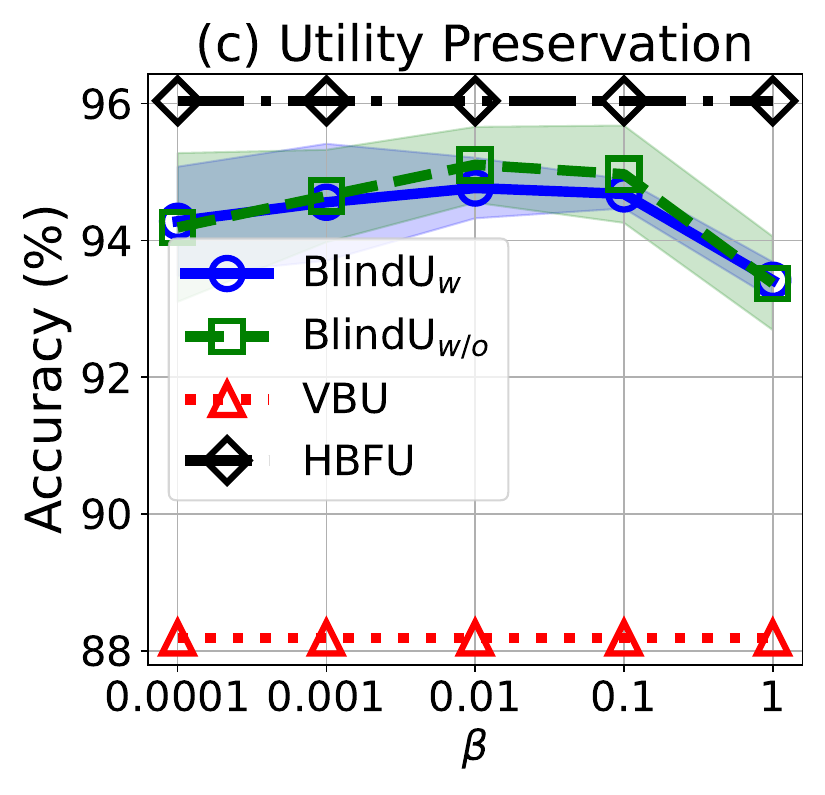}
	}\\
	\vspace{-3mm}
	\hspace{-5mm}
	\subfloat{ \label{fig:cifarmibetacurve}  \rotatebox{90}{\hspace{6mm}		\scriptsize{ On CIFAR10} }
		\includegraphics[scale=0.205]{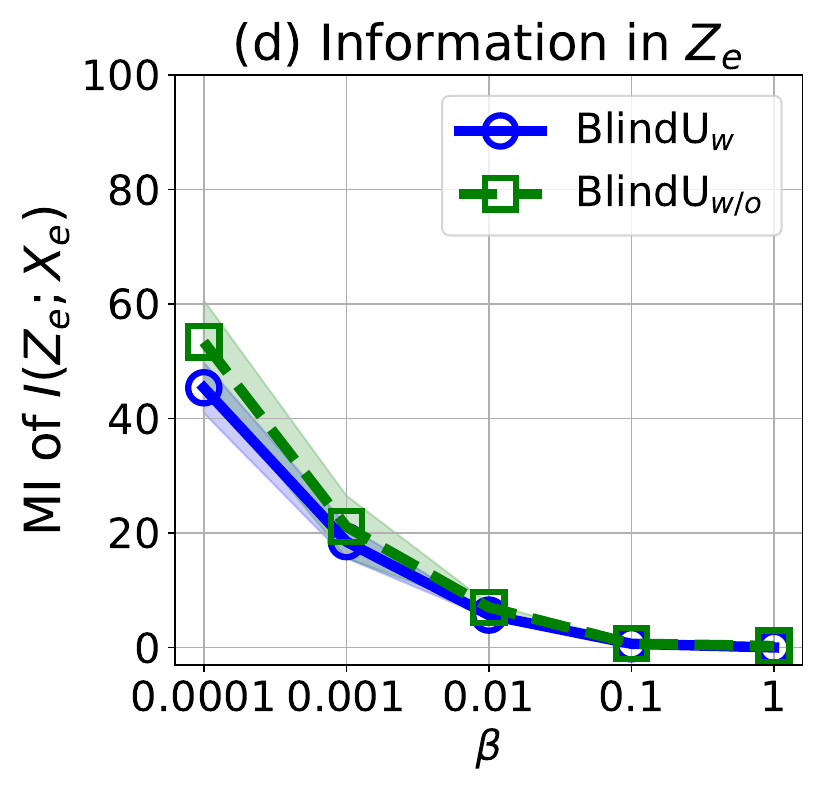}
	}
	\hspace{-5mm}
	\subfloat{ \label{fig:cifarbackaccbetacurve}
		\includegraphics[scale=0.205]{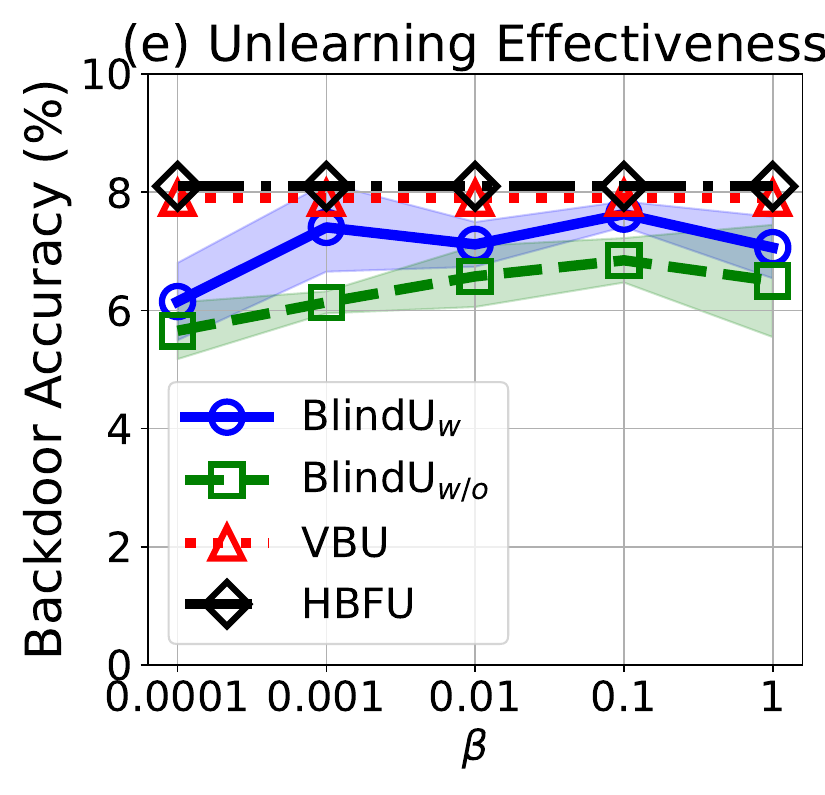}
	}
	\hspace{-5mm}
	\subfloat{ \label{fig:cifaraccbetacurve}
		\includegraphics[scale=0.205]{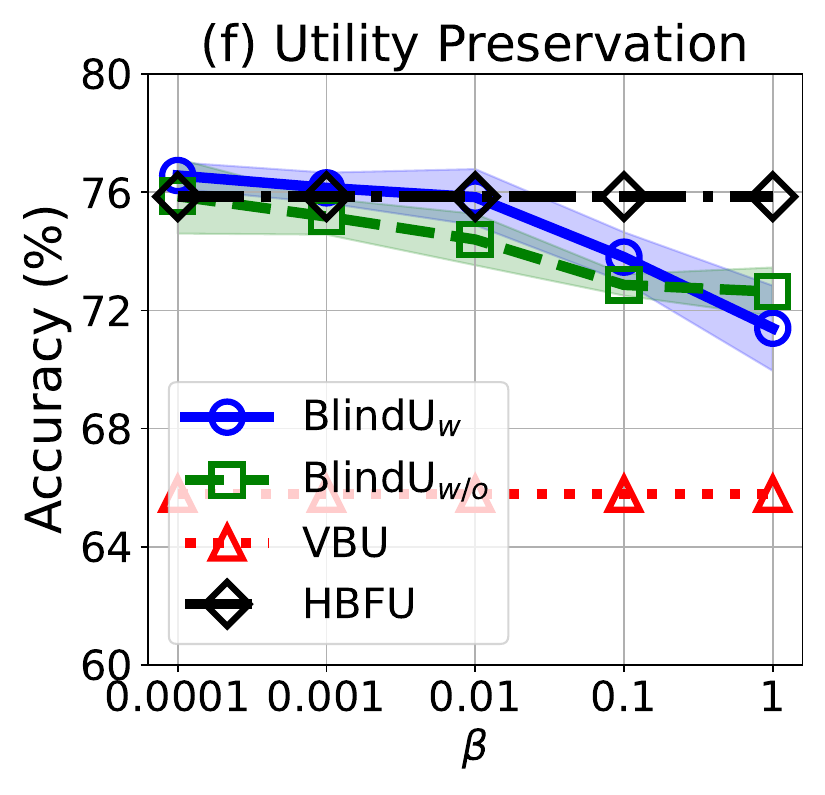}
	}\\
	\vspace{-3mm}
	\hspace{-5mm}
	\subfloat{ 	\label{fig:cifar100mibetacurve} \rotatebox{90}{\hspace{6mm}		\scriptsize{ On CIFAR100} }
		\includegraphics[scale=0.205]{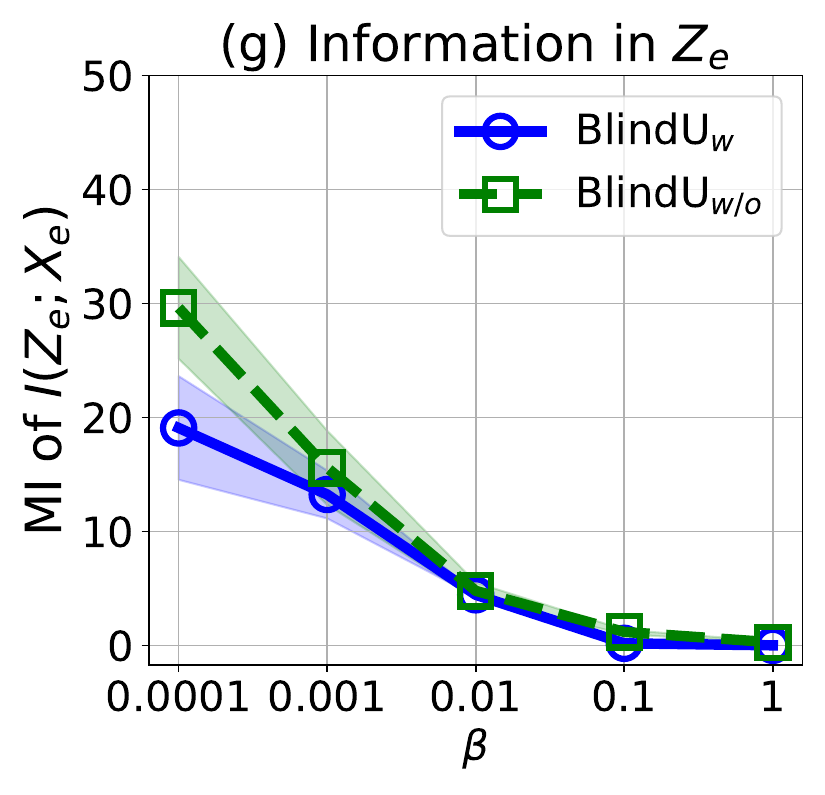}
	}
	\hspace{-5mm}
	\subfloat{ \label{fig:cifar100backaccbetacurve}
		\includegraphics[scale=0.205]{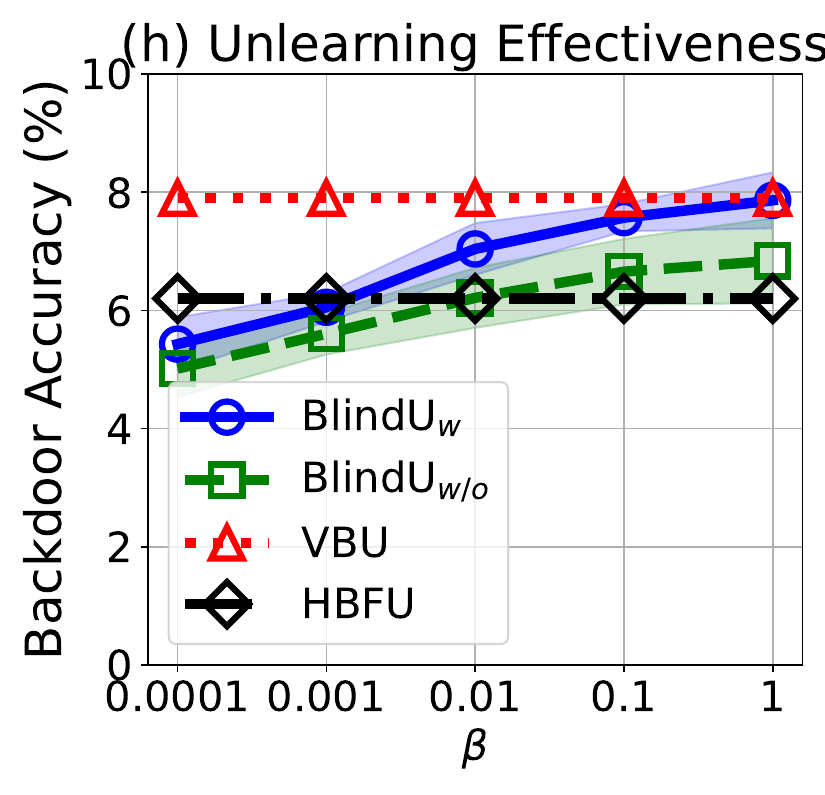}
	}
	\hspace{-5mm}
	\subfloat{ 	\label{fig:cifar100accbetacurve}
		\includegraphics[scale=0.205]{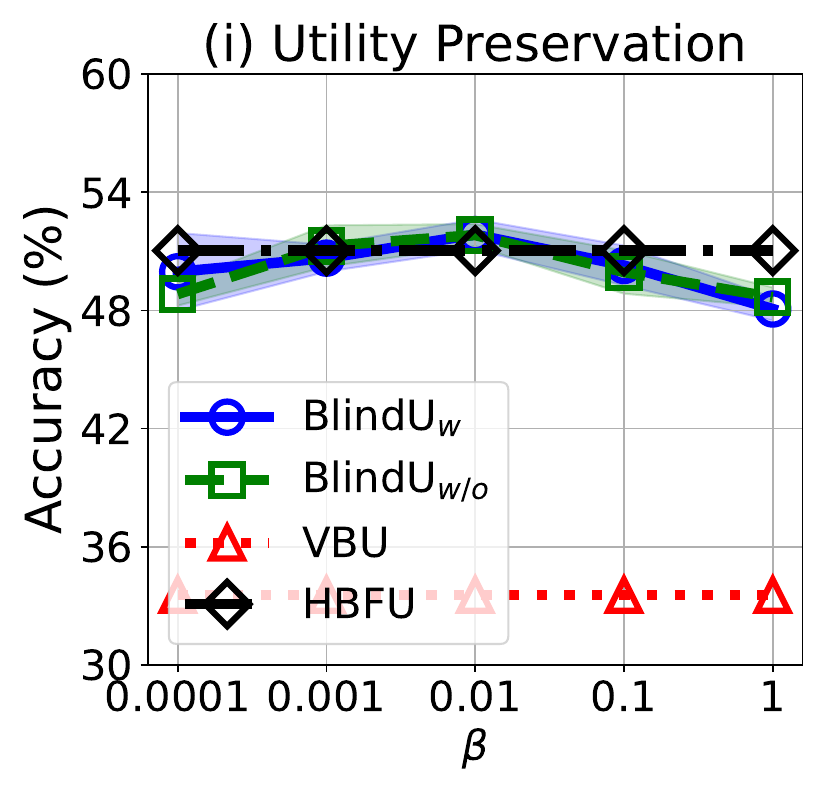}
	}\\
\vspace{-3mm}
\hspace{-6mm}
\subfloat{ 		\label{fig:tinyimibetacurve}   \rotatebox{90}{\hspace{2mm}		\scriptsize{On TinyImageNet}  }
\includegraphics[scale=0.205]{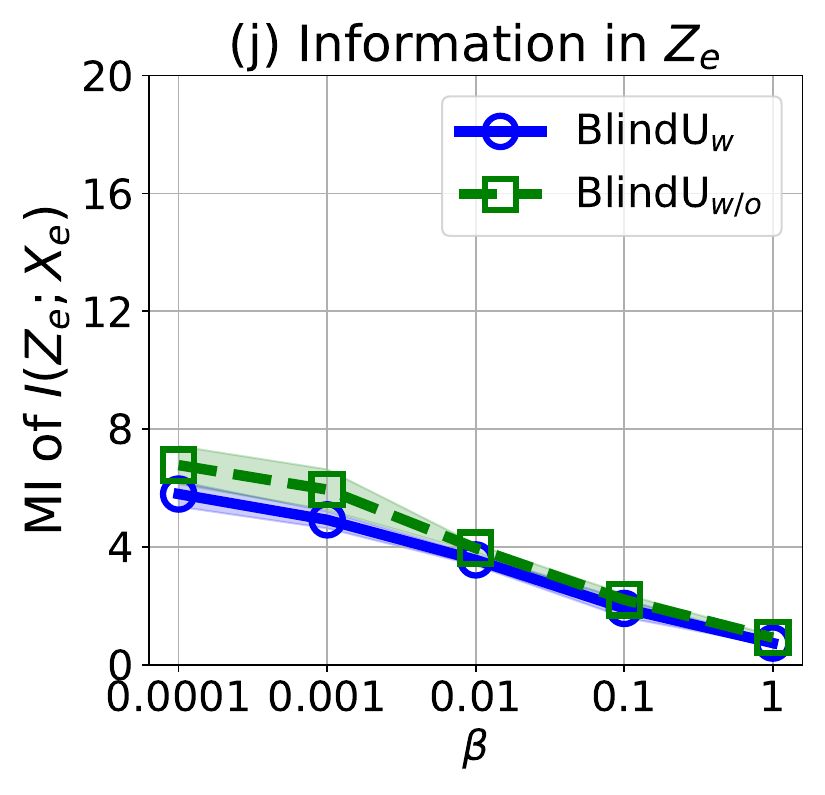}
}
\hspace{-5mm}
\subfloat{  	\label{fig:tinyibackaccbetacurve}
\includegraphics[scale=0.205]{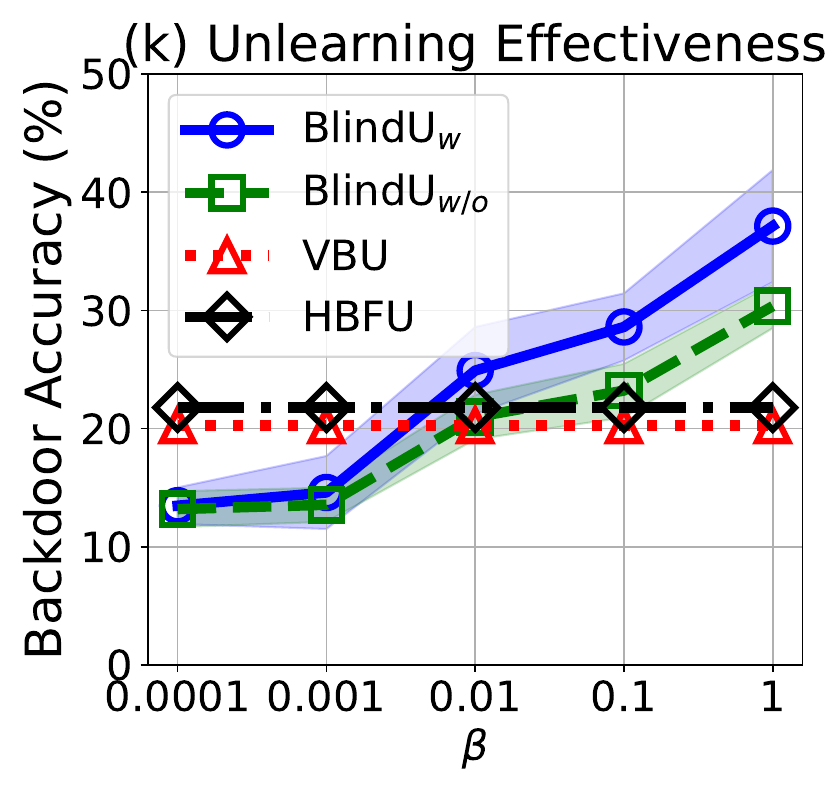}
}
\hspace{-5mm}
\subfloat{ 	\label{fig:tinyiaccbetacurve}
\includegraphics[scale=0.205]{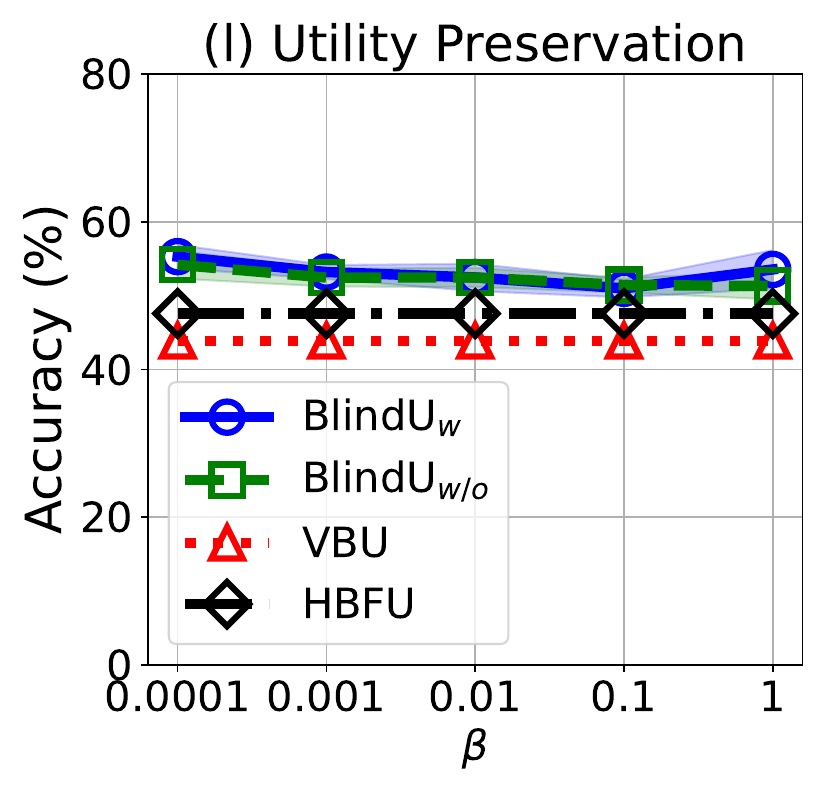}
}
\vspace{-3mm}
	\caption{{Evaluations of effect about different $\beta$}. \vspace{-4mm} } 
	\label{evaluation_of_beta} 
\end{figure}

\subsubsection{Impact of Compressive Rate $\beta$}



As we introduced before, a larger compressive rate $\beta$ will distort more input data information from the representation, making the uploaded compressive data expose less information. \Cref{evaluation_of_beta} shows the influence of $\beta$ in both BlindU$_{\text{w}}$ and BlindU$_{\text{w/o}}$ on MNIST, CIFAR10, CIFAR100, and TinyImageNet. When evaluating the influence of $\beta$ we fix the $\textit{SR}=60\%$ and $\textit{EDR}=6\%$. Since HBFU and VBU do not have this parameter, we keep a fixed performance of HBFU and VBU at $\textit{EDR}=6\%$.

\noindent
\textbf{Impact on Unlearning Effectiveness.}
{\Cref{evaluation_of_beta} shows the relationship between upload information and unlearning effectiveness.} In the first two columns in \Cref{evaluation_of_beta}, as $\beta$ increases, the higher compression makes the representation $Z_e$ contain lesser information about the erasing dataset. Hence, the MI decreases as $\beta$ increases, as shown in \Cref{fig:mnistmibetacurve,fig:cifarmibetacurve,fig:cifar100mibetacurve,fig:tinyimibetacurve} which guarantees better privacy protection. 
At the same time, when the compression rate $\beta$ increases, the model utility has a slight drop as shown in \Cref{fig:mnistaccbetacurve,fig:cifaraccbetacurve,fig:cifar100accbetacurve,fig:tinyiaccbetacurve}. In the middle column of \Cref{evaluation_of_beta}, it shows that the whole model becomes harder to unlearn because the compressor with a larger $\beta$ distorts the original data more, resulting in the compressed representation $Z_e$ contain less information for implementing unlearning. 



\begin{figure}[t]
	\centering
	\subfloat{ \label{fig:mnistrtbetabar} 
		\includegraphics[scale=0.25]{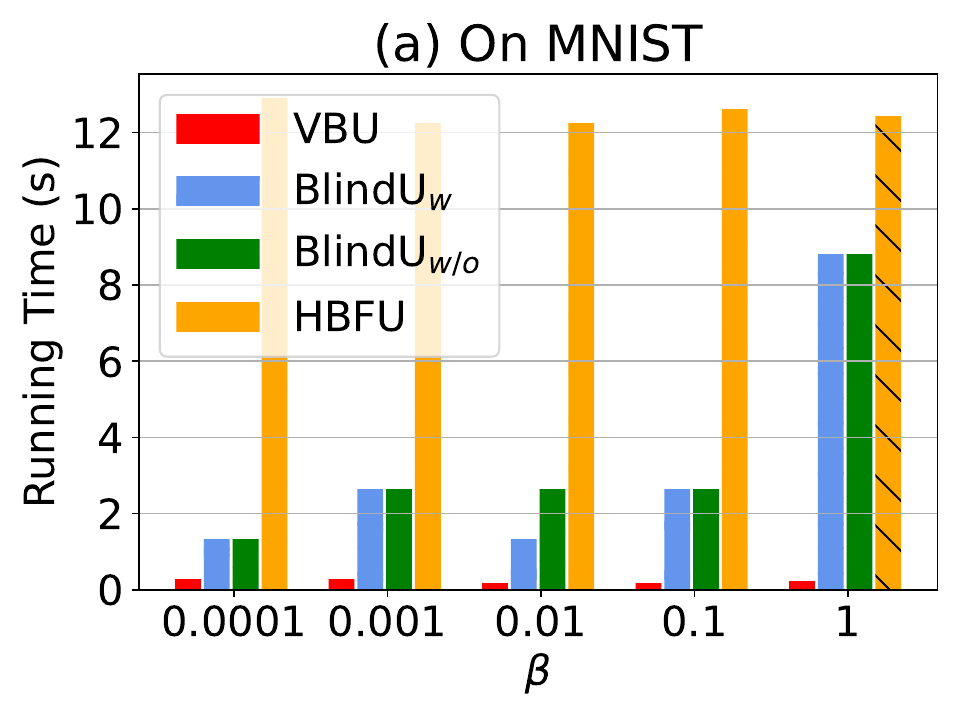}
	}
	\subfloat{ \label{fig:cifarrtbetabar} 
		\includegraphics[scale=0.25]{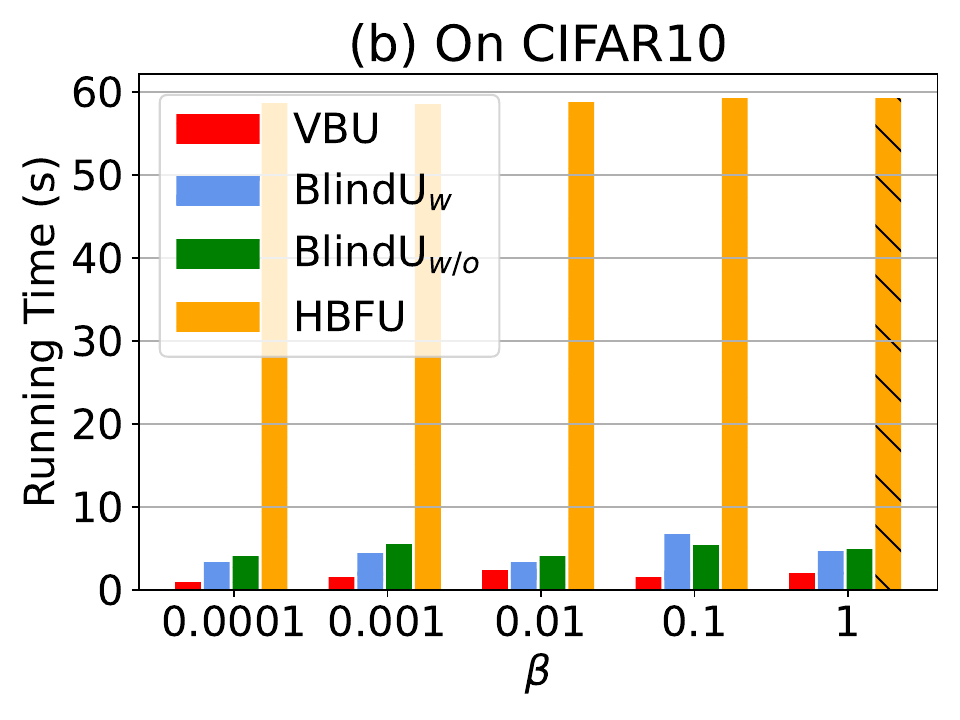}
	}
	\vspace{-2mm}
	\caption{Evaluations of efficiency about different $\beta$. \vspace{-2mm}}
	\label{fig_runingtime_beta}
\end{figure}

\noindent
\textbf{Impact on Unlearning Efficiency.} \Cref{fig_runingtime_beta} shows that HBFU consumes the maximum, and VBU consumes the minimum running time {because HBFU relies on training using the FL framework with other users while VBU implements unlearning only using the erased samples by the server.} BlindU takes more running time as $\beta$ increases, but it is always less than HBFU. {In BlindU, the running time slightly increases as the $\beta$ increases because the larger beta means stronger data compression, offering less information to implement unlearning. Therefore, it will take more computation costs to achieve a satisfactory unlearning effect.}


\begin{figure}[t]
	\centering
	\hspace{-5mm}
	\subfloat{ \label{fig:mnistmiepsiloncurve}  \rotatebox{90}{ \hspace{6mm}	\scriptsize{ On MNIST} }
		\includegraphics[scale=0.205]{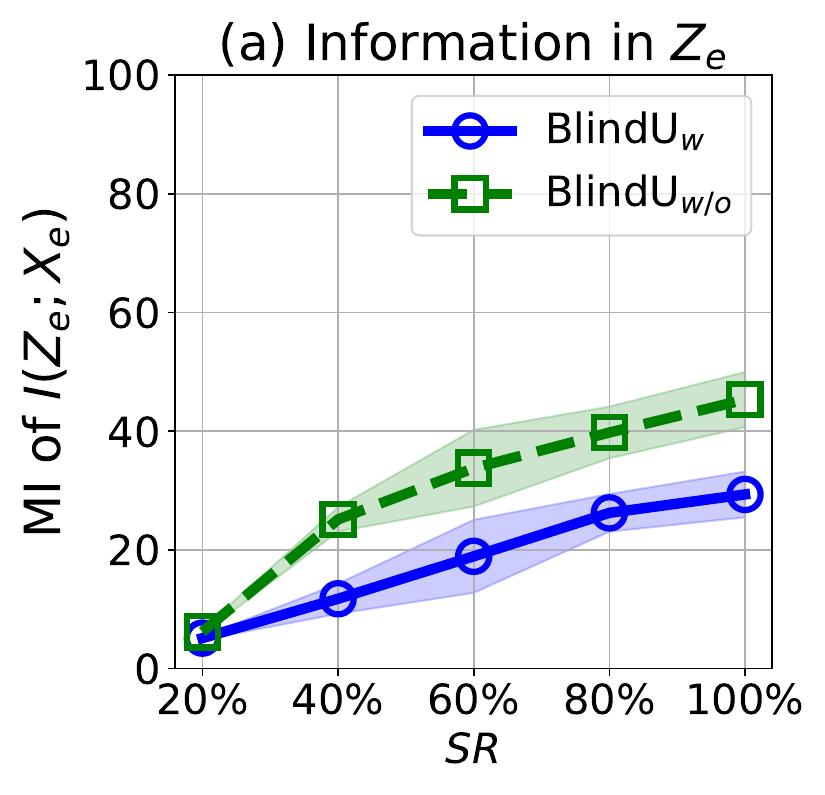}
	}
	\hspace{-5mm}
	\subfloat{ \label{fig:mnistbackaccepsiloncurve}
		\includegraphics[scale=0.205]{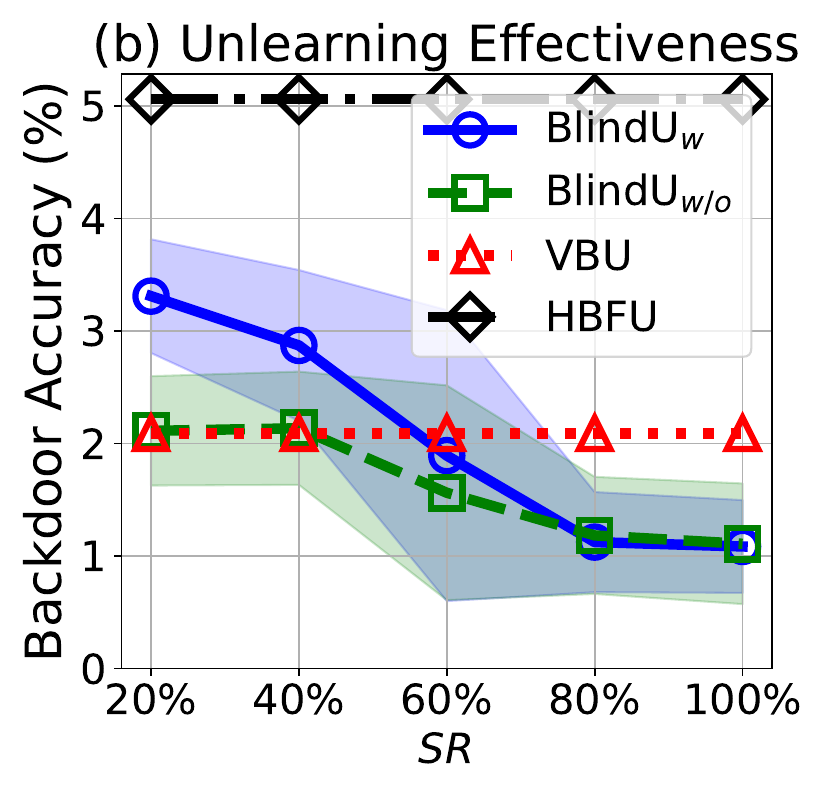}
	}
	\hspace{-5mm}
	\subfloat{ \label{fig:mnistaccepsiloncurve}
		\includegraphics[scale=0.205]{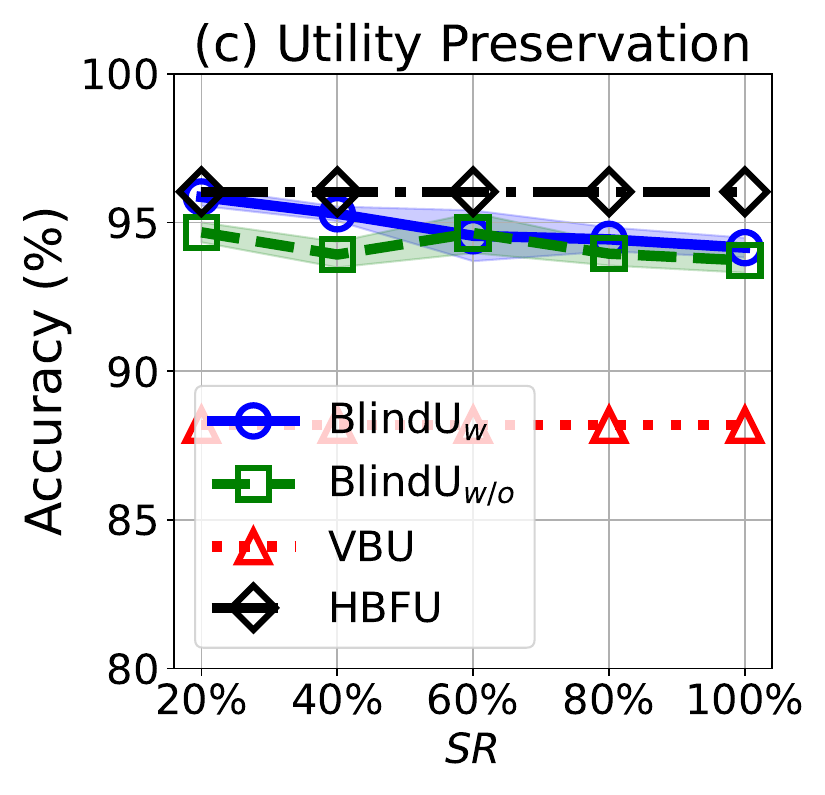}
	} \\
\vspace{-3mm}
	\hspace{-5mm}
	\subfloat{ \label{fig:cifarmiepsiloncurve} \rotatebox{90}{\hspace{7mm}\scriptsize{ On CIFAR10} }
		\includegraphics[scale=0.205]{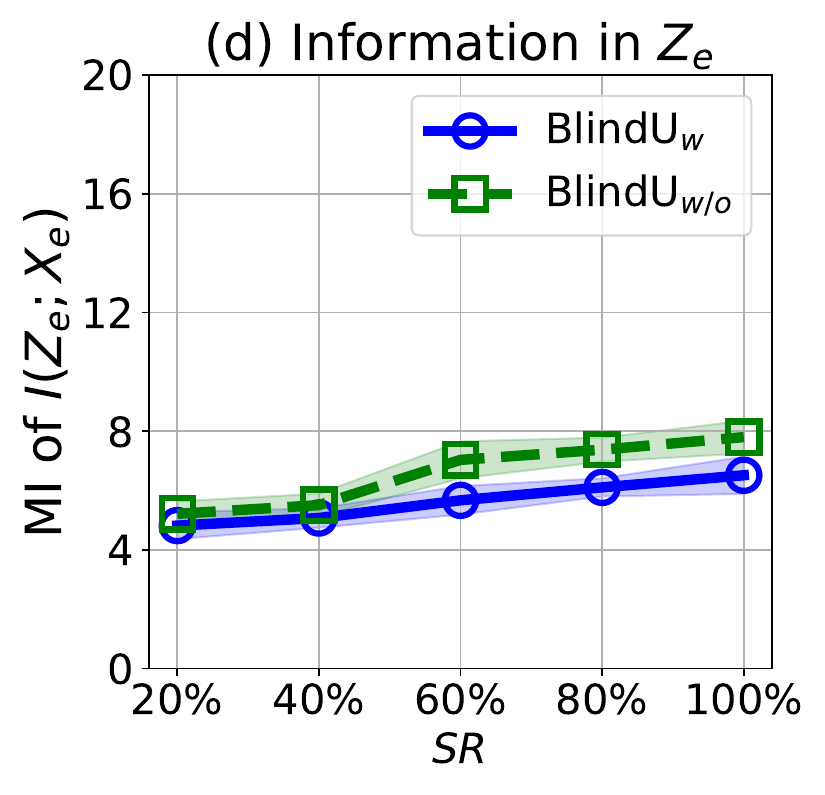}
	}
	\hspace{-5mm}
	\subfloat{ \label{fig:cifarbackaccepsiloncurve}
		\includegraphics[scale=0.205]{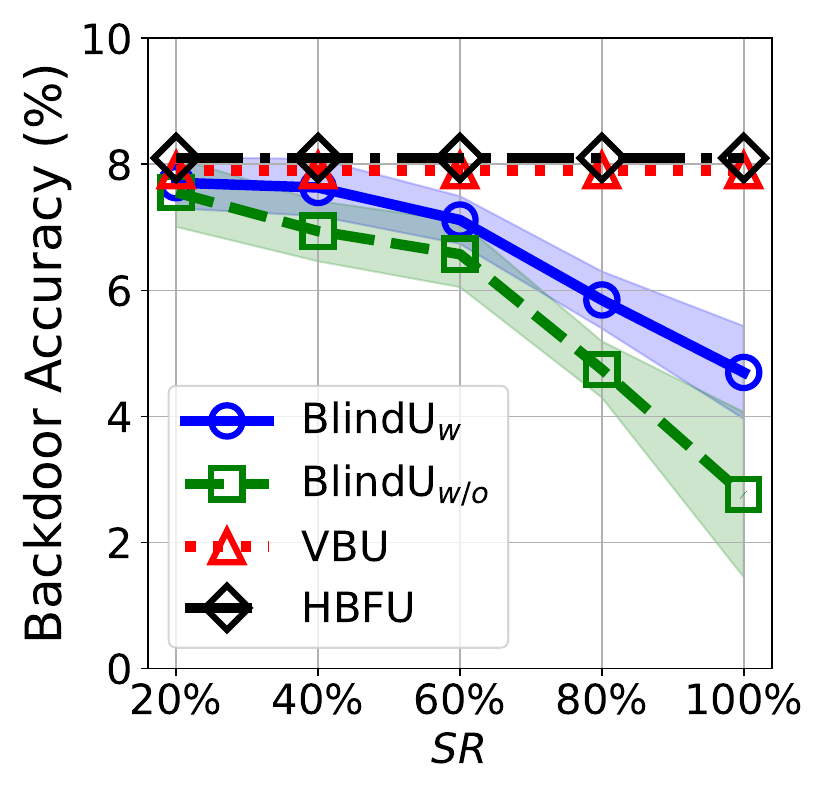}
	}
	\hspace{-5mm}
	\subfloat{ \label{fig:cifaraccepsiloncurve}
		\includegraphics[scale=0.205]{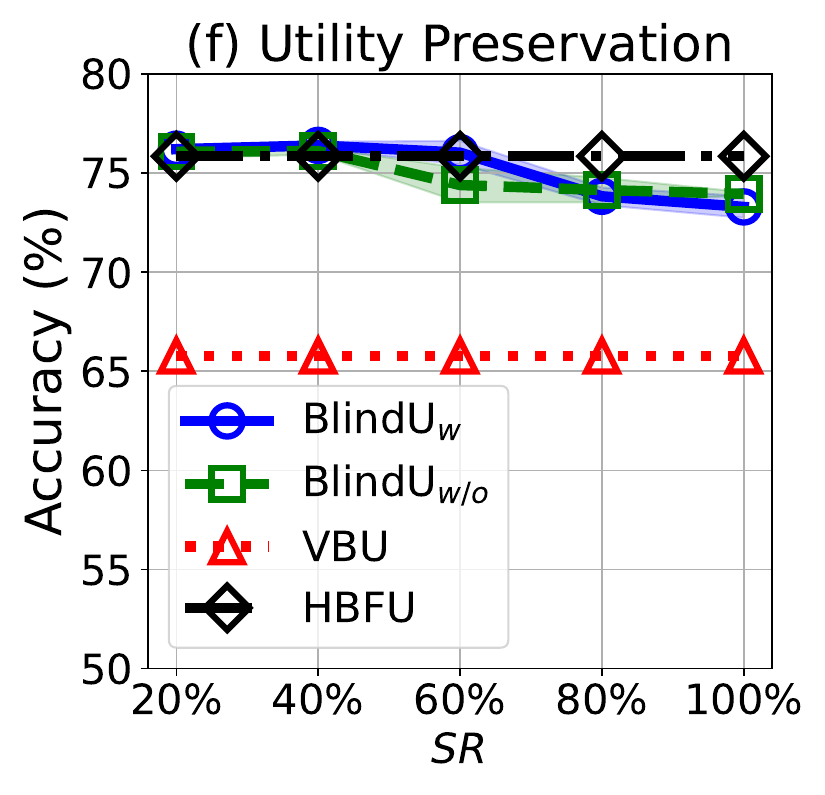}
	}\\
\vspace{-3mm}
	\hspace{-5mm}
	\subfloat{ \label{fig:cifar100miepsiloncurve} \rotatebox{90}{\hspace{6mm}	\scriptsize{ On CIFAR100} }
		\includegraphics[scale=0.205]{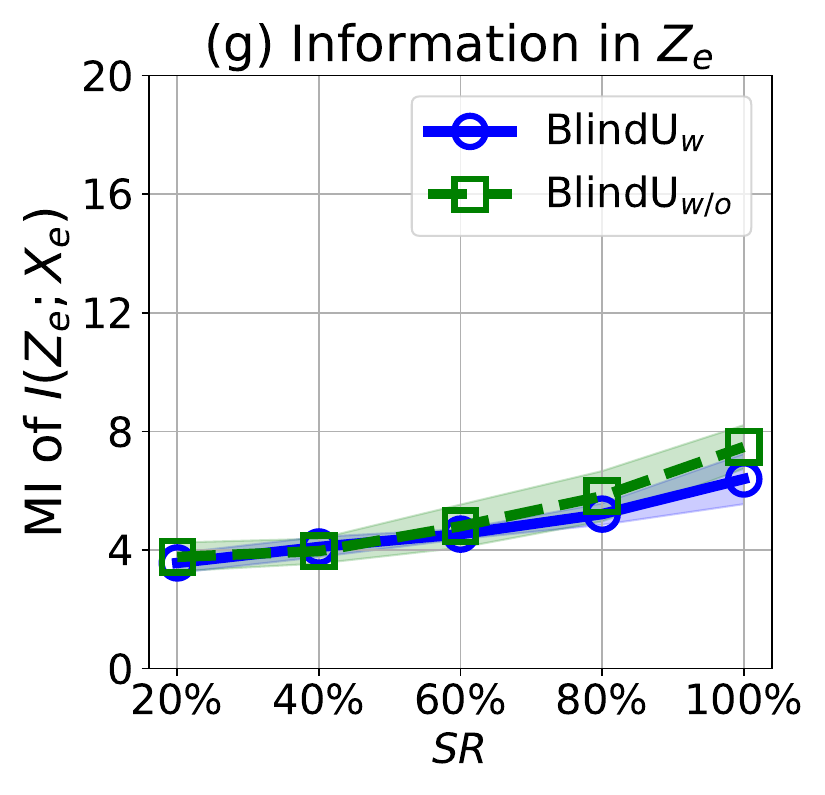}
	}
	\hspace{-5mm}
	\subfloat{ \label{fig:cifar100backaccepsiloncurve}
		\includegraphics[scale=0.205]{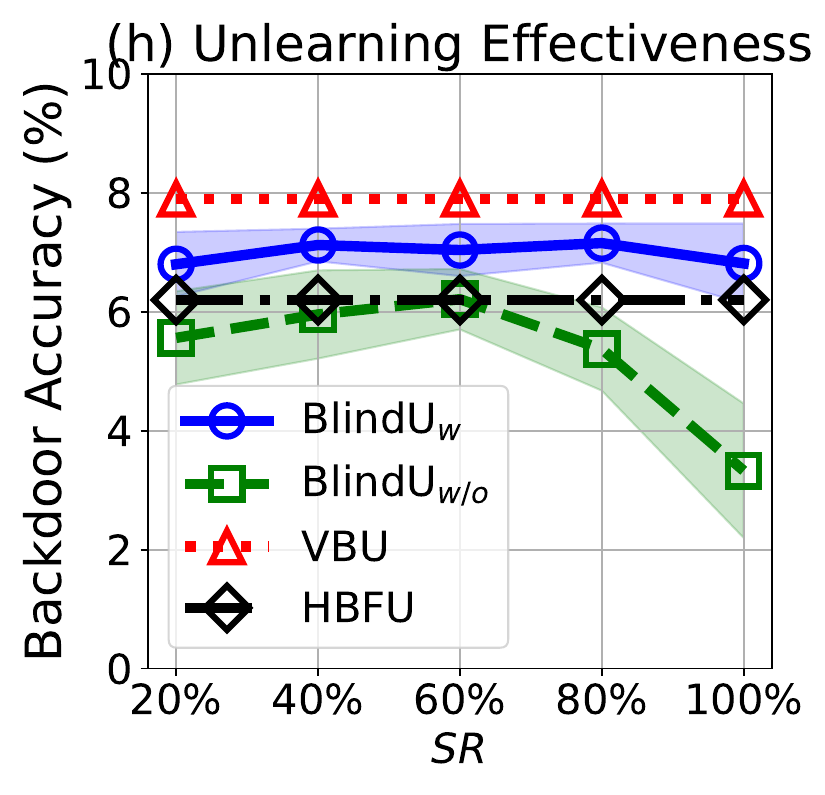}
	}
	\hspace{-5mm}
	\subfloat{ \label{fig:cifar100accepsiloncurve}
		\includegraphics[scale=0.205]{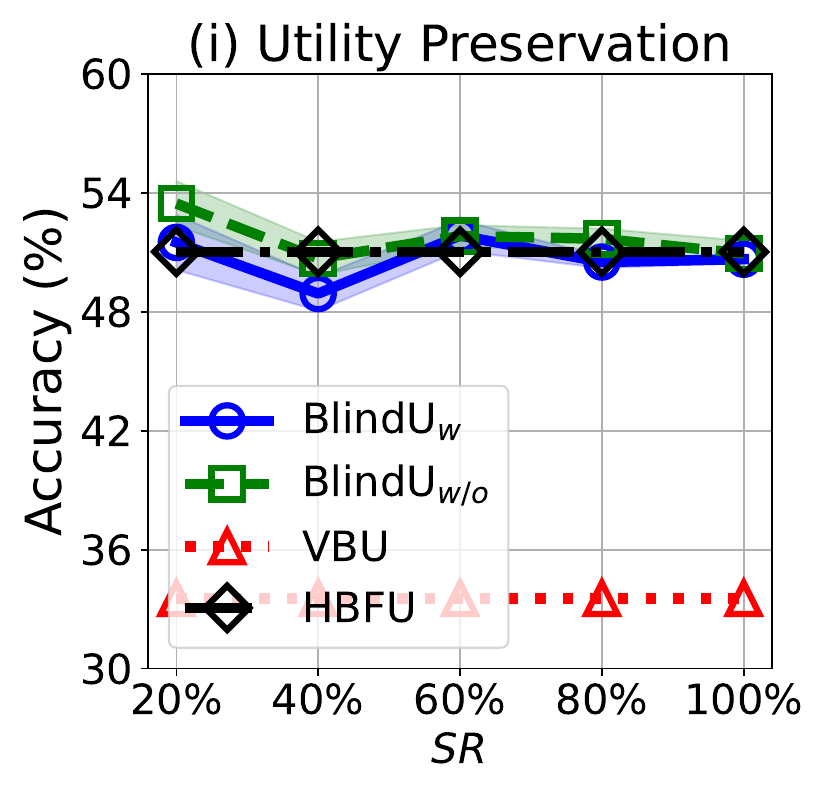}
	}\\
\vspace{-3mm}
\hspace{-6mm}
\subfloat{ 		\label{fig:tinyimiepsiloncurve}   \rotatebox{90}{\hspace{2mm}		\scriptsize{On TinyImageNet}  }
\includegraphics[scale=0.205]{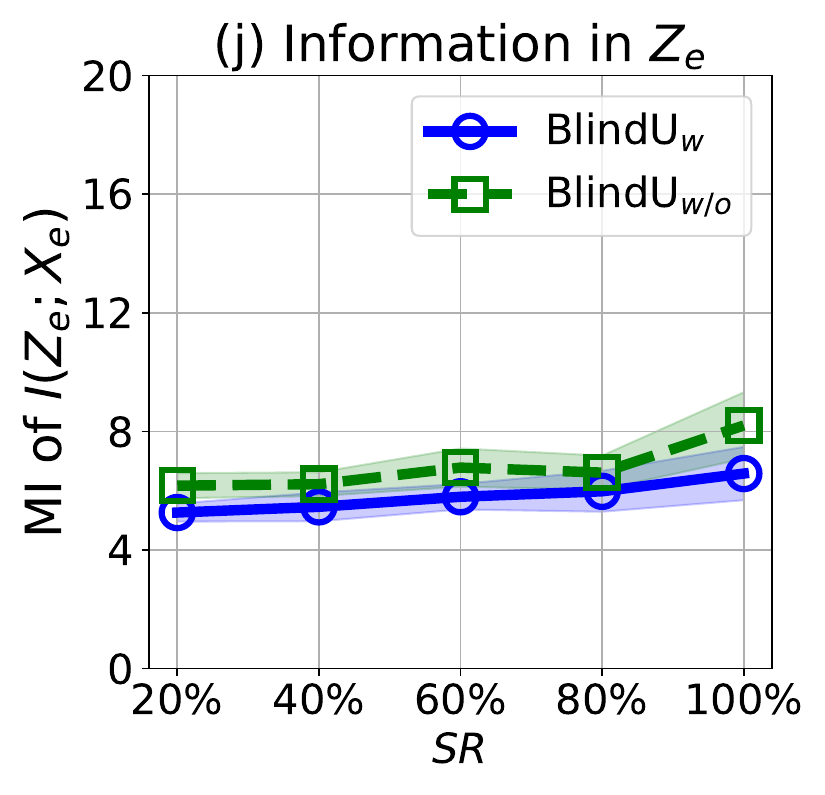}
}
\hspace{-5mm}
\subfloat{  	\label{fig:tinyibackaccepsiloncurve}
\includegraphics[scale=0.205]{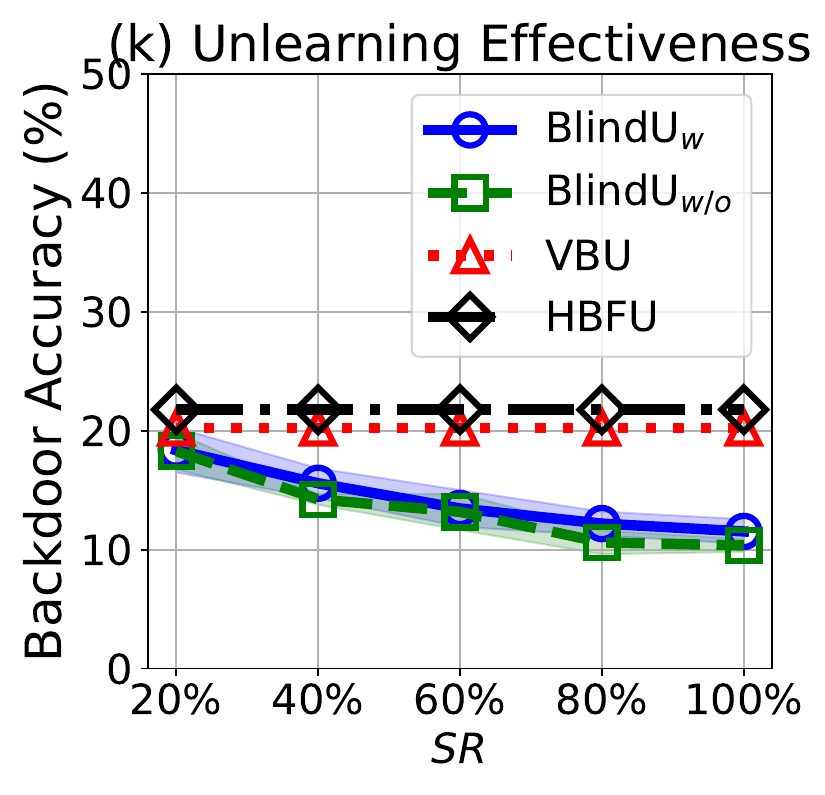}
}
\hspace{-5mm}
\subfloat{ 	\label{fig:tinyiaccepsiloncurve}
\includegraphics[scale=0.205]{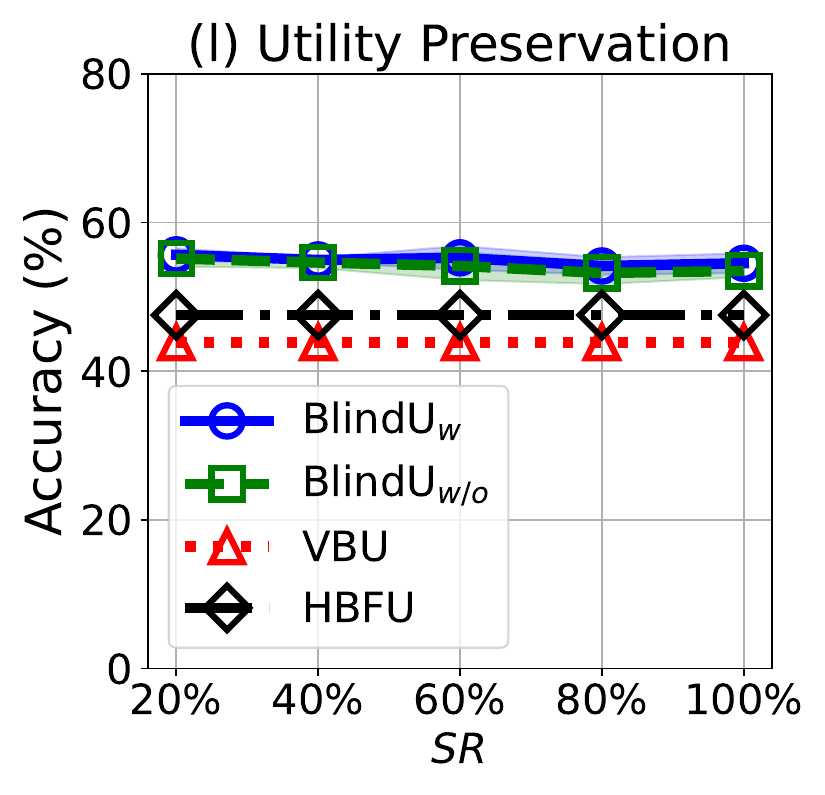}
}
\vspace{-3mm}
	\caption{{ Evaluations of effect about different $\textit{SR}$}. \vspace{-4mm}
	} 
	\label{evaluation_of_epsilon} 
\end{figure}

 \subsubsection{Impact of DP Masking with Sampling}

\Cref{evaluation_of_epsilon} presents effectiveness evaluation of BlindU with different sampling rate $\textit{SR}$ on MNIST, CIFAR10, CIFAR100 and TinyImageNet. Here, we fix the other variables and show the performance of HBFU and VBU at $\textit{EDR}=6\%$. When $\textit{SR}$ is small, the sampled features are fewer and masked features are more. Therefore, lesser information is contained in the input, and the MI will be smaller, as shown in \Cref{fig:mnistmiepsiloncurve,fig:cifarmiepsiloncurve,fig:cifar100miepsiloncurve,fig:tinyimiepsiloncurve}. As $\textit{SR}$ increases, more information is contained in the representation and the mutual information increases, too. At the same time, as shown in \Cref{fig:mnistbackaccepsiloncurve,fig:cifarbackaccepsiloncurve,fig:cifar100backaccepsiloncurve,fig:tinyibackaccepsiloncurve}, more information contained in the representation makes the forgetting effectiveness better. The backdoor accuracy decreases as the $\textit{SR}$ increases. Meanwhile, the model accuracy decreases slightly, as shown in the third column in \Cref{evaluation_of_epsilon}.

%% file: Contents/6_summary.tex
\section{Summary and Future Work}
\label{summary}


In this paper, we investigate the problem of privacy-preserving machine unlearning without revealing erasing data to the server. We propose the BlindU method to protect the privacy of erased data and implement effective unlearning. In BlindU, the unlearning user executes DP masking using two sampling strategies and compression using the compressor of the trained IB model, preparing the dual privacy-preserving representation. Then, the server solves the blind unlearning problem as a constrained retraining with two customized unlearning modules based on the received compressed representation. We provide a comprehensive theoretical analysis and conduct extensive experiments for BlindU, and the results demonstrate its superiority in both privacy protection for erasing data and unlearning effectiveness compared to existing approaches.

As machine unlearning grows in importance, our research lays a foundation for advancing privacy protection in the unlearning process. However, the privacy-preserving machine unlearning domain remains largely unexplored, offering significant opportunities for future investigation. One important open challenge lies in the potential privacy leakage that may occur when users interact with or download server-side model parameters during unlearning. Beyond this, it is promising to explore how a representation-based, privacy-preserving unlearning paradigm can be adapted to Large Language Models (LLMs) unlearning, where forgetting is typically defined over prompts and internal knowledge rather than fixed class labels \citep{wang2025rethinking,li2025llm,wang2025towards,zhang2024negative}. A potential adaptation is to compress token- and sequence-level hidden states and upload the resulting representations, then define forgetting objectives that reduce their dependence on the corresponding textual targets. In addition, real-world unlearning requests may involve concepts that may exhibit label and domain mismatch, in which case label-based forgetting objectives can be insufficient~\citep{zhu2024decoupling}. Extending BlindU to these scenarios may require augmenting the label-based constraint with a concept-targeted objective to explicitly forget the target concept.
